\documentclass[a4paper,11pt]{article}
\pdfoutput=1
\usepackage{jcappub}
\usepackage{mathtools}
\usepackage{multirow}

\graphicspath{{plots/}}

\newcommand{\Hrh}{H_\text{rh}}
\newcommand{\Trh}{T_\text{rh}}

\newcommand{\gs}{g_\star}
\newcommand{\gss}{g_{\star s}}
\newcommand{\arh}{a_\text{rh}}
\newcommand{\rp}{\rho_\chi}
\newcommand{\rR}{\rho_R}
\newcommand{\rgw}{\rho_\text{GW}}
\newcommand{\Ogw}{\Omega_\text{GW}}

\title{Primordial Gravitational Waves from Phase Transitions during Reheating }
\author[a,b]{Amitayus Banik,}
\author[c]{Nicolás Bernal}
\author[d]{and Fazlollah Hajkarim}

\affiliation[a]{Department of Physics, Chungbuk National University\\
Cheongju, Chungbuk 28644, Korea}
\affiliation[b]{Research Institute for Nanoscale Science and Technology,
Chungbuk National University\\
Cheongju, Chungbuk 28644, Korea}
\affiliation[c]{New York University Abu Dhabi\\
PO Box 129188, Saadiyat Island, Abu Dhabi, United Arab Emirates}
\affiliation[d]{Department of Physics and Astronomy, University of Oklahoma\\
Norman, OK 73019, USA}

\emailAdd{abanik@cbnu.ac.kr}
\emailAdd{nicolas.bernal@nyu.edu}
\emailAdd{fazlollah.hajkarim@ou.edu}

\abstract{We study primordial gravitational waves (GWs) generated from first-order phase transitions (PTs) during cosmic reheating. Using a minimal particle physics model, and a general parametrization of the inflaton energy density and the evolution of the Standard Model temperature, we explore the conditions under which PTs occur and determine the corresponding PT parameters (the PT temperature, duration and strength), which depend on the evolution of the background during reheating. We find that, in certain cosmological scenarios, PTs can be delayed and prolonged compared to the standard post-inflationary evolution. Incorporating these PT parameters, we compute the resulting GW spectrum generated from the various processes occurring during a first-order PT. We find that, in comparison to the standard cosmological history, the GW amplitude and peak frequency can be modified by several orders of magnitude due to the cosmological evolution during reheating. In particular, the GW spectra could be within the reach of next-generation GW observatories.}

\begin{document}
\def\lsim{\:\raisebox{-0.5ex}{$\stackrel{\textstyle<}{\sim}$}\:}
\def\gsim{\:\raisebox{-0.5ex}{$\stackrel{\textstyle>}{\sim}$}\:}

\begin{flushright}
\end{flushright}
\maketitle

\section{Introduction}
The detection of gravitational waves (GWs)~\cite{LIGOScientific:2016aoc, LIGOScientific:2017vwq, LIGOScientific:2017ync, LIGOScientific:2018mvr} has contributed to improving our understanding of the Universe. In particular, the detection of cosmological sources of GWs~\cite{Maggiore:1999vm, Caprini:2018mtu} -- including collapsing domain walls~\cite{Saikawa:2017hiv}, radiating cosmic strings~\cite{Damour:2004kw}, and phase transitions~\cite{Caprini:2019egz} -- by future detectors, such as LISA~\cite{LISA:2017pwj}, the Cosmic Explorer (CE)~\cite{Reitze:2019iox}, BBO~\cite{Crowder:2005nr, Corbin:2005ny, Harry:2006fi}, and the International Pulsar Timing Array (IPTA)~\cite{Antoniadis:2022pcn, IPTA:2023ero}, may offer valuable probes of the early Universe, thus providing a window into our so-far unexplored cosmological history. In this work, we study the dynamics of first-order phase transitions (PTs) occurring during the cosmic (re)heating era after inflation, and the impact on the generated GW spectrum.\footnote{Other sources of GW production during reheating can be perturbative, typically involving decays and annihilations of the inflaton~\cite{Ema:2015dka, Ema:2016hlw, Nakayama:2018ptw, Huang:2019lgd, Ema:2020ggo, Barman:2023ymn, Barman:2023rpg, Kanemura:2023pnv, Bernal:2023wus, Tokareva:2023mrt, Choi:2024ilx, Hu:2024awd, Choi:2024acs, Barman:2024htg, Xu:2024fjl, Inui:2024wgj, Jiang:2024akb, Bernal:2024jim, Bernal:2025lxp}, and non-perturbative~\cite{Caprini:2018mtu}.  Furthermore, the impact of non-standard cosmological scenarios {\it after} reheating on the primordial GW spectrum has recently received particular attention~\cite{Assadullahi:2009nf, Durrer:2011bi, Alabidi:2013lya, Cui:2018rwi, DEramo:2019tit, Bernal:2019lpc, Figueroa:2019paj, Bernal:2020ywq, Frey:2024jqy, Villa:2024jbf}.}

Gravitational waves from PTs (see e.g. Refs.~\cite{Caprini:2015zlo, Caprini:2018mtu, Caprini:2019egz, Athron:2023xlk, Croon:2024mde} for exhaustive reviews) have been well studied in the context of beyond the Standard Model (SM) extensions that modify the electroweak transition, rendering it strongly first order, see e.g. Refs.~\cite{Kamionkowski:1993fg, Apreda:2001us, Grojean:2006bp, Ashoorioon:2009nf, Kakizaki:2015wua, Vaskonen:2016yiu, Dorsch:2016nrg, Beniwal:2017eik, Ellis:2018mja, Chatterjee:2022pxf}, or hidden and dark sectors, cf. Refs.~\cite {Schwaller:2015tja, Jaeckel:2016jlh, Breitbach:2018ddu, Dev:2019njv, Dent:2022bcd, Morgante:2022zvc, Pasechnik:2023hwv, Koutroulis:2023wit, DiBari:2023upq, Feng:2024pab, Banik:2024zwj, Balan:2025uke}. While the general formalism governing GWs sourced from PTs within non-standard cosmological backgrounds is well established~\cite{Allahverdi:2020bys, Gouttenoire:2021jhk, LISACosmologyWorkingGroup:2022jok}, only a limited number of studies have explicitly tracked the time-dependent interplay between the background expansion and the PT dynamics~\cite{Barenboim:2016mjm, Guo:2020grp, Buen-Abad:2023hex, Kolesova:2023mno, Dent:2024bhi, Barni:2024lkj, Xiao:2024rsj}. As has been noted in these contexts and as we will demonstrate, the dynamics of the PT in general, and the produced spectrum of GWs in particular, are determined not only by the particle physics model but also by the cosmic evolution of the background on which these processes occur. 

Non-standard cosmic evolutions~\cite{Allahverdi:2020bys, Batell:2024dsi} naturally occur during cosmic reheating~\cite{Dolgov:1989us, Traschen:1990sw, Kofman:1994rk, Kofman:1997yn}. Following cosmic inflation, the energy stored in the inflaton field is transferred to SM degrees of freedom to (re)heat the Universe and establish a SM radiation-dominated epoch with a temperature of at least $\mathcal{O}$(MeV) to ensure successful Big Bang nucleosynthesis (BBN)~\cite{Kawasaki:2000en, Hannestad:2004px, Cyburt:2015mya, deSalas:2015glj}. A PT involves competition between the transition rate to the true phase and the expansion of the Universe~\cite{Coleman:1977py, Callan:1977pt, Linde:1980tt, Linde:1981zj}. The former is fixed by the underlying particle physics parameters, whereas the latter is the input from cosmology. Therefore, a PT occurring during the reheating era, where non-standard expansion rates can exist, can have its dynamics naturally modified.

Building upon this established foundation, in this work, we provide a systematic, parametric exploration of first-order PTs specifically within the context of different cosmic reheating histories. Using a minimal particle physics model, we study the PT dynamics occurring in a cosmological background with a general parameterization of the expansion rate of the Universe. Carefully making the distinction between the particle physics and cosmological parameters, we demonstrate that, due to the non-standard cosmological evolution during reheating, strong PTs can be delayed, prolonged, and further strengthened in certain scenarios by studying the parameters characterizing the PT: its temperature, inverse duration, and latent heat released. We then consider these implications for the GW spectrum sourced from the PT. Consistent with expectations from the literature on non-standard expansion eras, we find that the GW energy density observed today is generally suppressed, with our results explicitly mapping how this suppression behaves for the given reheating scenario, along with the shifting of the peak frequency of the spectra.

This manuscript is structured as follows: first, we parameterize the reheating scenario in Section~\ref{sec:reheating}, which sets the stage for studying PTs occurring for our model in Section~\ref{sec:fopt}. At the end of this section, we compute the PT parameters required for the GW spectrum in Section~\ref{sec:gw_fopt}, in which we appropriately generalize various formulae in the existing literature for the GW spectrum generated from a first-order PT. Finally, we summarize our results and conclude in Section~\ref{sec:concl}.

\section{Parameterizing Cosmic Reheating} \label{sec:reheating}
In minimal scenarios, cosmic reheating proceeds through decays or annihilations of the inflaton field $\chi$ to SM particles.\footnote{In general, the non-perturbative and non-linear dynamics of the background during reheating could be highly non-trivial. However, here we are interested in the last part of reheating, where the linear regime is generally a good approximation~\cite{Bassett:2005xm, Allahverdi:2010xz, Amin:2014eta, Lozanov:2019jxc, Barman:2025lvk}.} During reheating, the energy density of $\chi$ can evolve with an effective equation-of-state parameter $\omega$, such that $\rho_\chi(a) \propto a^{-3(1+\omega)}$. Furthermore, the energy stored in $\chi$ is transferred to SM radiation, whose energy density is given by
\begin{equation}
    \rR(T) \equiv \frac{\pi^2}{30}\, \gs(T)\, T^4,
\end{equation}
where $T$ is the temperature of the SM photons and $\gs(T)$ is the effective number of relativistic degrees of freedom contributing to SM radiation~\cite{Drees:2015exa}.

Reheating is completed at the cosmic scale factor $\arh \equiv a(\Trh)$, corresponding to the \textit{reheating temperature} $\Trh$, which is when the SM radiation density equals the inflaton energy density, 
\begin{equation}
    \rp (\arh) = \rR (\arh) = 3\, \Hrh^2\, M_P^2\,.
\end{equation} 
Here, $\Hrh$ is the Hubble rate at $T = \Trh$, and we have denoted the reduced Planck mass by $M_P \simeq 2.4 \times 10^{18}$~GeV. It follows that the Hubble expansion rate $H$ is~\cite{Bernal:2024yhu, Bernal:2024jim, Bernal:2024ndy, Bernal:2025fdr}
\begin{equation} \label{eq:Hubble}
    H(a) = \sqrt{\frac{\rp + \rR}{3\, M_P^2}} \simeq \Hrh \times
    \begin{dcases}
        \left(\frac{\arh}{a}\right)^\frac{3(1+\omega)}{2} &\text{ for } a \leq \arh,\\
        \left(\frac{\gs(T)}{\gs(\Trh)}\right)^\frac12 \left(\frac{\gss(\Trh)}{\gss(T)}\right)^\frac23 \left(\frac{\arh}{a}\right)^2 &\text{ for } \arh \leq a\,,
    \end{dcases}
\end{equation}
where the entropic degrees of freedom is denoted by  $\gss(T)$, the entropy density $s$ defined as~\cite{Drees:2015exa}
\begin{equation}
    s(T) = \frac{2 \pi^2}{45}\, \gss(T)\, T^3,
\end{equation}
and the Hubble rate when reheating completes is given by
\begin{equation} \label{eq:Hrh}
    \Hrh = \frac{\pi}{3} \sqrt{\frac{\gs(\Trh)}{10}}\, \frac{\Trh^2}{M_P}\,.
\end{equation}
As mentioned in the Introduction, the reheating temperature is subject to constraints $\Trh \gtrsim 4$~MeV~\cite{Kawasaki:2000en, Hannestad:2004px, Cyburt:2015mya, deSalas:2015glj} such that the Universe is radiation dominated before the onset of BBN.

Taking into account the energy transfer from $\chi$ to the SM thermal bath during reheating, the evolution of the SM temperature can be described by~\cite{Bernal:2024yhu, Bernal:2024jim, Bernal:2024ndy, Bernal:2025fdr}
\begin{equation} \label{eq:Tem}
    T(a) \simeq \Trh \times
    \begin{dcases}
        \left(\frac{\arh}{a}\right)^\xi &\text{ for } a_I \leq a \leq \arh,\\
        \left(\frac{\gss(\Trh)}{\gss(T)}\right)^\frac13 \frac{\arh}{a} &\text{ for } \arh \leq a\,,
    \end{dcases}
\end{equation}
where $a_I$ is the scale factor at the end of the inflationary era and $\xi$ captures the scaling of the SM temperature with the scale factor during reheating. During the reheating era, the SM temperature can remain constant if $\xi = 0$, or even increase in cases where $\xi < 0$~\cite{Co:2020xaf}. However, if $\xi > 0$, the SM temperature can reach a temperature $T_\text{max} > \Trh$~\cite{Giudice:2000ex}.\footnote{If the temperature increases slowly enough, for $\xi < 0$, then a PT can occur from the true to the false minimum, as an ``inverse PT''~\cite{Barni:2024lkj}. We focus here only on the case where this increase is almost instantaneous, such that a PT has time to occur only when the SM temperature decreases monotonically during reheating, indicated by $\xi > 0$.} In this work, we focus on the latter case and study PTs occurring during reheating, at a temperature $T_*$ in the range $T_\text{max} \gg T_* > \Trh$. For later convenience, we combine Eqs.~\eqref{eq:Hubble} and~\eqref{eq:Tem} to yield the Hubble parameter as a function of the temperature
\begin{equation} \label{eq:HvsT}
    H(T) \simeq H(\Trh) \times
\begin{dcases}
        \left(\dfrac{T}{\Trh} \right)^{\frac{3(1+\omega)}{2\xi}} & \text{for } T \geq \Trh\,,\\
         \left(\frac{\gs(T)}{\gs(\Trh)}\right)^\frac12 \left( \dfrac{T}{\Trh} \right)^2 & \text{for }  T \leq \Trh\,.
    \end{dcases}
\end{equation}

To close this section, we discuss some examples of the cosmic evolution of the Universe during reheating. The scenario most commonly studied, arising in Starobinsky inflation~\cite{Starobinsky:1980te} or polynomial inflation~\cite{Drees:2021wgd, Bernal:2021qrl, Drees:2022aea, Bernal:2024ykj}, has $\rho_\chi \sim a^{-3}$, that is, $\chi$ evolves as non-relativistic matter ($\omega = 0$) and undergoes perturbative decay into SM particles causing the temperature scaling as $\xi = 3/8$~\cite{Giudice:2000ex}. In the framework of $\alpha$-attractor inflation models~\cite{Kallosh:2013hoa, Kallosh:2013maa}, $\chi$ oscillates near the minimum of its potential $V(\chi) \propto \chi^p$ with $p \geq 2$ during reheating. The corresponding equation-of-state parameter is given by $\omega = (p - 2)/(p + 2)$~\cite{Turner:1983he}, while the value of $\xi$ depends on the mechanism of energy transfer from the inflaton to the radiation, determined by the inflaton-SM couplings~\cite{Co:2020xaf, Garcia:2020wiy, Xu:2023lxw, Barman:2024mqo}. For example, if $\chi$ decays primarily into fermions, we have $\xi = \min(3\, (p-1)/(2\, (p + 2)),\, 1)$~\cite{Shtanov:1994ce, Ichikawa:2008ne, Bernal:2022wck}. Considering a quartic potential with $p = 4$, we arrive at $(\omega,\,\xi) = (1/3,\,3/4)$, resulting in an alternate radiation-dominated scenario to standard cosmology, but with the temperature falling more slowly. If the inflaton energy density decreases faster than free radiation (that is, $\omega > 1/3$), $\chi$ may not need to decay or annihilate completely; an example of such a scenario is kination, where $\omega = \xi = 1$~\cite{Spokoiny:1993kt, Ferreira:1997hj}.

\begin{figure}[t!]
    \def\sepf{0.7}
    \centering
    \includegraphics[width=\sepf\columnwidth]{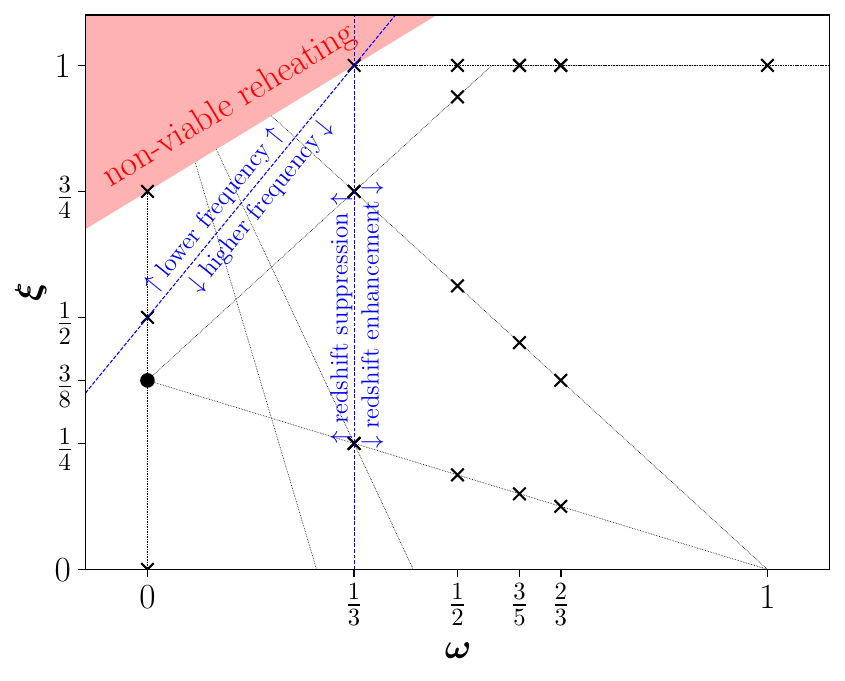}
    \caption{Summary of different reheating scenarios in the plane of $\omega$ and $\xi$. The black dot denotes the standard case where a non-relativistic inflaton decays into SM particles. Black crosses represent alternate scenarios, and are joined by thin dotted lines, due to correlations between $\omega$ and $\xi$, see the main text. The pink area in the upper left corner does not give rise to viable reheating scenarios. To the right (left) of the tilted blue dotted line ($1 + 3\, \omega = 2\, \xi$), the GW spectrum from first-order PTs is shifted to higher (lower) frequencies w.r.t. the case with a high reheating temperature, cf. Eq.~\eqref{eq:ftoday}. Additionally, to the right (left) of the vertical blue dotted line ($\omega = 1/3$), the redshift factor, cf. Eq.~\eqref{eq:R}, is amplified (suppressed).}
    \label{fig:alpha_omega}
\end{figure} 
Figure~\ref{fig:alpha_omega} presents various reheating scenarios in the $(\omega,\, \xi)$ plane, with thin dotted black lines that arise from their correlation in various reheating scenarios~\cite{Co:2020xaf, Barman:2022tzk, Chowdhury:2023jft, Cosme:2024ndc, Barman:2024mqo}. The vertical gray dotted line denotes the case $\omega = 0$. The pink area in the upper left corner, defined by $\xi > 3(1+\omega)/4$, gives configurations where $\rho_{R} < \rho_{\chi}$, thus leading to non-viable reheating scenarios as the energy density of SM radiation never exceeds that of the inflaton~\cite{Co:2020xaf}. Finally, we indicate some of the results of this work through blue-dotted lines: for various reheating scenarios, we find a shifting of the peak frequency of the GW spectra from first-order PTs. Although the GWs are suppressed due to the additional energy component $\rho_\chi$ diluting their energy budget, the degree of suppression can be less pronounced due to their evolution in a non-standard cosmological background; cf. Section~\ref{sec:gw_fopt}.

\section{First-Order Phase Transitions during Reheating} \label{sec:fopt}
A PT is often associated with spontaneous symmetry breaking, which occurs when at least one scalar degree of freedom acquires a vacuum expectation value (VEV). Additional bosonic degrees of freedom may increase the strength of the PT. Therefore, to realize a first-order PT, we consider the following Lagrangian density:
\begin{equation}
    \mathcal{L} \supset (D^{\mu}\Phi)^{\dag} (D_{\mu}\Phi) + \mu^2\, \Phi^{\dag}\Phi - \frac{\lambda}{2}\, (\Phi^{\dag}\Phi)^2,
    \label{eq:lagrangian}
\end{equation}
where the covariant derivative $D_{\mu}\Phi \equiv \partial_{\mu}\Phi + i\, g_X\, X_{\mu}\, \Phi$ couples the complex scalar field $\Phi$, which drives the PT, to the $U(1)_X$ dark gauge boson $X_{\mu}$ with gauge strength $g_X$.\footnote{We assume thermal contact with the SM radiation bath is achieved through couplings such as a Higgs portal interaction $\lambda'\,|H|^2||\Phi|^2$ or kinetic mixing with the SM photon $\epsilon\,F_{\mu\nu}X^{\mu\nu}$, to avoid a secluded sector with a temperature different than the one of the SM radiation. These will not affect our discussion of the PT associated with $\Phi$.}  This model is described by the three free parameters $\mu$, $\lambda$, and $g_X$. Taking $\mu^2 > 0$ and $\lambda > 0$ to ensure the stability of the potential, the minima lie at $\Phi = \pm v_0$ with
\begin{equation} \label{eq:VEV}
    v^2_0 = \frac{\mu^2}{\lambda}\,.
\end{equation}

\begin{figure}[t!]
    \centering
    \includegraphics[width=0.8\columnwidth]{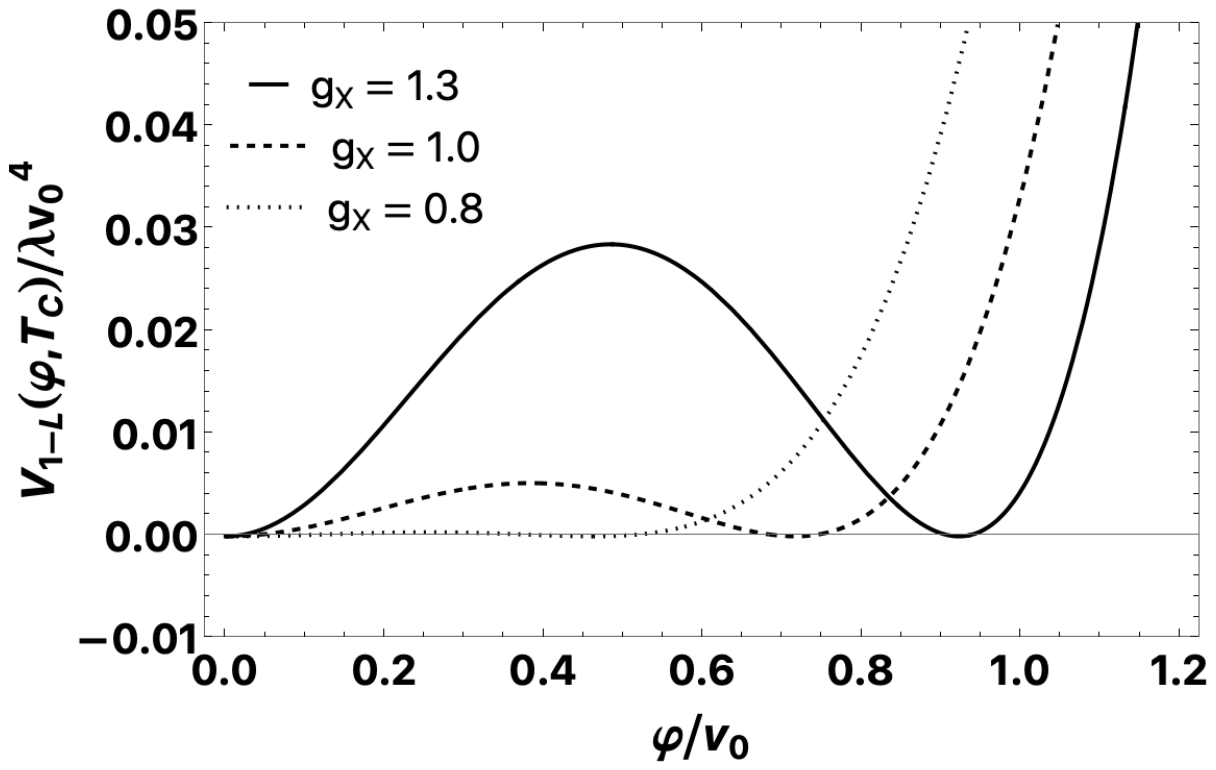}
    \caption{The effective potential for fixed $(\mu,\lambda) = (100\,{\rm MeV}, 0.05)$ and various gauge couplings $g_X$, at their respective critical temperatures. Increasing the gauge coupling increases the height of the potential barrier and the separation between the degenerate minima.}
    \label{fig:Vc}
\end{figure}

To study PTs, we follow the background field method, by introducing fluctuations $\varphi$ around the classical minimum. The thermal evolution of $\varphi$ is governed by the effective potential at finite temperature, which we calculate at one-loop in this work. The appearance of degenerate minima at the critical temperature $T_c$ can be used to classify a PT as first-order. For fixed $\mu$ and $\lambda$, $g_X$ determines the height of the potential barrier formed between the two minima, as shown in Fig.~\ref{fig:Vc}. This relates to properties such as the strength and duration of the PT, as we shall soon see. We review key aspects of this formalism in Appendix~\ref{app:FTEP}. Note that Eq.~\eqref{eq:lagrangian} provides an appropriate minimal setup to study first-order PTs in general cosmological scenarios.

First-order PTs proceed through the nucleation of bubbles of the true vacuum, occurring at a temperature $T <T_c$. This is determined by the thermal nucleation rate per Hubble volume,\footnote{At very low temperatures, it is possible that the transition occurs through quantum tunneling from the false to the true vacuum~\cite{Coleman:1977py}. For our subsequent parameter scans, we have checked that the thermal transition rate remains the dominant source.} given by~\cite{Coleman:1977py, Callan:1977pt, Linde:1980tt, Linde:1981zj}
\begin{equation}
    \Gamma_{N}(T) = T^4 \left(\frac{S_3}{2\pi\, T}\right)^{\frac{3}{2}} \exp\left(-\frac{S_3}{T}\right),
    \label{eq:GammaN}
\end{equation}
where $S_3$ is the $O(3)$-symmetric Euclidean action for the ``bounce'' configuration of the background field
\begin{equation}
    S_3 = 4\pi\int_0^\infty dr \, r^2\left[\frac{1}{2}\left(\frac{d\varphi_b}{dr}\right)^2+V_{\rm{1-L}}(\varphi_b,T)\right],
\end{equation}
where $V_{\rm{1-L}}$ is the finite-temperature effective potential at one-loop (see Appendix~\ref{app:FTEP}), with $\varphi_b$ being the solution of
\begin{equation}
    \frac{d^2 \varphi_b}{dr^2} + \frac{2}{r}\frac{d\varphi_b}{dr} = \frac{dV_{\rm{1-L}}}{d\varphi_b}(\varphi_b, T)\,,
\end{equation}
subject to the boundary conditions $\varphi_b (\infty) = 0$ and $\partial_r\varphi_b (0) = 0$. The solution is obtained using the \verb|Mathematica| package \verb|FindBounce|~\cite{Guada:2020xnz}. As the Universe expands, bubble nucleation occurs at the temperature $T_n$, signaling the onset of the PT. At this temperature, the nucleation rate in Eq.~\eqref{eq:GammaN} becomes comparable to the Hubble time, 
\begin{equation} \label{eq:nucl_cond}
    \Gamma_{N}(T_n) = H^4(T_n)\,.
\end{equation}
We emphasize that while $\Gamma_N$ depends only on the particle-physics parameters, $H$ is fully determined by the cosmological setup.

\begin{figure}[t!]
    \centering
    \includegraphics[width=0.82\columnwidth]{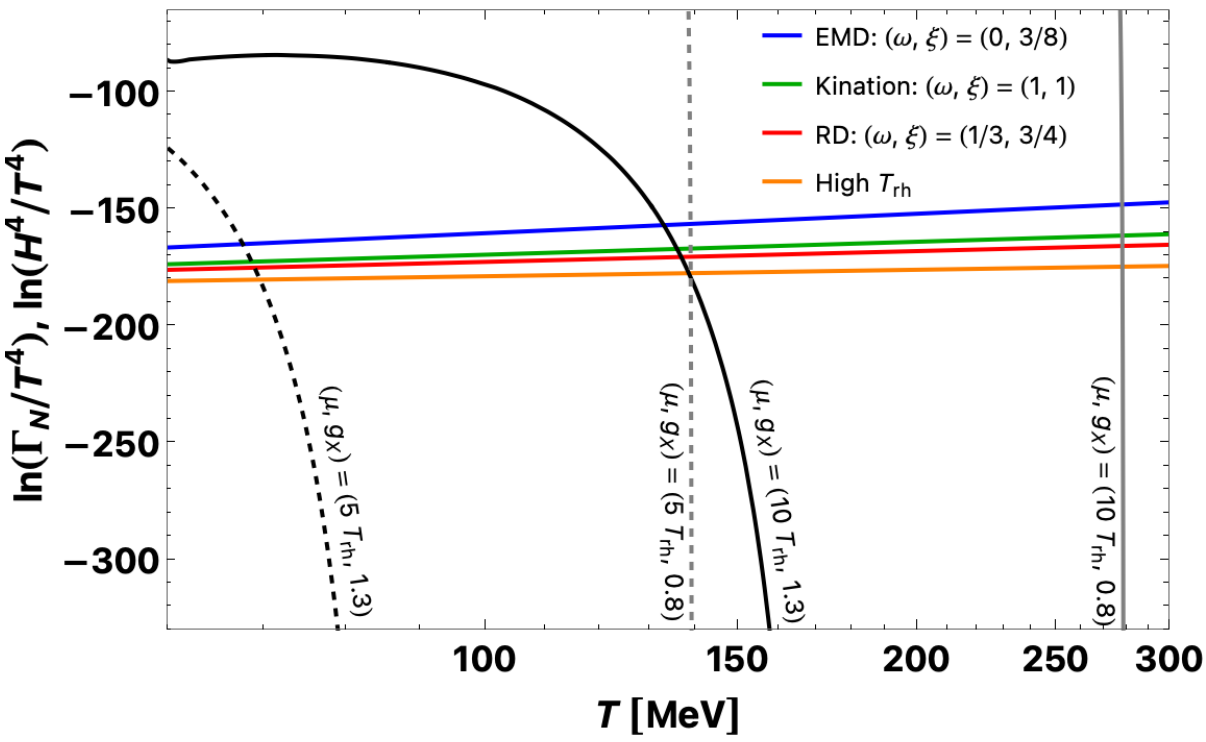}
    \caption{The nucleation rate per Hubble volume $\Gamma_N$ for different model parameters, and the Hubble expansion rate $H$ for different cosmologies, as a function of the temperature, with fixed $\Trh = 10$~MeV and $\lambda = 0.05$. Black and gray curves indicate different couplings, whereas solid and dashed lines indicate the value of the dimensionful $\mu$ parameter. Colored lines indicate the various cosmological scenarios. We divide out by $T^4$ to render the argument of the logarithm dimensionless.}
    \label{fig:Gamma_H_cosmo}
\end{figure} 
In Fig.~\ref{fig:Gamma_H_cosmo}, we compare the rates for nucleation and expansion. We first fix a relatively low $\Trh = 10$~MeV whenever required. We then consider the parameter in the scalar potential $\mu = 5~\Trh$ (black dashed lines) or $\mu = 10~\Trh$ (black solid lines), which sets different symmetry-breaking scales, and the quartic coupling $\lambda = 0.05$. Then, we choose $g_X = 1.3$ ($g_X = 0.8$) to correspond to a strong (weak) PT, see Appendix~\ref{app:FTEP} and Fig.~\ref{fig:Vc} for further details. Furthermore, for $H$ we assumed kination ($\omega = 1$, $\xi = 1$), early matter domination (EMD) ($\omega = 0$, $\xi = 3/8$), radiation domination ($\omega = 1/3$, $\xi = 3/4$), and the scenario with high-temperature reheating ($\Trh \gg T_n$, which effectively makes this case independent of $\omega$ and $\xi$). Finally, we take the effective relativistic degrees of freedom that contribute to the energy density and the entropy density at reheating $\gs(\Trh) = \gss(\Trh) = 10$ and ignore its temperature dependence.\footnote{We adhere to this for the high-temperature reheating scenario, in order to make a fair comparison.}

\begin{figure}[t!]
    \centering
    \includegraphics[width=0.82\columnwidth]{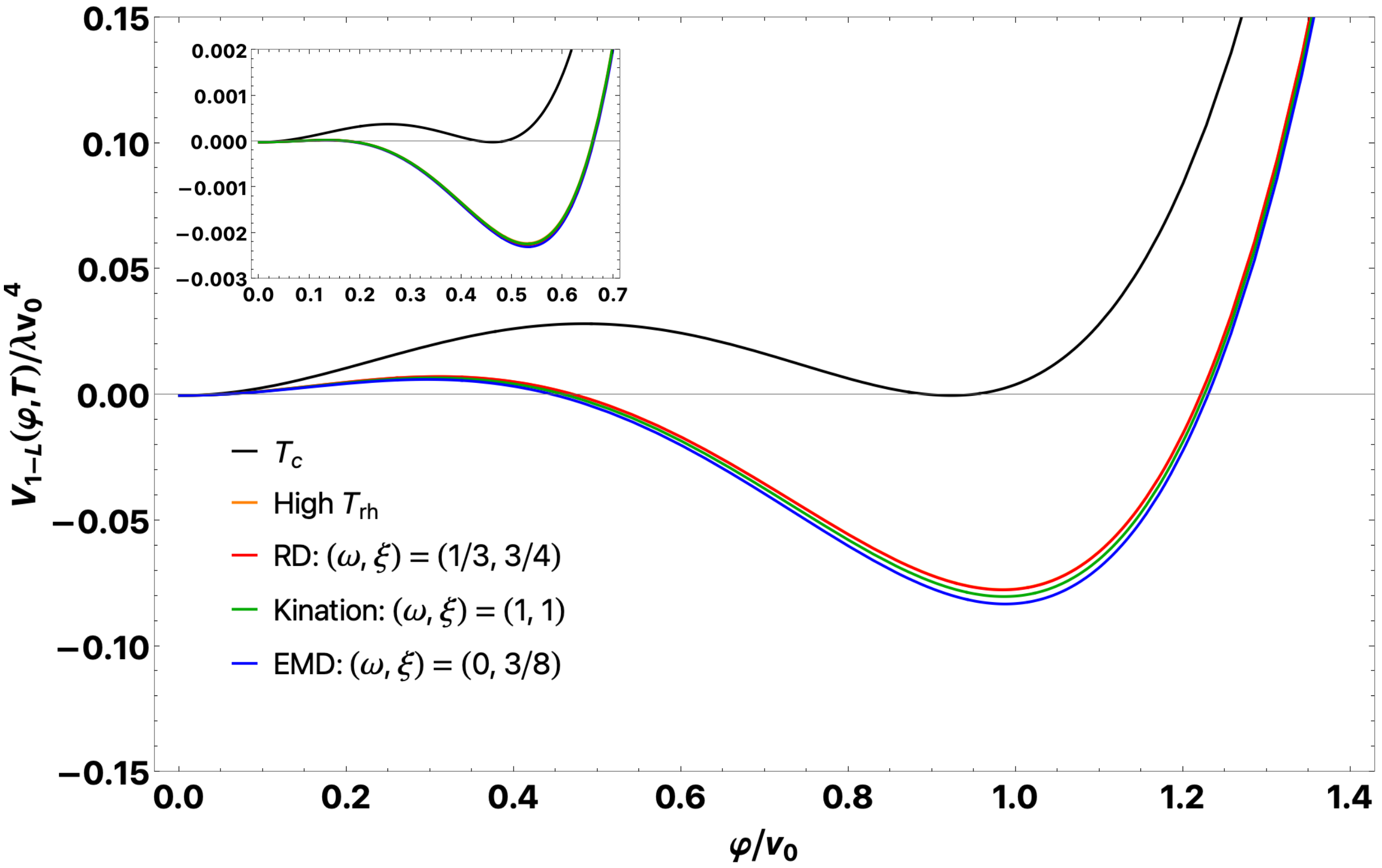}
    \caption{Effective potentials at the critical temperature $T_c$ (black line) for scalar potential parameters $(\mu,\,\lambda) = (10\,\Trh, 0.05)$ with $\Trh = 10$~MeV, and at the nucleation temperatures corresponding to various cosmologies (colored lines). We take $g_X = 1.3$ corresponding to a strong PT, resulting in a large thermally induced barrier between the minima. In comparison, for the inset plot, we have $g_X = 0.8$ corresponding to a weak PT, due to the comparatively smaller barrier.}
    \label{fig:Veff_cosmo}
\end{figure} 
The intersection points between the two rates in Fig.~\ref{fig:Gamma_H_cosmo} give the nucleation temperatures for each case. For weak PTs (almost-vertical gray lines), nucleation is easily achieved due to the relatively small loop-induced barrier formed between the degenerate minima, which can be easily traversed at high temperatures (see the inset plot of Fig.~\ref{fig:Veff_cosmo}). However, for strong PTs, the corresponding barrier height must first be reduced due to the falling temperature because of the expansion of the Universe. In this case, we stress that, compared to the standard cosmological scenario featuring high-temperature reheating, if the PT occurs {\it during reheating}, nucleation occurs at lower temperatures, which implies a delay in the PT.

In addition to nucleation, it is customary to consider the percolation temperature $T_p$ of the PT, which characterizes the completion of the PT~\cite{Ellis:2018mja, Ellis:2020nnr}, and is often taken as the temperature of the PT $T_*$. To this end, one considers the probability of finding a point still in the false vacuum at a given time
\begin{equation}
    P(t) = e^{-I(t)},
\end{equation}
with
\begin{equation}
    I(t) = \frac{4\pi}{3} \int_{t_c}^t d\bar{t}\, \Gamma_N(\bar{t})\, \left[a(\bar{t})\, r(t,\bar{t})\right]^3,
    \label{eq:vol_frac}
\end{equation}
where $t_c$ is the time at which the vacua are degenerate (corresponding to the critical temperature $T_c$). At the percolation time $t_p$, the probability $P$ drops to $\sim 70\%$, which implies 
\begin{equation}
    I(t_p) \simeq 0.34\,.
\end{equation}
Additionally, 
\begin{equation}
    r(t,\bar{t}) \equiv \int_{\bar{t}}^t d\hat{t}\, \frac{v_w}{a(\hat{t})}
\end{equation}
is the radius of a bubble, nucleated at $\bar{t}$ and growing until $t$, and $v_w$ is the wall velocity, which we will treat as independent of the cosmological evolution. Based on Eq.~\eqref{eq:HvsT}, one can then write Eq.~\eqref{eq:vol_frac} as a function of the temperature, and hence determine the associated percolation temperature. In addition, we require the false vacuum to shrink, whose volume is given by $V_{\rm false}(t) \equiv a^3(t)P(t)$. This gives the requirement
\begin{equation}
    3 H(t_p) - \frac{d}{dt}I(t)\bigg|_{t = t_p} < 0\,.
\end{equation}
We provide detailed expressions in Appendix~\ref{app:perc_temp}, where we also note, for a few benchmark points, deviations of the order $(T_n-T_p)/T_n \lesssim 10\%$ can arise for stronger PTs. In what follows, we will take $T_* = T_p$ and evaluate the remaining PT parameters at the percolation temperature.

\begin{figure}[t!]
    \centering
    \includegraphics[width=0.49\columnwidth]{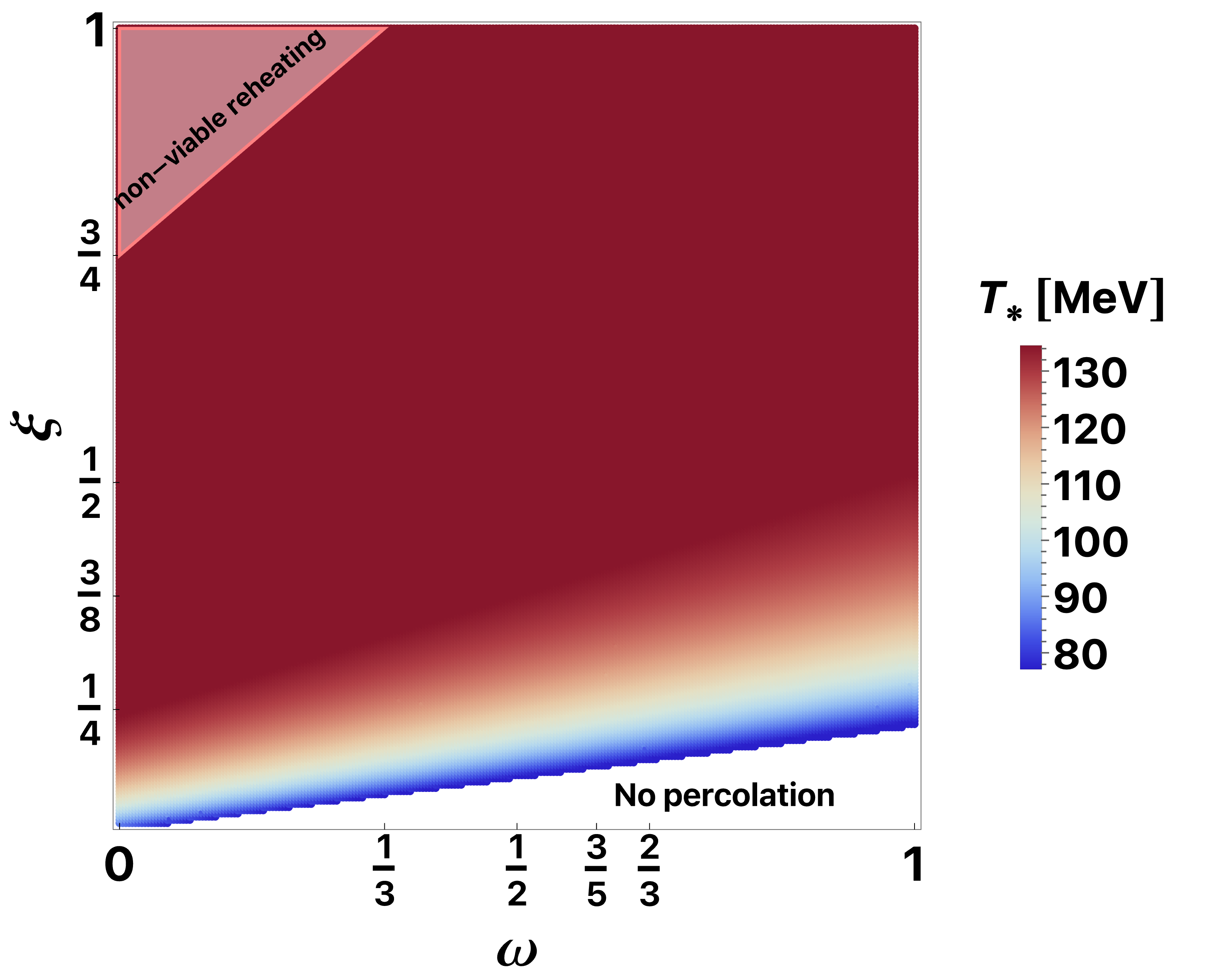}\\[1mm]
    \includegraphics[width=0.49\columnwidth]{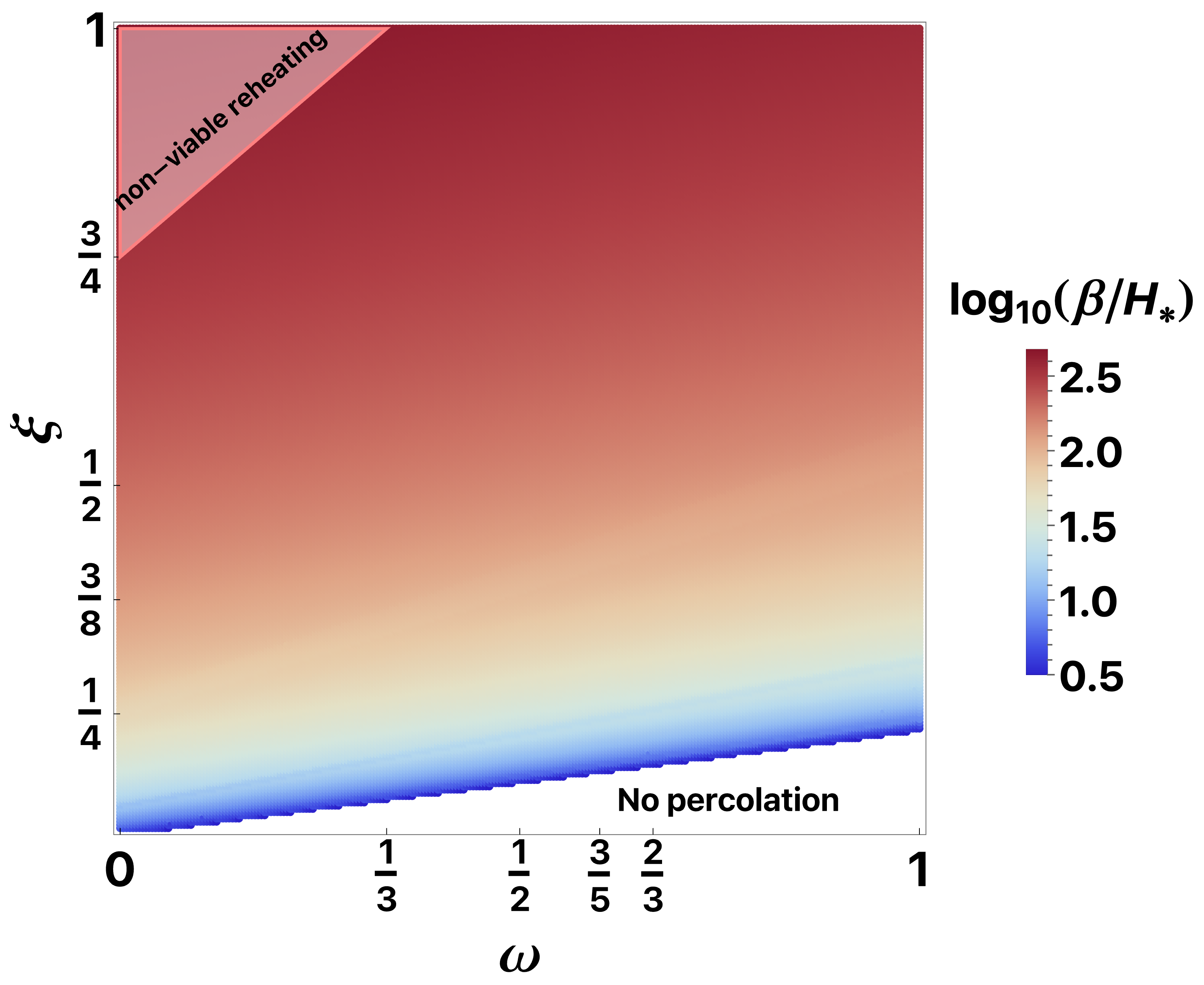}
    \includegraphics[width=0.49\columnwidth]{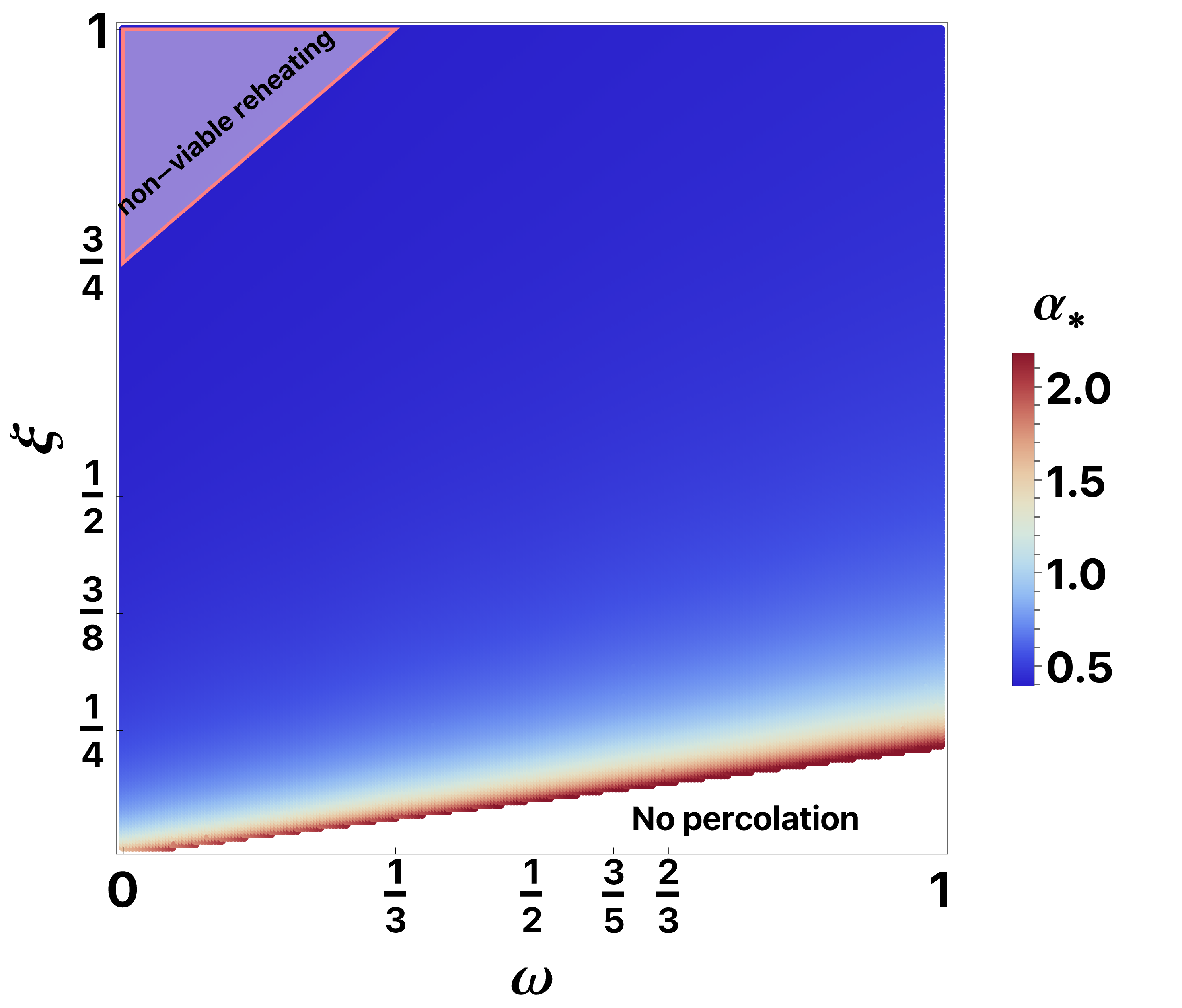}
    \caption{Values of $T_*$, $\beta/H_*$ and  $\alpha_*$ in the plane $(\omega, \xi)$, i.e. for different reheating scenarios. We assume $\Trh = 10$~MeV, $\mu = 10~\Trh$ and $(\lambda, g_X) = (0.05, 1.3)$, implying a relatively strong PT. The white regions indicate that the percolation (and implicitly, nucleation) criterion fails, while the pink shaded regions indicate the non-viable reheating scenarios.}
    \label{fig:contours_cosmo}
\end{figure} 

From Eq.~\eqref{eq:GammaN}, we see that the nucleation rate is exponentially suppressed by the factor $-S_3/T$. Close to the time of the PT $t_*$, one may write
\begin{equation}
    \Gamma_N(t) \simeq A_*\, e ^{\beta\, (t-t_*)},
\end{equation}
where we have Taylor expanded the exponent up to the first term. This then gives the inverse duration of the PT
\begin{equation}
    \beta \equiv -\frac{d}{dt}\left(\frac{S_3}{T}\right)\bigg|_{t=t_*}.
\end{equation}
We can then estimate the inverse duration in terms of the Hubble time at the time of the PT $H_* \equiv H(T_*)$
\begin{equation}
    \frac{\beta}{H_*} = \xi \left.\left[T\, \frac{d}{dT}\left(\frac{S_3}{T}\right)\right]\right|_{T_*}
    \label{eq:betaH}
\end{equation}
for $T_*>\Trh$. The previous result also applies to the case where $T_* < \Trh$, if we take $\xi = 1$. Notice the additional overall factor of $\xi$ arising from the non-standard scaling of the temperature with the scale factor. For well-motivated reheating scenarios, we have $\xi \leq 1$ (implying that SM radiation does not scale as free radiation due to the injection of entropy), thus decreasing the value of $\beta/H_*$. Furthermore, from Fig.~\ref{fig:Gamma_H_cosmo}, we note that the intersection point of the Hubble rate with the nucleation rate tends to move higher for smaller $\xi$. This implies a \textit{lower} point on the curve $S_3/T$, so that the derivative w.r.t. $T$, represented by $\beta/H_*$, decreases. The overall result is that PTs tend to be longer than in the standard case due to the smaller $\beta/H_*$.\footnote{In our subsequent parameter scans over cosmological scenarios, we exclude the cases where $\beta/H_* \lesssim 3$, following Refs.~\cite{Freese:2022qrl, Winkler:2024olr}, as this may lead to a phase of eternal inflation due to the PT never completing~\cite{Guth:1982pn}.} Finally, the strength $\alpha_*$ of the PT, defined as the latent heat released normalized to the radiation energy density, is given by
\begin{equation}
    \alpha_* \equiv \frac{1}{\rR(T_*)} \left.\left[\Delta V - T\, \frac{\partial \Delta V}{\partial T}\right]\right|_{T_*},
    \label{eq:alphastar}
\end{equation}
where $\Delta V$ is the potential difference between true and false vacua.\footnote{Notice that $\alpha_*$ is normalized to the SM energy density and not to the total energy density of the Universe. Furthermore, a large $\alpha_* \sim \mathcal{O}(10)$, may trigger a short period of vacuum energy domination due to the corresponding large $\Delta V$ released~\cite{Ellis:2018mja, Ellis:2020nnr, Kierkla:2022odc}. For our parameter scans, we have checked that such a scenario does not occur, with the maximal value of $\alpha_* \simeq 2$ occurring for $\xi \to 0$, see Fig.~\ref{fig:contours_cosmo}.} 

In Fig.~\ref{fig:contours_cosmo} we explore the explicit dependence of the different reheating scenarios with $\Trh = 10$~MeV on the characteristics of the PT ($T_*$, $\beta/H_*$, and $\alpha_*$) taking $\mu = 10\, \Trh$, $\lambda = 0.05$, and $g_X = 1.3$, corresponding to a strong PT at a symmetry breaking scale $\mathcal{O}(100)$~MeV. We observe that for certain reheating scenarios, the percolation criterion (where we also include the nucleation criterion, Eq.~\eqref{eq:nucl_cond}) is not met, implying that the Universe remains trapped in the false vacuum as the PT never takes place; cf. the lower right white corner. This corresponds to small values of $\xi \sim 0$, where the temperature during reheating tends to be constant.

Compared to the case where it takes place in the standard radiation-dominated era, PTs occurring during reheating are delayed (smaller values of $T_*$) due to the increase of the Hubble expansion rate and longer (smaller values of $\beta/H_*$). We also observe a slight increase from the value of $\alpha_*$. This increase can be associated with a growth in $\Delta V$, as can be discerned from Fig.~\ref{fig:Veff_cosmo}, for certain cosmologies, and also with a normalization of the radiation energy density, which decreases due to the delay in the PT. 

Additionally, Fig.~\ref{fig:contours_cosmo} exemplifies the fact that the nature of the PT depends not only on the parameters of the Lagrangian but also on the cosmological scenario considered. Finally, as we progress towards cosmological scenarios where the PT stops occurring during reheating, we observe that PTs are longer, occur later, and are stronger. In the next section, we study these implications for the associated GW spectrum.

\section{Gravitational Waves from Phase Transitions during Reheating} \label{sec:gw_fopt}
GWs from PTs have been extensively studied in the literature; see, e.g. Refs.~\cite{Caprini:2015zlo, Caprini:2018mtu, Caprini:2019egz, Athron:2023xlk, Croon:2024mde} for extensive reviews. Sources stem from collisions of nucleating bubbles~\cite{Kosowsky:1992rz, Kosowsky:1992vn, Caprini:2007xq, Huber:2008hg, Hindmarsh:2013xza, Weir:2016tov} of the true phase, acoustic GWs generated from the bulk motion of the plasma due to energy transferred from expanding bubbles~\cite{Hindmarsh:2013xza, Giblin:2014qia, Hindmarsh:2015qta, Hindmarsh:2016lnk, Hindmarsh:2017gnf, Hindmarsh:2019phv} and turbulent motion generated from shocks in the plasma~\cite{Kosowsky:2001xp, Dolgov:2002ra, Gogoberidze:2007an, Caprini:2009yp, Niksa:2018ofa, RoperPol:2019wvy, Kahniashvili:2020jgm, RoperPol:2021xnd,  Auclair:2022jod}.

In general, the spectral form of each contribution sourcing GWs can be parametrized as
\begin{equation} \label{eq:OmegaGW}
    h^2\,\Omega^{i}_{\rm GW}(f) = \mathcal{R}_{\omega,\xi}\, \widetilde{\Omega}^i_{\rm GW,*}\, \mathcal{S}_i\left(x\right),
\end{equation}
which is the GW energy density, normalized to the critical density today, from the source $i =$ bubble collisions (bc), sound waves (sw) and turbulence (turb), and $x \equiv f/f^i_0$. The factor $\mathcal{R}_{\omega,\xi}$ accounts for the change in the amplitude of the GWs due to the expansion of the Universe, and $f_0^i$ is the peak frequency of the GW from a particular source observed today, redshifted from the peak frequency emitted at the time of the PT $f_*^i$. Finally, $\widetilde{\Omega}^i_{\rm GW,*}$ is the scaling behavior of the amplitude of each source with the PT parameters, and $\mathcal{S}_i$ gives the corresponding spectral shape.\footnote{Since the duration of the PT is typically smaller than a Hubble time, as characterized by $\beta/H_* \gg 10$ in most well-motivated cosmological scenarios (cf. Fig.~\ref{fig:contours_cosmo}), we assume that the spectral shape from each source is unmodified to leading order from the usual results for standard cosmology, following  Refs.~\cite{Guo:2020grp, Xiao:2024rsj}. However, in some of the extreme cases where $\beta/H_* \sim 1$, modifications to the lower frequency-end of the spectra can occur~\cite{Sharma:2023mao, RoperPol:2023dzg}.}

For each source, the peak frequencies at the time of production are independent of the cosmological evolution and are given by~\cite{Caprini:2018mtu, Athron:2023xlk}
\begin{align}
    f^{\rm bc}_{*} &\simeq \frac{0.62\,\beta}{1.8 - 0.1 v_w + v_w^2}\,,\\
    f^{\rm sw}_{*} &\simeq \frac{\beta}{\sqrt{3}\,v_w}\,,\\
    f^{\rm turb}_{*} &\simeq 3\, f^{\rm sw}_*,
    \label{eq:fstar}
\end{align}
where $v_w$ is the wall velocity. In the so-called ``detonation regime''~\cite{Espinosa:2010hh, Ellis:2018mja, Hindmarsh:2019phv}, this can be approximated using the Jouguet velocity as
\begin{equation}\label{eq:wall_vel}
    v_w \simeq \frac{1/\sqrt{3}+\sqrt{\alpha^2_{\ast}+2\alpha_{\ast}/3}}{1+\alpha_{\ast}}\,.
\end{equation}

As we consider the evolution of GWs produced from PTs occurring during reheating, we account for the non-standard cosmological evolution through appropriate redshifting of the frequencies and the energy density, as described in Refs.~\cite{Allahverdi:2020bys, Gouttenoire:2021jhk, LISACosmologyWorkingGroup:2022jok}. In our particular scenario, we distinguish between the PTs that occur after and during reheating. Taking into account that the frequency of GWs always varies with the scale factor as $f \propto a^{-1}$, $f^i_0$ is related to $f^i_*$ through appropriate redshifting as 
\begin{equation}
    f^i_0 = f^i_* \times
    \begin{dcases}
        \left(\frac{\gss(T_0)}{\gss(\Trh)}\right)^\frac13 \left(\frac{\Trh}{T_*} \right)^{\frac{1}{\xi}} \frac{T_0}{\Trh} & \text{ for } T_* \geq \Trh, \\
        \left(\frac{\gss(T_0)}{\gss(T_*)}\right)^\frac13 \frac{T_0}{T_*} & \text{ for } \Trh \geq T_*,
    \end{dcases}
\end{equation}
where $T_0 \simeq 2.348 \times 10^{-10}$~MeV is the temperature of the CMB and $\gss(T_0) \simeq 3.94$~\cite{Planck:2018vyg}. It is convenient to introduce a factor of $H_*$ to recast this in the form 
\begin{equation} \label{eq:ftoday}
    f^i_0 = 1.14\times10^{-10}~{\rm Hz} \left(\frac{f^i_*}{H_*}\right) \times
    \begin{dcases}
        \left(\frac{10}{\gss(\Trh)}\right)^\frac13 \left(\frac{\gs(\Trh)}{10}\right)^\frac12\left(\frac{T_*}{1\,{\rm MeV}}\right) \left(\frac{T_*}{\Trh}\right)^{\frac{1 + 3\omega}{2 \xi}-1}\\
        \hspace{6.8cm} \text{for } T_* \geq \Trh, \\
        \left(\frac{10}{\gss(T_*)}\right)^\frac13 \left(\frac{\gs(T_*)}{10}\right)^\frac12 \left(\frac{T_*}{1\,\rm{MeV}}\right) \qquad \text{ for } \Trh \geq T_*.
    \end{dcases}
\end{equation}
Note that for $(\omega,\, \xi) = (1/3,\, 1)$ and, in general, for $1 + 3\, \omega = 2\, \xi$, the explicit dependence on the reheating temperature vanishes and only enters at subleading order through the relativistic and entropic degrees of freedom. For $1 + 3\, \omega > 2\, \xi$, the peak $f_0^i$ of the spectra moves to higher frequencies compared to the usual case with a high reheating temperature. In contrast, for $1 + 3\, \omega < 2\, \xi$, the spectrum moves to lower frequencies. The tilted blue dotted line in Fig.~\ref{fig:alpha_omega} shows these three regimes. For the reheating scenarios considered here, Eq.~\eqref{eq:ftoday} makes explicit the dependence of $f_0^i$ on $\beta/H_*$, $T_*$, and, assuming the relation in Eq.~\eqref{eq:wall_vel}, on $\alpha_*$ through the wall velocity.

Next, we consider the redshift factor $\mathcal{R}_{\omega,\xi}$ in Eq.~\eqref{eq:OmegaGW}. Adopting the standard scaling behavior for the background energy densities during non-standard epochs~\cite{Allahverdi:2020bys, Gouttenoire:2021jhk, LISACosmologyWorkingGroup:2022jok}, the total energy density of GWs scales as radiation, which redshifts between their time of production until present as
\begin{equation}
    \rgw(T_0) \simeq \rgw(T_*) \times
    \begin{dcases}
        \left(\frac{\gss(T_0)}{\gss(\Trh)}\right)^\frac43 \left(\frac{T_0}{\Trh}\right)^4 \left(\frac{\Trh}{T_*}\right)^\frac{4}{\xi} & \text{ for } T_* \geq \Trh, \\
        \left(\frac{\gss(T_0)}{\gss(T_*)}\right)^\frac43 \left(\frac{T_0}{T_*}\right)^4 & \text{ for } \Trh \geq T_*.
    \end{dcases}
\end{equation}
The fraction of the total energy density in GWs $\Ogw$ is defined with respect to the critical energy density of the Universe $\rho_c$
\begin{equation}
    \rho_\text{c}(T_0) \simeq \rho_c(T_*) \times
    \begin{dcases}
        \frac{90}{\pi^2\, \gs(\Trh)} \left(\frac{M_P\, H_0}{\Trh^2}\right)^2 \left(\frac{\Trh}{T_*}\right)^\frac{3(1+\omega)}{\xi} & \text{ for } T_* \geq \Trh, \\
        \frac{90}{\pi^2\, \gs(T_*)} \left(\frac{M_P\, H_0}{T_*^2}\right)^2 & \text{ for } \Trh \geq T_*,
    \end{dcases}
\end{equation}
with $H_0 = 100~h~{\rm km~s^{-1}~Mpc^{-1}}$, where $h$ encodes the uncertainty related to the measurement of the Hubble parameter today~\cite{Planck:2018vyg}. Taking into account that $\Ogw(T) \equiv \rgw(T) / \rho_c(T)$, it follows that $h^2\,\Ogw(T_0) \equiv \mathcal{R}_{\omega,\xi}\, \Ogw(T_*)$, which implies that
\begin{equation} \label{eq:R}
    \mathcal{R}_{\omega,\xi} \simeq 3.69\times 10^{-5} \times 
    \begin{dcases}
        \left(\frac{\gs(\Trh)}{10}\right)\left(\frac{10}{\gss(\Trh)}\right)^\frac43 \left(\frac{T_*}{\Trh}\right)^\frac{3\omega - 1}{\xi}& \text{ for } T_* \geq \Trh\,, \\
        \left(\frac{\gs(T_*)}{10}\right) \left(\frac{10}{\gss(T_*)}\right)^\frac43 & \text{ for } \Trh \geq T_*\,.
    \end{dcases}
\end{equation}
Consistent with previous analyses of GW propagation through non-standard cosmological eras, during reheating, if the inflaton scales as radiation $(\omega = 1/3)$, GWs do not receive an enhancement from redshifting. However, for $\omega > 1/3$, such as in kination $(\omega = 1)$, the redshift factor is enhanced, affecting the overall amplitude of GWs. A suppression from this factor occurs whenever $\omega < 1/3$, for example, in EMD $(\omega = 0)$. The vertical blue dotted line in Fig.~\ref{fig:alpha_omega} shows these three regimes.

We find that the contributions from bubble collisions are negligible, as we consider a thermal PT with no significant supercooling~\cite{Ellis:2018mja, Ellis:2019oqb, Lewicki:2019gmv, Lewicki:2020azd, Lewicki:2020jiv, Lewicki:2022pdb}. Therefore, we focus on the GWs arising from the plasma motion. In this context, we exclude the contribution from the turbulence, due to uncertainties in determining the onset of turbulence and in disentangling it from the sound wave source; see, e.g. Refs.~\cite{Caprini:2024gyk, Correia:2025qif}. Therefore, in our analysis, we focus on the acoustic GWs. However, we emphasize that the parametric dependence of the redshift factor and frequency on the cosmology presented here can be systematically applied as a targeted refinement once better estimates and closed-form expressions for this contribution are computed in the future.

For the sound-wave source of GWs, one needs to consider the production and propagation of GWs in the background with a general equation-of-state parameter. Consequently, from the fitting to simulations~\cite{Hindmarsh:2015qta, Hindmarsh:2017gnf}, the scaling and shape of the spectrum, widely used in the literature, the following formulae can be adopted~\cite{Weir:2017wfa, Caprini:2018mtu, Caprini:2019egz}
\begin{align} 
    \label{eq:sw_spec_sim}
    \mathcal{S}_{\rm sw}(x) &= x^3 \left(\frac{7}{4 + 3\, x^2}\right)^{7/2},\\
    \label{eq:sw_sim_amp}
    \widetilde{\Omega}^{\rm sw}_{{\rm GW},*} &\simeq 7.244 \times10^{-2}\, v_w \left(\frac{\kappa_{\rm sw}\, \alpha_*}{1+\alpha_* + R_{\chi,*}}\right)^2 \left(\frac{\beta}{H_*}\right)^{-1} \Upsilon(y, \omega)\,,
\end{align}
where the factor
\begin{equation}
    R_{\chi,*} \equiv \frac{\rho_\chi(a_*)}{\rho_R(a_*)} = \frac{\gs(\Trh)}{\gs(T_*)} \left(\frac{T_*}{\Trh}\right)^\frac{3 + 3 \omega - 4 \xi}{\xi}
\end{equation}
accounts for the energy density of the inflaton, which contributes to the expansion rate, cf. Eq.~\eqref{eq:Hubble}, for PTs occurring during reheating. Accordingly, this term in the denominator corrects the normalization for the abundance of GWs during this non-standard cosmological evolution, consistent with the treatment found in Refs.~\cite{Allahverdi:2020bys, Gouttenoire:2021jhk, LISACosmologyWorkingGroup:2022jok}. Note that to recover the case of standard cosmology, one requires $\Trh \gg T_*$, which implies $R_{\chi,*} \to 0$. The efficiency factor for conversion of the latent heat released in the PT to the bulk motion of the plasma can be approximated using the ``bag model'' as~\cite{Steinhardt:1981ct, Espinosa:2010hh}
\begin{equation}
    \kappa_{\rm{sw}} \simeq \frac{\alpha_{\ast}}{0.73 + 0.083\, \sqrt{\alpha_{\ast}} + \alpha_{\ast}}\,.
    \label{eq:kappasw}
\end{equation}
Note that $\kappa_{\rm sw} \equiv \rho^{\rm sw}_{\rm GW}/\rho_{R}$ is inherently independent of the non-standard cosmological evolution from its definition~\cite{Gouttenoire:2021jhk}. To account for the finite lifetime of the source~\cite{Ellis:2020awk, Guo:2020grp}, we include a suppression factor $\Upsilon$,  given for general cosmological scenarios as~\cite{Xiao:2024rsj, Hindmarsh:2019phv}
\begin{align}
    \Upsilon(y, \omega) &= \frac{2}{3\, (1 - \omega)} \left[1 - y^{-\frac{3\, (1-\omega)}{2}}\right],
    \label{eq:upsilon}
    \\
    y &= \left[1 + \frac{3\, (1+\omega)}{2}\, \tau_\text{sw}\, H_*\right]^\frac{2}{3\, (1+\omega)},
    \label{eq:upsilon_argy}
    \\
    \tau_{\rm sw}\, H_* &\simeq \frac{(8\pi)^\frac13\, v_w}{\overline{U}_f} \,\left(\frac{\beta}{H_*}\right)^{-1} \,.
    \label{eq:lifetime_sw}
\end{align}
Here, we have followed Ref.~\cite{Ellis:2020awk} in estimating the lifetime of the source (cf. Eq.~\eqref{eq:lifetime_sw}) by considering the characteristic fluid length (comparable to the mean bubble separation at the time of the PT) divided by $\overline{U}_f$, the root-mean-square fluid-velocity~\cite{Hindmarsh:2017gnf}, given by
\begin{equation}
    \overline{U}^2_f = \frac{\kappa_{\rm sw}\alpha_*}{\bar{w}}\rho_R \approx \frac34\left[\frac {\kappa_{\rm sw}\alpha_*(1+R_{\chi,*})}{1+\alpha_*+R_{\chi,*}}\right]\left[1+\frac{(1+\omega)\,R_{\chi,*}}{4}\right]^{-1}\,,
    \label{eq:rms_vel}
\end{equation}
where $\bar{w}$ is the total average enthalpy, given by the sum of the average pressure and energy, and we account for the general equation of state of the fluid. Equation~\eqref{eq:lifetime_sw} confirms that shorter PTs (characterized by larger values of $\beta/H_*$) suffer greater suppression due to their shorter lifetime. Furthermore, it is interesting to note that for $\omega > 1/3$, the suppression factor $\Upsilon$ could turn into an ``enhancement'' factor, as $\Upsilon(y \to \infty, \omega > 1/3) >1$. Finally, for $\omega = 1$, as in kination, Eq.~\eqref{eq:upsilon} is not well defined. However, in the limit, one can obtain $\Upsilon(y,1) = \ln y$~\cite{Xiao:2024rsj}.

Various improvements have been made in understanding acoustic GWs from first-order PTs~\cite{Hindmarsh:2019phv, Guo:2020grp, Gowling:2021gcy, Guo:2024gmu}. We adopt the following precise fit expression for the GWs from sound waves, obtained within the sound-shell model~\cite{Guo:2024gmu}
\begin{equation} \label{eq:sw_new_fit}
    \Omega_{\text{GW}}^{\rm sw, fit} h^2 \simeq \frac{\mathcal{R}_{\omega,\xi}}{\mathcal{R}_{1/3,1}} \, \Omega_p \left( \frac{f}{\tilde{s}_0} \right)^9 \frac{2 + \tilde{r}_b^{-12 + \tilde{b}}}{\left( \frac{f}{\tilde{s}_0} \right)^{\tilde{a}} + \left( \frac{f}{\tilde{s}_0} \right)^{\tilde{b}} + \tilde{r}_b^{-12 + \tilde{b}} \left( \frac{f}{\tilde{s}_0} \right)^{12}} \, \frac{\Upsilon(y,\omega)}{\Upsilon(y,1/3)} \left(\frac{1 + \alpha_*}{1+\alpha_* + R_{\chi,*}}\right)^2,
\end{equation}
where we incorporate the effect due to the equation-of-state parameter during reheating, discussed earlier, using the first and last factors. Equation~\eqref{eq:sw_new_fit} shows a broken power-law with fitting parameters, $\Omega_p$, $\tilde{s}_0$, $\tilde{a}$, $\tilde{b}$, and $\tilde{r}_b = f_b / f_p$. These are fixed by the various physical processes that contribute to acoustic GWs. Here, $\tilde{r}_b$ sets the ratio of the double peak spectrum, $\tilde{b}$ controls the scaling in the intermediate regime, and $\tilde{a}$ represents the low-frequency scaling. Each of these parameters can be obtained in terms of $\alpha_*$, $T_*$, and $\beta/H_*$, using the package provided in Ref.~\cite{Guo:2024gmu}. In our subsequent plots, we use Eq.~\eqref{eq:sw_new_fit}, while the formula based on the fitting to the simulations, Eqs.~\eqref{eq:sw_spec_sim} and~\eqref{eq:sw_sim_amp}, will be used for analytic insight. A comparison of the spectra and the corresponding fit parameters for our spectra are presented in Appendix~\ref{app:sw_spectra}, where we note that the amplitudes of the spectra are comparable for strong PTs, while for weak PTs, the sound-shell model typically gives a higher amplitude compared to the simulation results. The (prominent) peak frequency in the sound-shell model is typically to the left (i.e., at lower frequencies) than the one resulting from the simulation fit.   

\begin{table}[t!]
    \centering
    \begin{tabular}{|cccc|cc||ccc|}
        \hline
        $\boldsymbol \mu$~[MeV] & $\boldsymbol \lambda$ & $\boldsymbol{g_X}$ & $\boldsymbol{v_c/T_c}$ & $\boldsymbol \omega$ & $\boldsymbol \xi$ & $\boldsymbol{T_*}$~[MeV] & $\boldsymbol{\alpha_*}$ & $\boldsymbol{\beta/H_*}$ \\
        \hline \hline
        \multirow{4}{*}{$10^2$} & \multirow{4}{*}{0.05} & \multirow{4}{*}{1.3} & \multirow{4}{*}{2.191} & 1/3 & 1     & 133.1 & 0.4049 & 446  \\

                              & &                      &                          & 1/3 & 3/4   & 131.4 & 0.4284 & 306  \\
                              & &                      &                          & 0   & 3/8   & 127.9 & 0.4490 & 128  \\
                              & &                      &                          & 1   & 1     & 130.0 & 0.4304 & 379  \\
        \hline
        \multirow{4}{*}{$10^2$} & \multirow{4}{*}{0.05} & \multirow{4}{*}{0.8} & \multirow{4}{*}{0.739} & 1/3 & 1     & 274.4 & 0.0251 & 8264 \\

                              & &                      &                          & 1/3 & 3/4   & 274.4 & 0.0251 & 6029 \\
                              & &                      &                          & 0   & 3/8   & 274.4 & 0.0251 & 3023 \\
                              & &                      &                          & 1   & 1     & 274.4 & 0.0251 & 8082 \\
        \hline
            \multirow{4}{*}{$10^4$} & \multirow{4}{*}{0.05} & \multirow{4}{*}{1.3} & \multirow{4}{*}{2.191} & 1/3 & 1     & $1.27 \times 10^4$ & 0.4553 & 330 \\
                              & &                      &                          & 1/3   & 3/4   & $1.20 \times 10^4$ & 0.5330 & 175 \\
                              & &                      &                          & 0   & 3/8   & $9.8 \times 10^3$ & 0.9666 & 32 \\
                              & &                      &                          & 1   & 1     & $1.15 \times 10^4$ & 0.6050 & 183 \\
        \hline
    \end{tabular}
    \caption{Benchmark points, with $\Trh = 10$~MeV, used to obtain the spectra in Figs.~\ref{fig:GWspec_cosmo} ($\mu = 10^2$~MeV) and~\ref{fig:GWspec_cosmo-1e5} ($\mu = 10^4$~MeV).}
    \label{tab:benchmark}
\end{table}

One additionally needs to account for the GW modes that lie beyond the Hubble horizon at the time of the PT~\cite{Allahverdi:2020bys, Gouttenoire:2021jhk, LISACosmologyWorkingGroup:2022jok}. To distinguish these super-horizon modes, we define the frequency corresponding to this time
\begin{equation}
    \label{eq:fSH}
    f_{\rm hor} \equiv \frac{a_*}{a_0}\frac{H_*}{2\pi} \simeq 1.82\times 10^{-11}~{\rm Hz} \left(\frac{10}{\gss(\Trh)}\right)^\frac13 \left(\frac{\gs(\Trh)}{10}\right)^\frac12\left(\frac{T_*}{1\,{\rm MeV}}\right) \left(\frac{T_*}{\Trh}\right)^{\frac{1 + 3\omega}{2 \xi}-1}.
\end{equation}
Then, the spectral shape for modes that re-enter the horizon is given by~\cite{Hook:2020phx, Domenech:2020kqm, Gouttenoire:2021jhk} 
\begin{equation}
    \label{eq:GW_superhor}
     h^2\Omega _{\rm GW} \simeq h^2\Omega_{\text{GW}}^{\rm sw, fit}(f_{\rm hor})\times\left(\frac{f}{f_{\rm hor}}\right)^{\frac{1+15\omega}{1+3\omega}} \quad  \text{ for } f < f_{\rm hor}.
\end{equation}
These super-horizon modes are sensitive to the equation-of-state parameter of the inflaton, $\rho_\chi \propto a^{3(1 + \omega)}$. In particular, for radiation domination $\omega = 1/3$, the scaling is $h^2\Omega_{\rm GW} \sim f^3$ as usual~\cite{Caprini:2009fx, Cai:2019cdl}, while for EMD $(\omega = 0)$ and kination $(\omega = 1)$, we have $\sim f$ and $\sim f^4$, respectively. The factor $h^2\Omega_{\text{GW}}^{\rm sw, fit}(f_{\rm hor})$ allows for continuity with the sub-horizon modes, $f \geq f_{\rm hor}$, which are given by Eq.~\eqref{eq:sw_new_fit}.

\begin{figure}[t!]
    \centering
    \includegraphics[width=0.49\columnwidth]{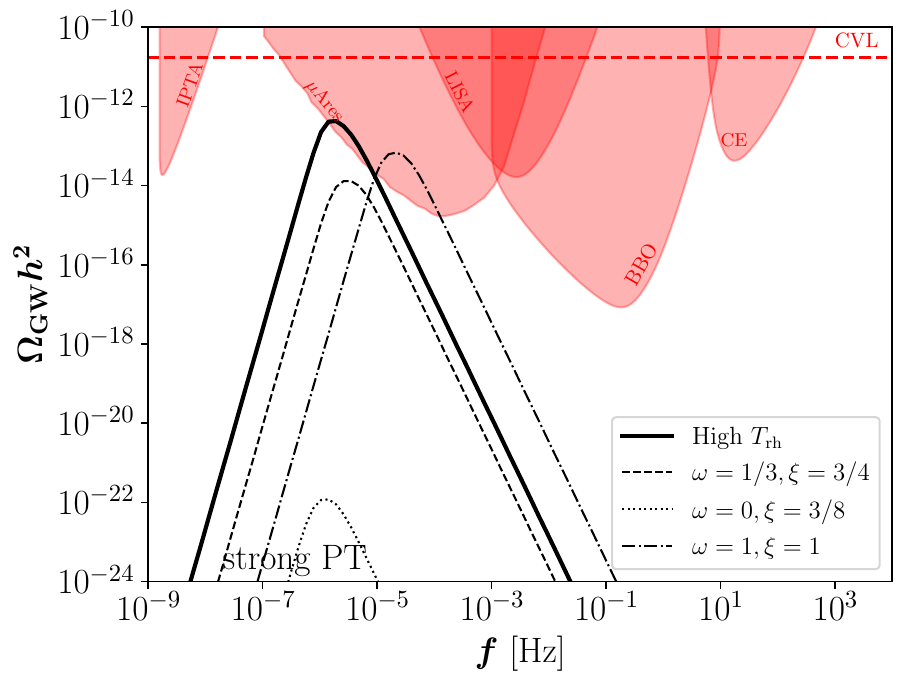}
    \includegraphics[width=0.49\columnwidth]{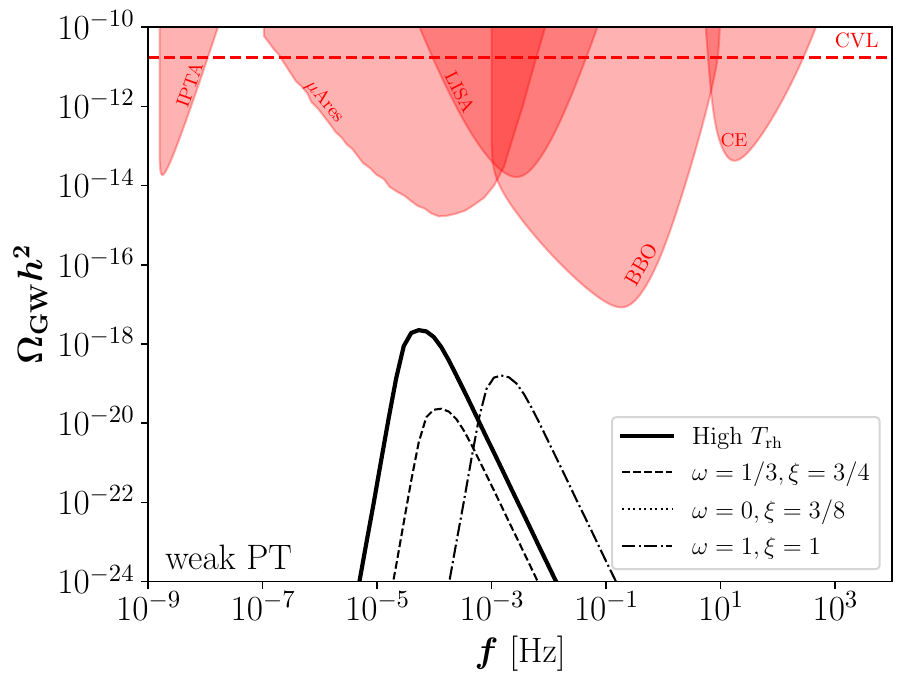}
    \caption{Examples of GW spectra arising from PTs occurring during various reheating scenarios. We have set $\Trh = 10$~MeV and $\mu = 10~\Trh$, with $(\lambda,\, g_X) = (0.05,\, 1.3)$ for a strong first-order PT (left panel) and $(\lambda,\, g_X) = (0.05,\, 0.8)$ for a weak first-order PT (right panel); see Table~\ref{tab:benchmark}. The spectrum corresponding to EMD is suppressed out of the scale of the $y$-axis and therefore absent in the right plot.
    }
    \label{fig:GWspec_cosmo}
\end{figure} 

Figure~\ref{fig:GWspec_cosmo} presents the GW spectra sourced from the sound waves, determined from the PT and the cosmological parameters defined in Table~\ref{tab:benchmark}. The left panel corresponds to a strong first-order PT with $g_X = 1.3$, while the right panel corresponds to a weak first-order PT with $g_X = 0.8$. We have also taken $\lambda = 0.05$ and $\mu = 10~\Trh$. For these two cases, we report the value of the ratio $v_c/T_c$, which is a proxy for the PT strength. Four cosmological scenarios were used: $(\omega,\,\ \xi) = (1/3,\, 1)$ which is equivalent to the case with a high reheating temperature, $(1/3,\, 3/4)$ corresponding to the case of a relativistic inflaton decaying into fermions, $(0,\, 3/8)$ EMD, and $(1,\, 1)$ kination. For each point, the values of $T_*$, $\alpha_*$ and $\beta/H_*$ are reported in Table~\ref{tab:benchmark} and used to obtain the spectra. In Fig.~\ref{fig:GWspec_cosmo} we also present with red-shaded areas the power-law integrated sensitivity (PLIS) curves of several proposed GW detectors, including the Laser Interferometer Space Antenna (LISA)~\cite{LISA:2017pwj}, the Cosmic Explorer (CE)~\cite{Reitze:2019iox}, the Big Bang Observer (BBO)~\cite{Crowder:2005nr, Corbin:2005ny, Harry:2006fi}, $\mu$ARES~\cite{Sesana:2019vho} and the International Pulsar Timing Array (IPTA)~\cite{Antoniadis:2022pcn, IPTA:2023ero}. Furthermore, the energy stored in GWs behaves similarly to dark radiation, contributing to the effective number of neutrino species, $N_\text{eff}$. We therefore include a limit of $\Delta N_\text{eff} \lesssim 3 \times 10^{-6}$, reported in Ref.~\cite{Ben-Dayan:2019gll}, based on a hypothetical cosmic-variance-limited (CVL) CMB polarization experiment for completeness.

Compared to the standard case with high temperature reheating (thick solid black lines), we observe in Fig.~\ref{fig:GWspec_cosmo} that despite having longer PTs (smaller $\beta/H_*$) and stronger PTs in various reheating scenarios such as radiation-domination (dashed lines) or EMD (black dotted lines), the overall effect is a suppression of the amplitude. This is expected due to the additional energy density of the component $R_{\chi}$ in \eqref{eq:sw_sim_amp}, resulting in a dilution of the energy budget of the GWs~\cite{Allahverdi:2020bys, Gouttenoire:2021jhk, LISACosmologyWorkingGroup:2022jok}. For EMD, the suppression is more pronounced due to the redshift factor~\eqref{eq:R}. For the radiation-dominated scenario, this is absent $(\omega =1/3)$, and the enhancement from the PT parameters results in the overall amplitude dropping by approximately 2 orders of magnitude, with the spectrum being shifted to higher frequencies by a factor of approximately 10. In particular, the GW spectrum arising from a PT occurring during kination (black dashed-dotted lines) has an enhancement to the redshift factor, with the net result being a modest suppression of $\sim 10$ to the peak amplitude, and the peak frequency of the spectrum shifting to within the PLIS of $\mu$ARES. Quantitatively for detection prospects, based on a calculation of the signal-to-noise ratio (SNR, see Appendix~\ref{app:snr}), we find the GW spectrum from kination cosmology is now reported with an ${\rm SNR} \simeq 650$, as opposed to the original spectrum from high temperature reheating, a proxy for standard cosmology, which reports ${\rm SNR} \simeq 50$. Therefore, if parameters of the potential were probed in laboratory experiments, such as through dark photon searches~\cite{Fabbrichesi:2020wbt}, which is the relevant case for this work, the complementary detection of GWs in the future could provide insight into the cosmological history before BBN.

\begin{figure}[t]
    \centering
    \includegraphics[width=0.49\columnwidth]{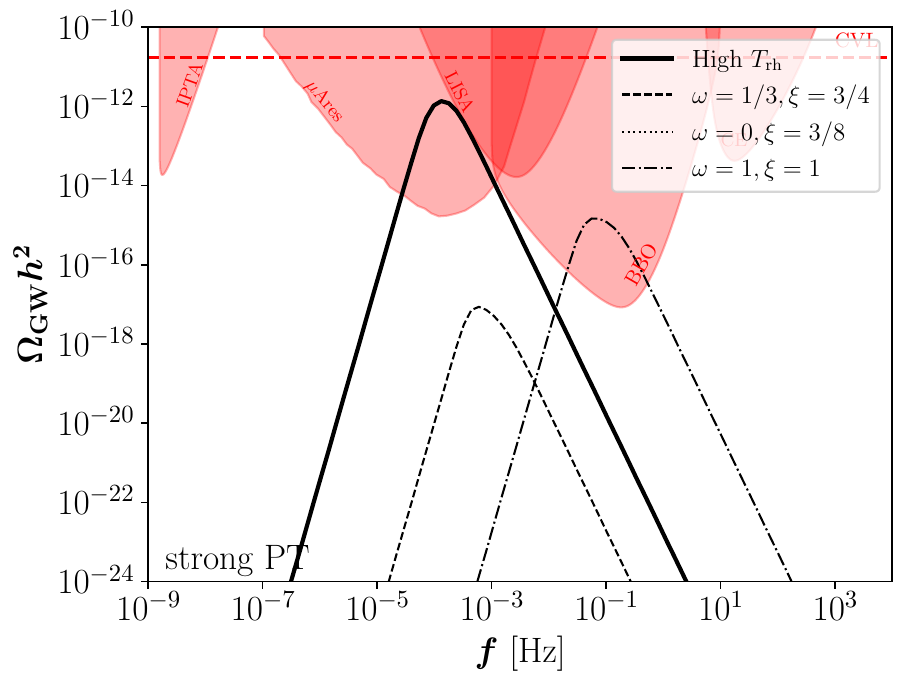}
    \caption{Same as the left panel of Fig.~\ref{fig:GWspec_cosmo}, but for $\mu = 10^3~\Trh$.}
    \label{fig:GWspec_cosmo-1e5}
\end{figure} 

The impact of the cosmological evolution of the background is further demonstrated in cases with a larger hierarchy between $\mu$ and $\Trh$, as the PT has more time to evolve in the reheating era. Figure~\ref{fig:GWspec_cosmo-1e5} compares to the left panel of Fig.~\ref{fig:GWspec_cosmo}, but for $\mu = 10^3~\Trh$. In EMD, the spectrum is again strongly suppressed by several orders of magnitude, whereas for kination, the hierarchy between $\Trh$ and $T_*$ shifts the spectrum from the frequency band of $\mu$ARES to the PLIS of BBO. Specifically, the original spectrum with ${\rm SNR} \simeq 9.5 \times 10^4$ in $\mu$ARES now appears with an appreciable  ${\rm SNR} \simeq 110$ in BBO. 

In principle, these modifications to GW spectra may be mimicked by varying the model parameters, but in a polynomial-like potential~\cite{Caprini:2019egz} we have considered in this work~\eqref{eq:lagrangian}, this would require a degree of fine-tuning of the couplings due to the monotonic dependence of the PT parameters on the Lagrangian parameters, see e.g. Refs.~\cite{Baldes:2018emh, Madge:2018gfl}. We note that the spectral break on account of the transition from super-horizon modes, given by Eq.~\eqref{eq:GW_superhor} to sub-horizon modes, given by Eq.~\eqref{eq:sw_new_fit}, lies well below the sensitivity of the future GW observatories, and would provide a characteristic signature of a PT occurring during reheating. To observe this feature would require a separate model-dependent investigation to generate PT parameters enhancing the overall GW amplitude. This may be achievable in models featuring a large degree of supercooling, such as in Ref.~\cite{Ellis:2020nnr}, and we defer these studies to future work.

\begin{figure}[t!]
    \centering
    \includegraphics[width=0.49\columnwidth]{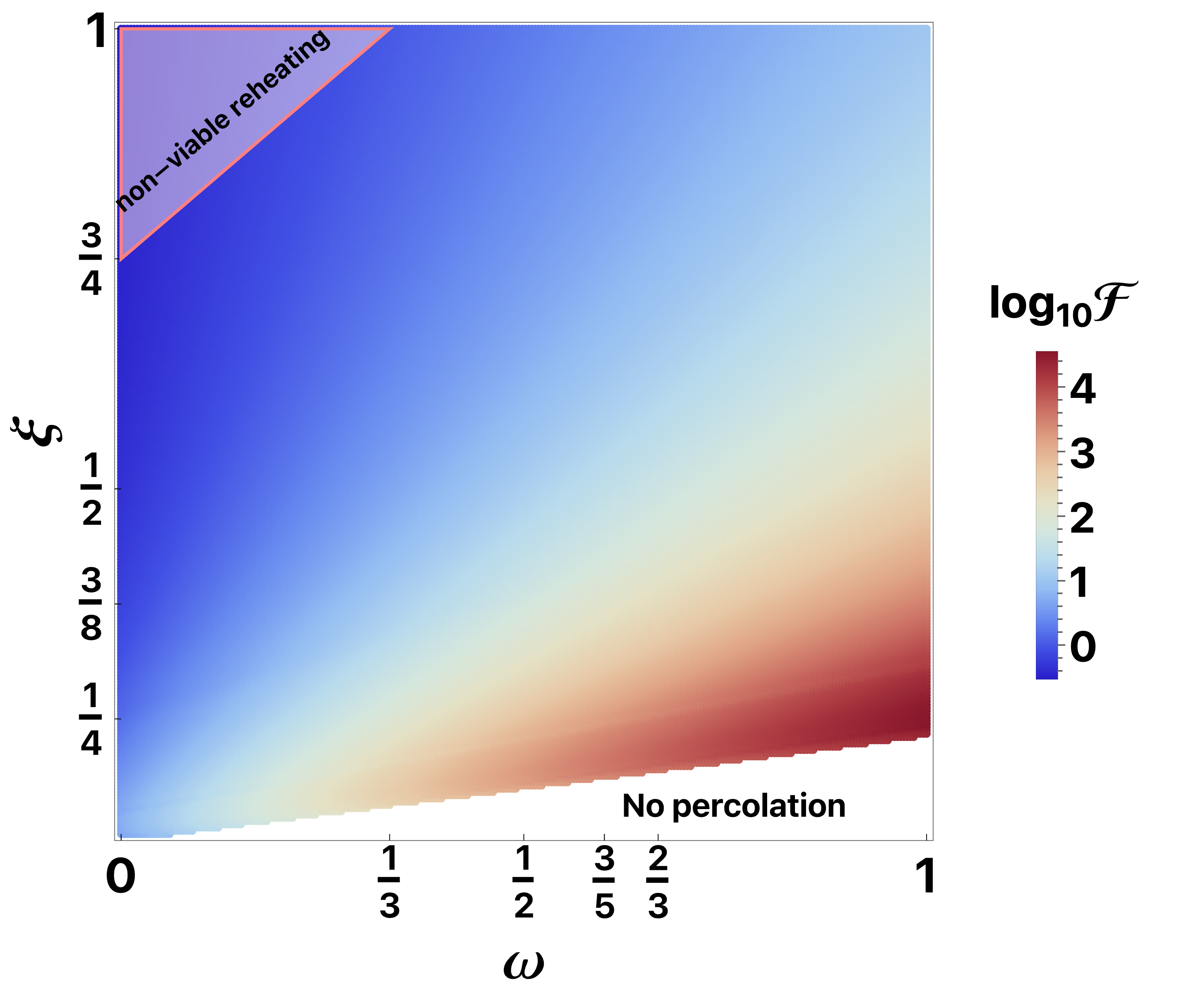}
    \includegraphics[width=0.49\columnwidth]{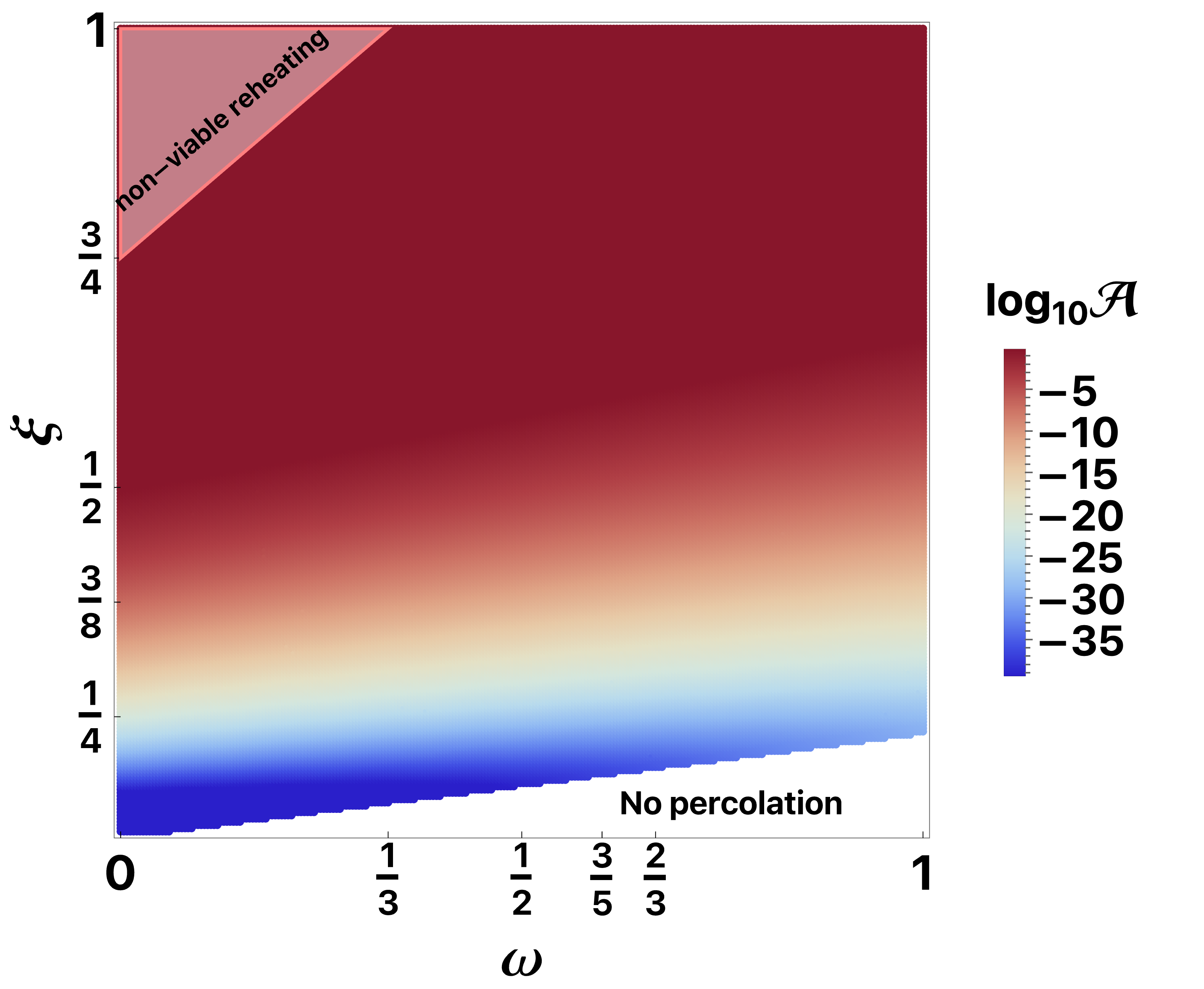}
    \caption{The peak frequency and peak amplitude of acoustic GWs from a PT $(\mu,\, \lambda,\, g_X) = (10\, \Trh,\, 0.05,\, 1.3)$ occurring in different cosmologies, where we have set $\Trh = 10$~MeV. The GWs are blue-shifted for PTs occurring in cosmological scenarios indicated by the red shades of the left plot. The least suppression to the GW amplitude occurs for $\xi \to 1$ and $\omega > 1/3$, such as in kination cosmology.}
    \label{fig:SWpeak_cosmo}
\end{figure} 

To further study the cosmological impact of reheating, we take acoustic GWs as an example and consider in Fig.~\ref{fig:SWpeak_cosmo} the position $(f^{\rm sw}_0)$ and the amplitude of the peak of the spectra (cf. Eq.~\eqref{eq:sw_sim_amp}, with the redshift factor Eq.~\eqref{eq:R}), normalized to the case with a high reheating temperature. We consider a strong PT with the parameters $(\mu,\, \lambda,\, g_X) = (100~{\rm MeV},\, 0.05,\, 1.3)$, and we define
\begin{align}
    \mathcal{F} &\equiv \frac{f^{\rm sw}_0(\omega, \xi)}{f^{\rm sw}_0(\Trh \gg T_*)}\,,\\
    \mathcal{A} &\equiv \frac{\Omega^{\rm sw}_{\rm GW}(f = f^{\rm sw}_0(\omega, \xi))}{\Omega^{\rm sw}_{\rm GW}(f = f^{\rm sw}_0(\Trh \gg T_*))}\,,
\end{align}
which allows us to quantify the total effect on the GW spectra due to the PT occurring during reheating. Although the additional energy density of the inflaton suppresses the overall amplitude, we observe that for $\omega > 1/3$, this suppression becomes less pronounced due to the redshift factor in Eq.~\eqref{eq:R}, and the converse for $\omega < 1/3$. Furthermore, for $\xi \to 0$, on account of the larger inflaton energy density, parameterized as $R_{\chi}$ in \eqref{eq:sw_sim_amp}, the GW spectrum is suppressed the most. Thus, in scenarios where $\xi \gsim 1/2$ and $\omega > 1/3$, the combination of the PT parameters being enhanced and the larger redshift factor results in the least suppression to the amplitude. We note that the peak frequency of the GW spectrum is shifted in directions orthogonal to the line $1+3\omega = 2 \xi$; specifically, blue-shifting in the direction towards the red shades in the left of Fig.~\ref{fig:SWpeak_cosmo} and red-shifting in the opposite direction, towards the blue shades. This can result in the spectrum generated from a PT with the same particle physics parameters, but a different cosmological scenario appearing in a different GW observatory, as discussed previously. Figure~\ref{fig:SWpeak_cosmo}, therefore, demonstrates the full effect of cosmology on the PT dynamics and the GW spectrum. To demonstrate how to disentangle these effects, we provide details in Appendix~\ref{app:details} through our benchmark points in Table~\ref{tab:benchmark}.

\section{Conclusions} \label{sec:concl}
In this work, we present a systematic study of the dynamics of first-order phase transitions (PT) occurring during cosmic reheating and the corresponding gravitational wave (GW) spectrum sourced from sound waves.  The (not instantaneous!) reheating era is described using a general parameterization that assumes $i)$ an effective equation-of-state parameter $\omega$ of the inflaton field, $ii)$ a scaling of the Standard Model (SM) temperature $T(a) \propto a^{-\xi}$, where $a$ is the cosmic scale factor, and $iii)$ a reheating temperature $\Trh$ setting the end of reheating and the beginning of the SM radiation-dominated era.

Compared to the case where PTs occur after reheating, in the radiation-dominated era, the non-standard cosmological evolution \textit{during} reheating implies that PTs are typically delayed (lower nucleation and percolation temperatures $T_n$ and $T_*$, respectively), prolonged in time (smaller $\beta/H_*$) and further strengthened (higher $\alpha_*$), as summarized in our Fig.~\ref{fig:contours_cosmo}. The corresponding GW spectrum, computed using standard formulae in the literature, is immediately modified due to the change of the parameters characterizing the PT: $T_*$, $\beta/H_*$, and $\alpha_*$, induced by the non-standard cosmological evolution during reheating. However, since the GWs are produced during reheating (and not in the standard radiation-dominated era), the peak frequency of the spectra can be shifted to lower or higher values, whereas the GW amplitude is generally suppressed, the extent of which depends on the precise reheating scenario. We allude to this in Fig.~\ref{fig:alpha_omega} and demonstrate this explicitly in Fig.~\ref{fig:SWpeak_cosmo}, by considering the peak frequency and peak amplitude of acoustic GWs. 

Finally, in this work, to highlight the impact of the cosmological background on the PT dynamics, we have considered in Eq.~\eqref{eq:lagrangian} a minimal particle physics model with a spontaneously broken dark $U(1)$ symmetry. However, we expect our conclusions regarding the PT dynamics to carry over to more complex or well-motivated models beyond the SM that are capable of generating first-order PTs. We also emphasize that our application of the expressions for the frequency redshift~\eqref{eq:ftoday}, the amplitude redshift~\eqref{eq:R}, and our implementation of the sound-shell fit ~\eqref{eq:sw_new_fit} within the reheating scenarios considered here provide a systematic implementation of existing non-standard cosmology results for PTs within the reheating scenarios studied here.

\acknowledgments
The authors thank Rouzbeh Allahverdi, Kimmo Kainulainen, and Kuver Sinha for useful discussions. AB is supported by the Basic Science Research Program through the National Research Foundation of Korea (NRF) funded by the Ministry of Education under grant numbers [NRF-2021R1C1C1005076, NRF-2020R1I1A3068803, RS-2026-25484206]. NB received funding from the grants PID2023-151418NB-I00 funded by MCIU/AEI/10.13039/501100011033/ FEDER and PID2022-139841NB-I00 by MICIU/AEI/10.13039/501100011033 and FEDER, UE. FH thanks the organizers of the Mitchell Conference in May 2025 at  Texas A\&M University for their hospitality and support during the final stages of this project. FH is also grateful to the organizers of the workshop of Center for Theoretical Underground Physics and Related Areas (CETUP* - 2025), The Institute for Underground Science at Sanford Underground Research Facility (SURF), Lead, South Dakota, for their hospitality and financial support.

\appendix
\section{Finite-Temperature Effective Potential} \label{app:FTEP}
From our model Lagrangian in Eq.~\eqref{eq:lagrangian}, it is possible to obtain the following scalar potential in terms of a background field $\varphi$~\cite{Anderson:1991zb, Espinosa:1992kf}
\begin{equation}
    V_0(\varphi) = -\frac12\, \mu^2\, \varphi^2 + \frac14\, \lambda\, \varphi^4,
    \label{eq:V0}
\end{equation}
with the tree-level VEV occurring at the minima of this potential, $v_0^2 = \mu^2/\lambda $. The field-dependent squared masses for the physical scalar (i.e. the dark Higgs) $\phi$, the Goldstone boson $\eta$, and the gauge boson $X$ are
\begin{align}
    m^2_\phi(\varphi) &= 3\, \lambda\, \varphi^2 - \mu^2,\label{eq:masses_phi1}\\
    m^2_\eta(\varphi) &= \lambda\, \varphi^2 - \mu^2,\label{eq:masses_phi2}\\
    m^2_{X}(\varphi) &= g^2_X\, \varphi^2.\label{eq:masses_phi3}
\end{align}
The finite-temperature effective potential at one-loop is given by~\cite{Dolan:1973qd, Quiros:1999jp, Laine:2016hma} 
\begin{equation}
    V_{\rm{1-L}} (\varphi,T) = V_{0}(\varphi) + V_{\rm{CW}}(\varphi) + V_T(\varphi,T)\,.
    \label{eq:V1L}
\end{equation}
We work in Landau gauge and employ the on-shell renormalization scheme, which preserves the tree-level relations for $v_0$ and $m^2_\phi(v_0)$. There is a slight subtlety regarding the contribution of the Goldstone boson, which needs to be accounted for separately as it is massless at $v_0$~\cite{Anderson:1991zb, Espinosa:1992kf}. The result for the Coleman-Weinberg potential is accordingly
\begin{align}
    V_{\rm{CW}}(\varphi) &= \displaystyle \sum_{i\neq \eta} \frac{n_i}{64\pi^2} \bigg\{m_i^4(\varphi) \left[\ln\left(\frac{m^2_i(\varphi)}{m^2_i(v_0)}\right) - \frac32\right] + 2 m^2_i(\varphi)\, m^2_i(v_0)\bigg\} \nonumber\\
    &\qquad+ \frac{1}{64\pi^2} m_\eta^4(\varphi) \left[\ln\left(\frac{m^2_\eta(\varphi)}{m^2_\phi(v_0)}\right) - \frac32\right],
    \label{eq:VCW}
\end{align}
where $n_i$ refers to the internal degrees of freedom of the $i^\text{th}$ particle; specifically, $n_{\phi,\eta} = 1,\,n_{A'} = 3$.\footnote{Notably, one ``double counts'' the Goldstone boson by considering its contribution as well as the contribution from the massive gauge boson, giving it 3 degrees of freedom; this procedure has been shown to be correct; see, e.g. Ref.~\cite{Delaunay:2007wb}.} Next, the finite temperature corrections are given by~\cite{Dolan:1973qd, Quiros:1999jp, Laine:2016hma}
\begin{equation}
    V_T(\varphi,T) =\frac{T^4}{2\pi^2}\displaystyle \sum_i n_i \int_0^{\infty} dx\, x^2\,\ln\left[1 - \exp\left(-\sqrt{x^2+ \frac{m^2_i(\varphi)}{T^2}}\right)\right].
    \label{eq:VT}
\end{equation}

Typically, one needs to consider resummation of the bosonic masses in Eqs.~\eqref{eq:masses_phi1} to~\eqref{eq:masses_phi3}, to account for infrared divergences arising at small field values (see, e.g., Ref.~\cite{Laine:2016hma}). The most common methods for such a procedure, widely used in the literature, involve the resummation of daisy diagrams (also known as the Arnold-Espinosa method)~\cite{Carrington:1991hz, Arnold:1992rz}, or the Parwani approximation~\cite{Parwani:1991gq}, whereby one uses the leading-order thermal masses for the bosons in Eqs.~\eqref{eq:VCW} and~\eqref{eq:VT}. This leads to an \textit{improved} thermal effective potential. In this work, we employ the Parwani approximation that allows for continuation between both low- and high-temperature regimes; see, e.g., Refs.~\cite{Cline:2011mm, Laine:2017hdk, Kainulainen:2019kyp, Kainulainen:2021eki}, to consistently track the thermal evolution. This amounts to performing the following replacements for the bosonic masses $m^2_i(\varphi) \to m^2_i(\varphi,T)$ where
\begin{align}
    m^2_{\phi,\,\eta} (\varphi,T) &= m^2_{\phi,\,\eta}(\varphi) + \frac{4 \lambda + 3\, g_X^2}{12}\, T^2, \\
    m^2_X(\varphi,T) &= m^2_X(\varphi) + \frac13\, g^2_X\, T^2.
\end{align}
The requirement for a first-order PT can be met through degenerate minima at $\varphi = 0$ and $\varphi = v_c$ at the critical temperature $T_c$. We can realize this with the following conditions
\begin{align}
    &V_{\rm {1-L}}(v_c,T_c) = V_{\rm {1-L}}(0,T_c)\,,\\
    &\frac{\partial V_{\rm {1-L}}(\varphi,T)}{\partial \varphi}\bigg|_{v_c,T_c} = 0\,,
\end{align}
and for stable minima $\partial^2_{\varphi}V_{\rm {1-L}}(v_c,T_c) >0$.

Although we use the full expression in Eq.~\eqref{eq:VT} to accurately track the thermal evolution, we can gain some insight by considering the high-temperature expansion~\cite{Laine:2016hma}
\begin{equation}
    V_T(\varphi,T) \simeq \displaystyle \sum_{i} n_i \,T^4\left[-\frac{\pi^2}{90} +\frac{m^2_i(\varphi)}{24\,T^2}-\frac{m^3_i(\varphi)}{12\pi\,T^3} + \dots\right].
    \label{eq:VT_highT}
\end{equation}
The cubic term is responsible for generating a loop-induced barrier, thus rendering the PT first-order, by separating the two degenerate minima. The height of this barrier, which is related to the strength of the transition, is directly proportional to the strength of the bosonic couplings and serves as a criterion to classify a first-order PT as strong or weak.

\section{Phase Transition Temperatures} \label{app:perc_temp}
Firstly, Eq.~\eqref{eq:nucl_cond} allows us to derive an analytical estimate of the ratio $S_3/T$ when $T = T_n$. Using Eqs.~\eqref{eq:Hrh} and~\eqref{eq:HvsT} one obtains, during reheating,
\begin{align}
    \left.\frac{S_3}{T}\right|_{T = T_n} &= -2 \ln\left[(2\pi)^\frac34\, \left(\frac{\Hrh}{\Trh}\right)^2\right] + 4\left[1 - \frac{3 (1 + \omega)}{2\xi}\right] \ln\left(\frac{T_n}{\Trh}\right) + \frac{3}{2} \ln\left(\frac{S_3}{T}\right)\bigg|_{T=T_n} \nonumber\\
    &\simeq 193 - 2\ln\left(\frac{\gs(\Trh)}{10}\right ) - 4 \ln\left(\frac{\Trh}{10~\text{MeV}}\right) + 4 \left[1 - \frac{3 (1+\omega)}{2\xi}\right] \ln\left(\frac{T_n}{\Trh}\right)\,.
    \label{eq:S3T_cosmo}
\end{align}
Setting $(\omega, \xi) = (1/3,1)$, which is a proxy for the high $\Trh$ scenario discussed in the main text, gives
\begin{equation}
    \left.\frac{S_3}{T}\right|_{T = T_n} \simeq 193 - 2\ln\left(\frac{\gs(\Trh)}{10}\right ) - 4 \ln\left(\frac{T_n}{10~\text{MeV}}\right) 
\end{equation}
whereby the explicit dependence on the reheat temperature vanishes and enters only through $\gs$. Note that this expression is familiar when PTs occur in radiation domination, with the replacement $\gs(\Trh) \to \gs(T_*)$; see, e.g. Ref.~\cite{Breitbach:2018ddu}.

\begin{table}[t!]
    \centering
    \begin{tabular}{|cccc|cc||ccc|}
        \hline
        $\boldsymbol \mu$~[MeV] & $\boldsymbol \lambda$ & $\boldsymbol{g_X}$ & $\boldsymbol{T_c}$~[MeV] & $\boldsymbol \omega$ & $\boldsymbol \xi$ & $\boldsymbol{T_n}$~[MeV] & $\boldsymbol{T_p}$~[MeV] & $\boldsymbol{\frac{T_n-T_p}{T_n}}~[\%]$ \\
        \hline \hline
        \multirow{4}{*}{$10^2$} & \multirow{4}{*}{0.05} & \multirow{4}{*}{1.3} & \multirow{4}{*}{188.3} & 1/3 & 1     & 138.5 & 133.1 & 3.94  \\
                              & &                      &                          & 1/3 & 3/4   & 136.8 & 131.4 & 3.97   \\
                              & &                      &                          & 0   & 3/8   & 133.0 & 127.9 & 3.89  \\
                              & &                      &                          & 1   & 1     & 135.9 & 130.0 & 4.38  \\
        \hline
        \multirow{4}{*}{$10^2$} & \multirow{4}{*}{0.05} & \multirow{4}{*}{0.8} & \multirow{4}{*}{279.1} & 1/3 & 1     & 276.2 & 274.4 & 0.67 \\
                                & &                      &                          & 1/3 & 3/4   & 276.1 & 274.4 & 0.67 \\
                              & &                      &                          & 0   & 3/8   & 276.2 & 274.4 & 0.66 \\
                              & &                      &                          & 1   & 1     & 276.2 & 274.4 & 0.67 \\
                \hline
        \multirow{4}{*}{$10^4$} & \multirow{4}{*}{0.05} & \multirow{4}{*}{1.3} & \multirow{4}{*}{$1.88 \times 10^4$} & 1/3 & 1     & $1.34 \times 10^4$ &  $1.27 \times 10^4$ & 4.87 \\
                              & &                      &                          & 1/3 & 3/4   & $1.27 \times 10^4$ & $1.20 \times 10^4$ & 5.82 \\
                              & &                      &                          & 0   & 3/8   & $1.07 \times 10^4$ &  $9.81 \times 10^3$ & 8.50 \\
                              & &                      &                          & 1   & 1     &  $1.24 \times 10^4$ &  $1.15 \times 10^4$ & 7.18 \\
        \hline
    \end{tabular}
    \caption{Critical temperatures for a strong and weak PT, with the nucleation and percolation temperatures for the benchmark scenarios used in Table~\ref{tab:benchmark}, for $\Trh = 10$~MeV.}
    \label{tab:Tperc}
\end{table}

For the percolation temperature, we provide additional details regarding its computation for PTs occurring during reheating. Starting from the quantity entering the volume fraction of the false vacuum in Eq.~\eqref{eq:vol_frac}, based on our cosmological setup, we discuss the following scenarios:
\begin{equation}
    \frac{a(\overline{T})\, r(T,\overline{T})}{v_w} 
    \simeq
    \begin{dcases}
        \frac{1}{\Hrh}\, \frac{\Trh^2}{T\, \overline{T}} \left[1 - \frac{T}{\overline{T}}\right] \hspace{5cm} \text{for } \Trh > T_c > \overline{T} > T,\\
        \frac{2}{1 + 3\omega}\, \frac{1}{\Hrh} \left(\frac{\Trh}{\overline{T}}\right)^\frac{1}{\xi} \left[1 - \left(\frac{\Trh}{\overline{T}}\right)^\frac{1+3\omega}{2\xi}\right] + \frac{1}{\Hrh}\, \frac{\Trh^2}{T\, \overline{T}} \left[1 - \frac{T}{\Trh}\right]\\
        \hspace{8cm} \text{for } T_c > \overline{T} > \Trh > T,\\
        \frac{2}{1 + 3\omega}\, \frac{1}{\Hrh} \left(\frac{\Trh}{T}\right)^\frac{1+3\omega}{2\xi} \left(\frac{\Trh}{\overline{T}}\right)^\frac{1}{\xi} \left[1 - \left(\frac{T}{\overline{T}}\right)^\frac{1+3\omega}{2\xi}\right]\\
        \hspace{8cm} \text{for } T_c >\overline{T} > T > \Trh.
    \end{dcases}
\end{equation}
We can then write down 
\begin{equation}
    I(T)  \simeq
    \begin{dcases}
        \frac{4\pi}{3}\int_{T}^{T_c} \frac{d\overline{T}\,}{\overline{T}} \frac{\Gamma_N(\overline{T})}{H(\overline{T})} \left[a(\overline{T})\, r(T,\overline{T})\right]^3 \hspace{4.2cm}\text{for } \Trh > T_c > T,\\[1mm]
        \frac{4\pi}{3}\left[\int_{T}^{\Trh} \frac{d\overline{T}\,}{\overline{T}} \frac{\Gamma_N(\overline{T})}{H(\overline{T})} \left[a(\overline{T})\, r(T,\overline{T})\right]^3 +\int_{\Trh}^{T_c} \frac{d\overline{T}\,}{\xi \overline{T}} \frac{\Gamma_N(\overline{T})}{H(\overline{T})} \left[a(\overline{T})\, r(T,\overline{T})\right]^3\right]\\
        \hspace{10cm}\text{for } T_c > \Trh > T,\\[1mm]
       \frac{4\pi}{3\xi}\int_{T}^{T_c} \frac{d\overline{T}\,}{\overline{T}} \frac{\Gamma_N(\overline{T})}{H(\overline{T})} \left[a(\overline{T})\, r(T,\overline{T})\right]^3\\
       \hspace{10cm}\text{for } T_c > T > \Trh.
    \end{dcases}
\end{equation}

In this work, we have focused mainly on the GWs sourced from the PT, and the impact of the non-standard cosmological evolution during reheating on the GW spectrum, and therefore have considered only the case $T_c > T > \Trh$. The first case $\Trh > T_c > T$ is the expression considered for standard cosmological scenarios with radiation domination (such as used in Ref.~\cite{Ellis:2020nnr}). Finally, the condition of having the false vacuum shrinking requires
\begin{equation}
    \left[T\,\frac{d I(T)}{dT}\right]\Bigg|_{T = T_p} \nonumber <
    \begin{dcases}
        3 &\text{for } \Trh > T_p\,,\\
        \frac{3}{\xi} &\text{for }  T_p > \Trh\,.
    \end{dcases}
\end{equation}
In Table~\ref{tab:Tperc}, we give the nucleation and percolation temperatures for a strong and weak PT in the benchmark scenarios used in Table~\ref{tab:benchmark}.

\section{Comparison of the Sound Wave Spectra} \label{app:sw_spectra}
In this appendix, we show a comparison between the acoustic GW spectra using the precise fitting formula obtained in the sound-shell model~\cite{Guo:2024gmu}, cf. Eq.~\eqref{eq:sw_new_fit}, and the widely used formula~\cite{Weir:2017wfa, Caprini:2018mtu, Caprini:2019egz}, based on fitting to simulations~\cite{Hindmarsh:2015qta, Hindmarsh:2017gnf}, cf. Eqs.~\eqref{eq:sw_spec_sim} and~\eqref{eq:sw_sim_amp}. 

\begin{table}[t!]
    \centering
    \begin{tabular}{|ccc|cc||ccccc|}
        \hline
        $\boldsymbol \mu$~[MeV] & $\boldsymbol \lambda$ & $\boldsymbol{g_X}$ & $\boldsymbol \omega$ & $\boldsymbol \xi$ & $\boldsymbol{\Omega_p}$ & $\boldsymbol{\tilde{s}_0}$~[Hz] & $\boldsymbol{\tilde{r}_b}$ & $\boldsymbol{\tilde{a}}$ & $\boldsymbol{\tilde{b}}$ \\
        \hline \hline
        \multirow{4}{*}{$10^2$} & \multirow{4}{*}{0.05} & \multirow{4}{*}{1.3} & 1/3 & 1     & $5.59\times10^{-13}$ & $1.20\times10^{-6}$ & 2.45 & 3.94 & 9.26 \\
                              &                        &                      & 1/3 & 3/4   & $5.12\times10^{-11}$ & $8.51\times10^{-5}$ & 7.55 & 3.98 & 9.52 \\
                              &                        &                      & 0   & 3/8   & $5.18\times10^{-12}$ & $7.39\times10^{-7}$ & 2.37 & 3.96 & 8.51 \\
                              &                        &                      & 1   & 1     & $5.57\times10^{-13}$ & $1.17\times10^{-5}$ & 2.76 & 3.95 & 8.70 \\
        \hline
        \multirow{4}{*}{$10^2$} & \multirow{4}{*}{0.05} & \multirow{4}{*}{0.8} & 1/3 & 1     & $1.34\times10^{-18}$ & $2.98\times10^{-5}$ & 2.96 & 0.92 & 8.47 \\
                              &                        &                      & 1/3 & 3/4   & $1.10\times10^{-17}$ & $1.29\times10^{-1}$ & 3.97 & 1.10 & 7.95 \\
                              &                        &                      & 0   & 3/8   & $9.63\times10^{-18}$ & $3.28\times10^{-5}$ & 2.96 & 0.92 & 8.47 \\
                              &                        &                      & 1   & 1     & $1.40\times10^{-18}$ & $8.00\times10^{-4}$ & 2.96 & 0.92 & 8.47 \\
        \hline
        \multirow{3}{*}{$10^4$} & \multirow{3}{*}{0.05} & \multirow{3}{*}{1.3} & 1/3 & 1     & $8.41\times10^{-13}$ & $8.12\times10^{-5}$ & 2.37 & 3.96 & 8.51 \\
                              &                        &                      & 0   & 3/8   & $1.79\times10^{-10}$ & $3.92\times10^{-5}$ & 6.00 & 4.00 & 9.08 \\
                              &                        &                      & 1   & 1     & $4.46\times10^{-12}$ & $4.03\times10^{-2}$ & 3.82 & 3.98 & 9.23 \\
        \hline
    \end{tabular}
    \caption{Fit parameters obtained for the benchmark points in Table~\ref{tab:benchmark}, for the sound wave spectrum in Eq.~\eqref{eq:sw_new_fit}.}
    \label{tab:fit_param}
\end{table}

\begin{figure}[t!]
    \centering
    \includegraphics[width=0.49\columnwidth]{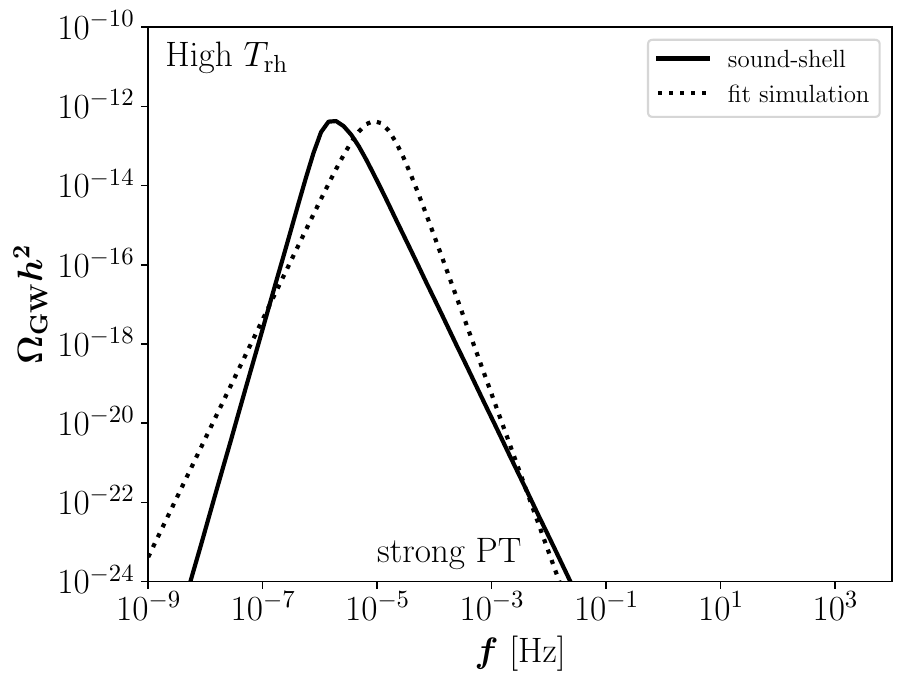}
    \includegraphics[width=0.49\columnwidth]{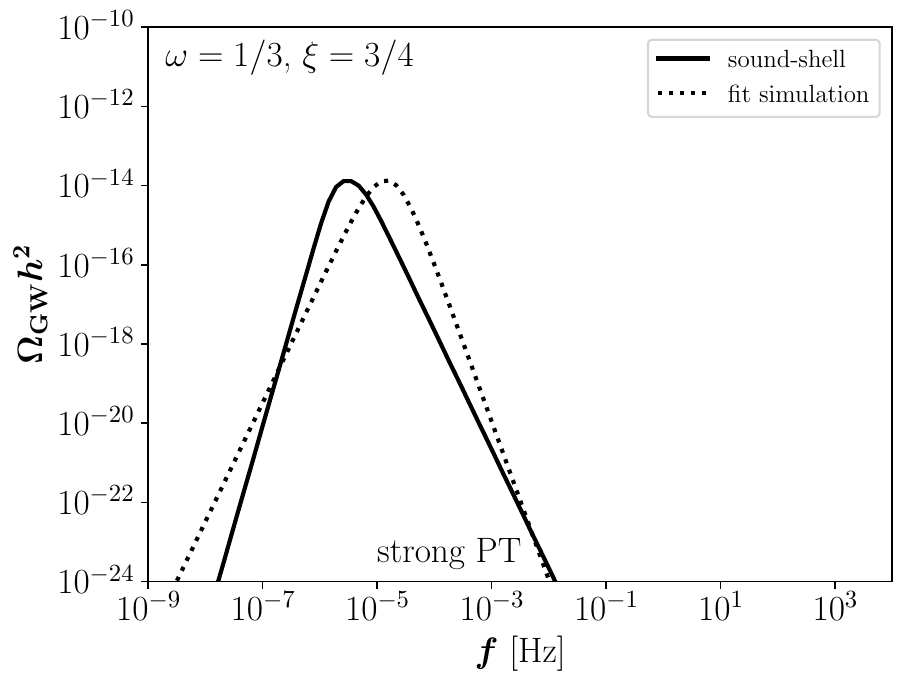}
    \includegraphics[width=0.49\columnwidth]{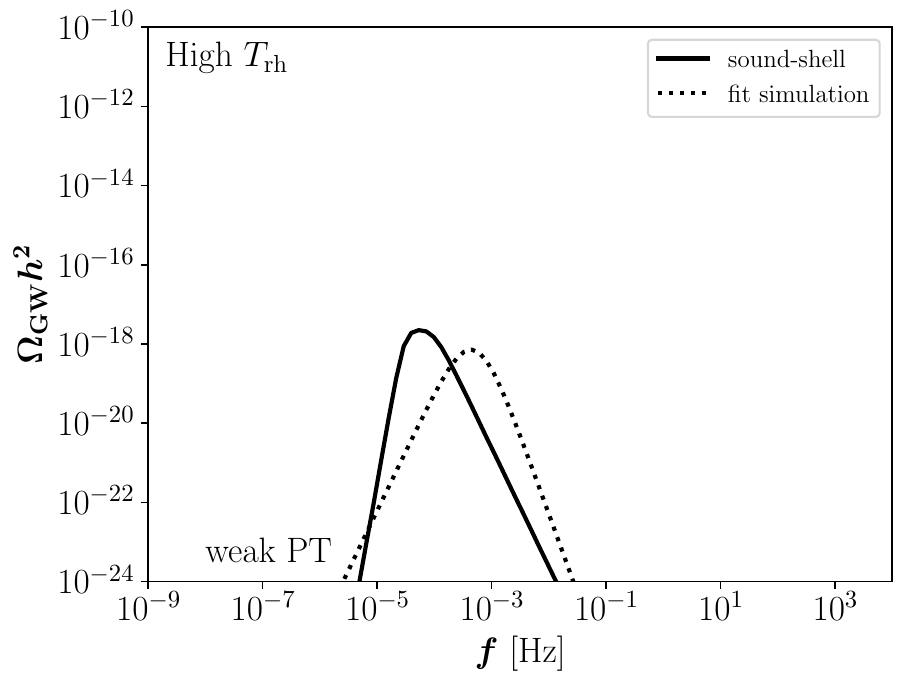}
    \includegraphics[width=0.49\columnwidth]{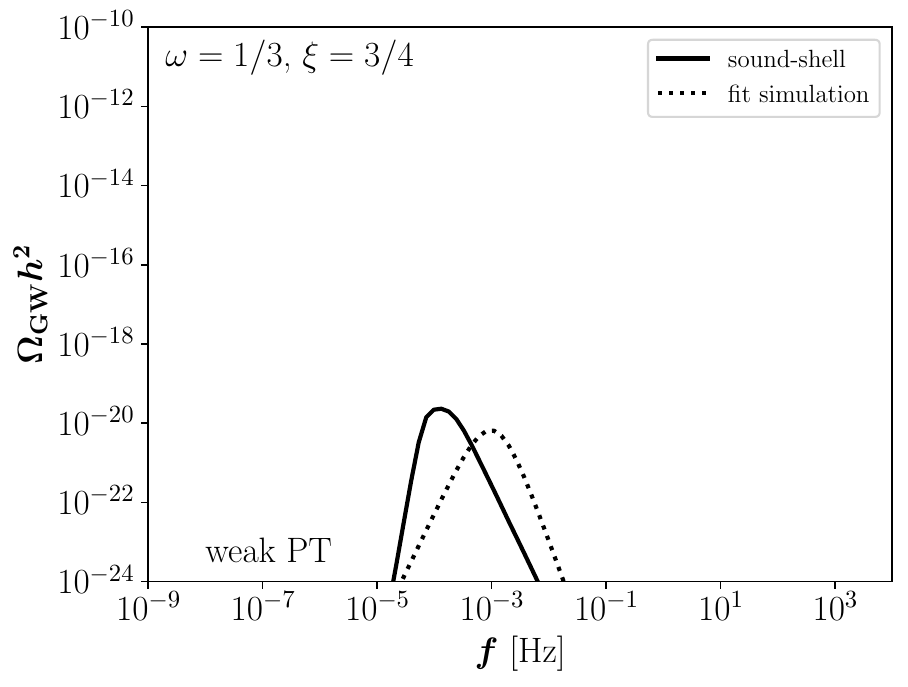}
    \caption{Comparison of the spectra of acoustic GWs from the sound-shell model (solid lines) and from the formulae based on fitting to simulations (dotted lines). The top (bottom) row depicts spectra arising from a strong (weak) PT, whereas the left (right) column depicts a PT occurring in standard cosmology (radiation domination).
    }
    \label{fig:GWsw}
\end{figure}

In Fig.~\ref{fig:GWsw}, we show the different spectra arising from PTs with $(\mu, \lambda) = (100~ {\rm MeV}, 0.05)$, and choosing gauge couplings $g_X = 1.3$ and $g_X = 0.8$, corresponding to a strong and a weak PT, respectively. The cosmology does not play a primary role in comparing the spectra, so we restrict ourselves to the case of high temperature reheating (equivalent to standard cosmology) and the radiation-dominated scenario $(\omega = 1/3,\, \xi = 3/4)$. The PT parameters used to generate these spectra are obtained from Table~\ref{tab:benchmark}. 

We observe that the sound-shell model notably has its peak shifted to lower frequencies by about an order of magnitude in comparison to the simulation. This results in the peak aligning closer to the peak frequency from bubble collisions, c.f. Eq.~\eqref{eq:fstar}. Furthermore, the peak is no longer as prominent as in the simulation fit; the sound-shell model involves a double-broken power law with two characteristic frequencies, arising from considering the spacing between bubbles and thickness of sound shells~\cite{Xiao:2024rsj}. Finally, we note that, although the amplitudes are comparable in the case of strong PTs, there is a slight enhancement of the amplitude for the weak PTs.

Finally, in Table~\ref{tab:fit_param} we give the parameters used to fit the GW spectra obtained for the benchmark points in Table~\ref{tab:benchmark}, for the sound wave spectrum in Eq.~\eqref{eq:sw_new_fit}.

\section{Signal-to-Noise Ratio of a GW Signal} \label{app:snr}
The potential for detecting a signal in a given experiment can be quantified through the signal-to-noise ratio (SNR), given by \cite{Thrane:2013oya,Caprini:2015zlo}
\begin{equation}
    {\rm SNR} \equiv 
    \sqrt{
    \tau_{\rm obs} \int_{f_{\rm min}}^{f_{\rm max}} df
    \left(\frac{h^2\Omega_{\rm GW}(f)}{h^2\Omega^{\rm noise}_{\rm det}(f)}\right)
    }\,.
\end{equation}
Here, $\tau_{\rm obs}$ refers to the observation time of the detector under consideration, $\Omega_{\rm GW}$ is the prediction of the spectrum from theory (first-order PT), and we have quantified the expected sensitivity of the detector through:
\begin{equation}
    \Omega^{\rm noise}_{\rm det}(f) \equiv \frac{4\pi^2}{3H_0^2}f^3 S_{\rm det}(f)
\end{equation}
where $S_{\rm det}(f)$ refers to the noise power spectral density of the detector. If the calculated SNR is above a threshold value, often taken as ${\rm SNR_{\rm thr}} = 10$, then the signal is expected to be detectable by the corresponding experiment.

\section{Reheating Strikes the GW Spectrum Twice} \label{app:details}
The impact on the GW spectrum due to the change of background is two-fold. On the one hand, for fixed values in the Lagrangian density ($\lambda$, $g_X$ and $\mu$), the parameters characterizing the PT (that is, $T_*$, $\beta/H_*$ and $\alpha_*$) and controlling the GW spectrum change if $\omega$, $\xi$ and $\Trh$ are modified. This impact may be understood, for example, by examining the amplitudes, see Eqs.~\eqref{eq:sw_sim_amp}. On the other hand, from a phenomenological perspective, even if one uses $T_*$, $\beta/H_*$ and $\alpha_*$ as the initial input parameters, the GW spectrum also strongly depends on $\omega$, $\xi$ and $\Trh$, such as through the various redshift factors, cf. Eqs.~\eqref{eq:ftoday} and~\eqref{eq:R}.

\begin{figure}[t!]
    \def\sepf{0.32}
    \centering
    \includegraphics[width=\sepf\columnwidth]{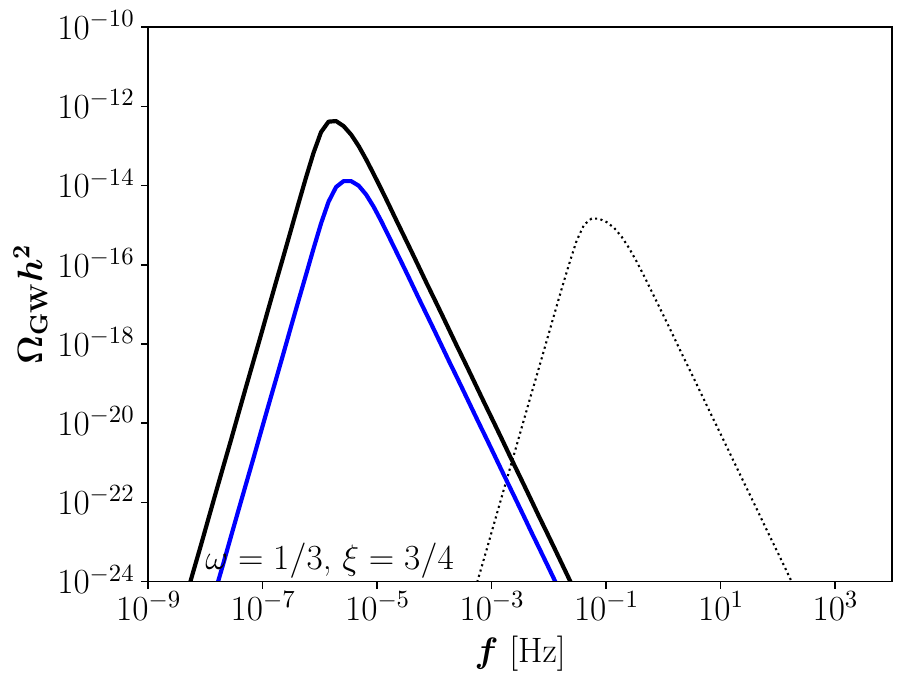}
    \includegraphics[width=\sepf\columnwidth]{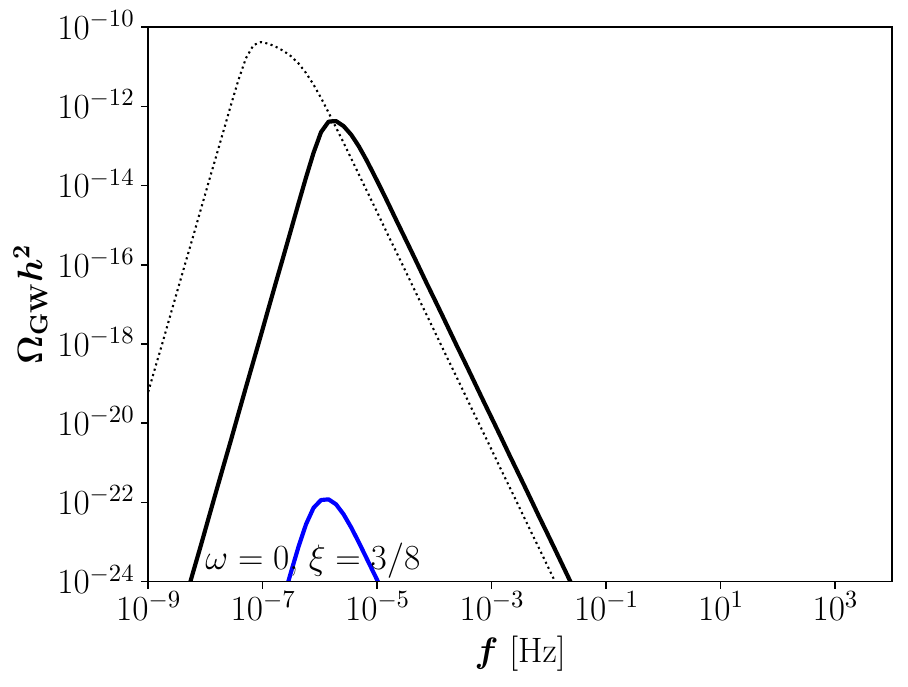}
    \includegraphics[width=\sepf\columnwidth]{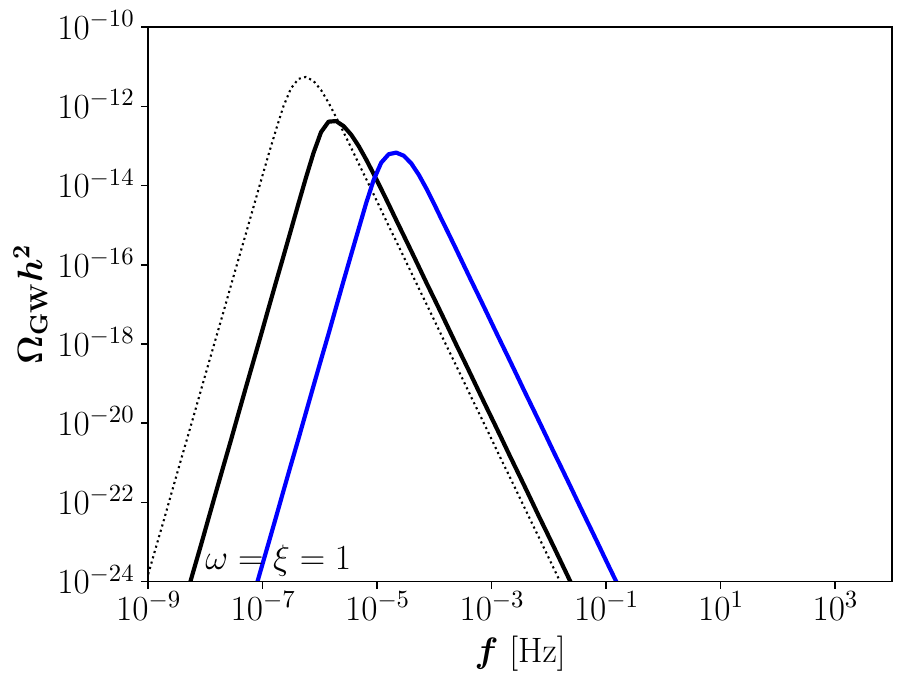}
    \caption{GW spectra for the first benchmark point in Table~\ref{tab:benchmark}. The thick black line corresponds to the standard case with high reheating-temperature, while the blue lines to $\Trh = 10$~MeV with $\omega = 1/3$ and $\xi = 3/4$ (left panel), $\omega = 0$ and $\xi = 3/8$ (central panel), and $\omega = \xi = 1$ (right panel). The thin dotted lines represent the {\it unphysical} intermediate state described in the text.}
    \label{fig:GWs}
\end{figure}

This double effect is illustrated in Fig.~\ref{fig:GWs}, where the thick black lines correspond to the GW spectrum for the first benchmark point in Table~\ref{tab:benchmark} ($\mu = 10^2$~MeV, $\lambda = 0.05$, and $g_X = 1.3$), with a large $\Trh$. The blue lines in the three panels correspond to three different cosmologies: $\omega = 1/3$ and $\xi = 3/4$ (left panel), $\omega = 0$ and $\xi = 3/8$ (central panel) and $\omega = \xi = 1$ (right panel). To understand the impact of the background, we also plot with thin dotted lines the spectra corresponding to $T_*$, $\beta/H_*$ and $\alpha_*$ from the PT during reheating, but {\it artificially} keeping $\omega = 1/3$ and $\xi =1$, as in the standard radiation-dominated era, i.e., ignoring essentially the effect of the cosmology, and focusing on the enhancement from the PT dynamics.

For the left panel in Fig.~\ref{fig:GWs}, with respect to the case where the PT occurs after reheating, $T_*$ decreases, $\alpha_*$ increases, and $\beta/H_*$ decreases, which implies a reduction in the peak frequency, an increase in the amplitude, and a double effect of reduction of the peak frequency with an increase in the amplitude, respectively, corresponding to the thin black dotted line. Now, $\omega = 1/3$ implies that there is no change in the amplitude, but $\xi = 3/4$ induces an increase in the peak frequency. All in all, the overall effect is a decrease in amplitude from the additional energy density component and an increase in the frequency of the GW spectrum, as shown with the blue line. The central panel in Fig.~\ref{fig:GWs}, corresponding to $\omega = 0$ and $\xi = 3/8$, resembles the previous case, the main difference being a large suppression of the amplitude due to the value of the equation-of-state parameter. Finally, in the right panel $\omega = \xi = 1$ and the values of $T_*$, $\alpha_*$, and $\beta/H_*$ are almost unaffected by the change of background, which is reflected in the fact that the solid and dotted black lines almost overlap. However, as $1 + 3\, \omega > 2\, \xi$ and $\omega > 1/3$, the final GW spectrum suffers the least suppression because the larger redshift factor somewhat compensates for the additional energy density of the inflation, and the spectrum is shifted to higher frequencies.

\bibliographystyle{JHEP}
\bibliography{biblio}

@article{LIGOScientific:2016aoc,
    author = "Abbott, B. P. and others",
    collaboration = "LIGO Scientific, Virgo",
    title = "{Observation of Gravitational Waves from a Binary Black Hole Merger}",
    eprint = "1602.03837",
    archivePrefix = "arXiv",
    primaryClass = "gr-qc",
    reportNumber = "LIGO-P150914",
    doi = "10.1103/PhysRevLett.116.061102",
    journal = "Phys. Rev. Lett.",
    volume = "116",
    number = "6",
    pages = "061102",
    year = "2016"
}

@article{LIGOScientific:2017vwq,
    author = "Abbott, B. P. and others",
    collaboration = "LIGO Scientific, Virgo",
    title = "{GW170817: Observation of Gravitational Waves from a Binary Neutron Star Inspiral}",
    eprint = "1710.05832",
    archivePrefix = "arXiv",
    primaryClass = "gr-qc",
    reportNumber = "LIGO-P170817",
    doi = "10.1103/PhysRevLett.119.161101",
    journal = "Phys. Rev. Lett.",
    volume = "119",
    number = "16",
    pages = "161101",
    year = "2017"
}

@article{LIGOScientific:2017ync,
    author = "Abbott, B. P. and others",
    collaboration = "LIGO Scientific, Virgo, Fermi GBM, INTEGRAL, IceCube, AstroSat Cadmium Zinc Telluride Imager Team, IPN, Insight-Hxmt, ANTARES, Swift, AGILE Team, 1M2H Team, Dark Energy Camera GW-EM, DES, DLT40, GRAWITA, Fermi-LAT, ATCA, ASKAP, Las Cumbres Observatory Group, OzGrav, DWF (Deeper Wider Faster Program), AST3, CAASTRO, VINROUGE, MASTER, J-GEM, GROWTH, JAGWAR, CaltechNRAO, TTU-NRAO, NuSTAR, Pan-STARRS, MAXI Team, TZAC Consortium, KU, Nordic Optical Telescope, ePESSTO, GROND, Texas Tech University, SALT Group, TOROS, BOOTES, MWA, CALET, IKI-GW Follow-up, H.E.S.S., LOFAR, LWA, HAWC, Pierre Auger, ALMA, Euro VLBI Team, Pi of Sky, Chandra Team at McGill University, DFN, ATLAS Telescopes, High Time Resolution Universe Survey, RIMAS, RATIR, SKA South Africa/MeerKAT",
    title = "{Multi-messenger Observations of a Binary Neutron Star Merger}",
    eprint = "1710.05833",
    archivePrefix = "arXiv",
    primaryClass = "astro-ph.HE",
    reportNumber = "LIGO-P1700294, VIR-0802A-17, FERMILAB-PUB-17-478-A-AE-CD",
    doi = "10.3847/2041-8213/aa91c9",
    journal = "Astrophys. J. Lett.",
    volume = "848",
    number = "2",
    pages = "L12",
    year = "2017"
}

@article{LIGOScientific:2018mvr,
    author = "Abbott, B. P. and others",
    collaboration = "LIGO Scientific, Virgo",
    title = "{GWTC-1: A Gravitational-Wave Transient Catalog of Compact Binary Mergers Observed by LIGO and Virgo during the First and Second Observing Runs}",
    eprint = "1811.12907",
    archivePrefix = "arXiv",
    primaryClass = "astro-ph.HE",
    reportNumber = "LIGO-P1800307",
    doi = "10.1103/PhysRevX.9.031040",
    journal = "Phys. Rev. X",
    volume = "9",
    number = "3",
    pages = "031040",
    year = "2019"
}

@article{LISA:2017pwj,
    author = "Amaro-Seoane, Pau and others",
    collaboration = "LISA",
    title = "{Laser Interferometer Space Antenna}",
    eprint = "1702.00786",
    archivePrefix = "arXiv",
    primaryClass = "astro-ph.IM",
    month = "2",
    year = "2017"
}

@article{Reitze:2019iox,
    author = "Reitze, David and others",
    title = "{Cosmic Explorer: The U.S. Contribution to Gravitational-Wave Astronomy beyond LIGO}",
    eprint = "1907.04833",
    archivePrefix = "arXiv",
    primaryClass = "astro-ph.IM",
    reportNumber = "LIGO-P1900316",
    journal = "Bull. Am. Astron. Soc.",
    volume = "51",
    number = "7",
    pages = "035",
    year = "2019"
}

@article{Crowder:2005nr,
    author = "Crowder, Jeff and Cornish, Neil J.",
    title = "{Beyond LISA: Exploring future gravitational wave missions}",
    eprint = "gr-qc/0506015",
    archivePrefix = "arXiv",
    doi = "10.1103/PhysRevD.72.083005",
    journal = "Phys. Rev. D",
    volume = "72",
    pages = "083005",
    year = "2005"
}

@article{Corbin:2005ny,
    author = "Corbin, Vincent and Cornish, Neil J.",
    title = "{Detecting the cosmic gravitational wave background with the big bang observer}",
    eprint = "gr-qc/0512039",
    archivePrefix = "arXiv",
    doi = "10.1088/0264-9381/23/7/014",
    journal = "Class. Quant. Grav.",
    volume = "23",
    pages = "2435--2446",
    year = "2006"
}

@article{Harry:2006fi,
    author = "Harry, G. M. and Fritschel, P. and Shaddock, D. A. and Folkner, W. and Phinney, E. S.",
    title = "{Laser interferometry for the big bang observer}",
    doi = "10.1088/0264-9381/23/15/008",
    journal = "Class. Quant. Grav.",
    volume = "23",
    pages = "4887--4894",
    year = "2006",
    note = "[Erratum: Class.Quant.Grav. 23, 7361 (2006)]"
}

@article{Antoniadis:2022pcn,
    author = "Antoniadis, J. and others",
    title = "{The International Pulsar Timing Array second data release: Search for an isotropic gravitational wave background}",
    eprint = "2201.03980",
    archivePrefix = "arXiv",
    primaryClass = "astro-ph.HE",
    doi = "10.1093/mnras/stab3418",
    journal = "Mon. Not. Roy. Astron. Soc.",
    volume = "510",
    number = "4",
    pages = "4873--4887",
    year = "2022"
}

@article{IPTA:2023ero,
    author = "Falxa, M. and others",
    collaboration = "IPTA",
    title = "{Searching for continuous Gravitational Waves in the second data release of the International Pulsar Timing Array}",
    eprint = "2303.10767",
    archivePrefix = "arXiv",
    primaryClass = "gr-qc",
    doi = "10.1093/mnras/stad812",
    journal = "Mon. Not. Roy. Astron. Soc.",
    volume = "521",
    number = "4",
    pages = "5077--5086",
    year = "2023"
}

@article{Maggiore:1999vm,
    author = "Maggiore, Michele",
    title = "{Gravitational wave experiments and early universe cosmology}",
    eprint = "gr-qc/9909001",
    archivePrefix = "arXiv",
    reportNumber = "IFUP-TH-20-99",
    doi = "10.1016/S0370-1573(99)00102-7",
    journal = "Phys. Rept.",
    volume = "331",
    pages = "283--367",
    year = "2000"
}

@article{Saikawa:2017hiv,
    author = "Saikawa, Ken'ichi",
    title = "{A review of gravitational waves from cosmic domain walls}",
    eprint = "1703.02576",
    archivePrefix = "arXiv",
    primaryClass = "hep-ph",
    reportNumber = "DESY-17-036",
    doi = "10.3390/universe3020040",
    journal = "Universe",
    volume = "3",
    number = "2",
    pages = "40",
    year = "2017"
}

@article{Damour:2004kw,
    author = "Damour, Thibault and Vilenkin, Alexander",
    title = "{Gravitational radiation from cosmic (super)strings: Bursts, stochastic background, and observational windows}",
    eprint = "hep-th/0410222",
    archivePrefix = "arXiv",
    doi = "10.1103/PhysRevD.71.063510",
    journal = "Phys. Rev. D",
    volume = "71",
    pages = "063510",
    year = "2005"
}

@article{Ema:2015dka,
    author = "Ema, Yohei and Jinno, Ryusuke and Mukaida, Kyohei and Nakayama, Kazunori",
    title = "{Gravitational Effects on Inflaton Decay}",
    eprint = "1502.02475",
    archivePrefix = "arXiv",
    primaryClass = "hep-ph",
    reportNumber = "UT-15-03",
    doi = "10.1088/1475-7516/2015/05/038",
    journal = "JCAP",
    volume = "05",
    pages = "038",
    year = "2015"
}

@article{Ema:2016hlw,
    author = "Ema, Yohei and Jinno, Ryusuke and Mukaida, Kyohei and Nakayama, Kazunori",
    title = "{Gravitational particle production in oscillating backgrounds and its cosmological implications}",
    eprint = "1604.08898",
    archivePrefix = "arXiv",
    primaryClass = "hep-ph",
    reportNumber = "UT-16-20, KEK-TH-1898, IPMU-16-0062",
    doi = "10.1103/PhysRevD.94.063517",
    journal = "Phys. Rev. D",
    volume = "94",
    number = "6",
    pages = "063517",
    year = "2016"
}

@article{Nakayama:2018ptw,
    author = "Nakayama, Kazunori and Tang, Yong",
    title = "{Stochastic Gravitational Waves from Particle Origin}",
    eprint = "1810.04975",
    archivePrefix = "arXiv",
    primaryClass = "hep-ph",
    reportNumber = "UT-18-20",
    doi = "10.1016/j.physletb.2018.11.023",
    journal = "Phys. Lett. B",
    volume = "788",
    pages = "341--346",
    year = "2019"
}

@article{Huang:2019lgd,
    author = "Huang, Da and Yin, Lu",
    title = "{Stochastic Gravitational Waves from Inflaton Decays}",
    eprint = "1905.08510",
    archivePrefix = "arXiv",
    primaryClass = "hep-ph",
    doi = "10.1103/PhysRevD.100.043538",
    journal = "Phys. Rev. D",
    volume = "100",
    number = "4",
    pages = "043538",
    year = "2019"
}

@article{Ema:2020ggo,
    author = "Ema, Yohei and Jinno, Ryusuke and Nakayama, Kazunori",
    title = "{High-frequency Graviton from Inflaton Oscillation}",
    eprint = "2006.09972",
    archivePrefix = "arXiv",
    primaryClass = "astro-ph.CO",
    reportNumber = "DESY-20-104",
    doi = "10.1088/1475-7516/2020/09/015",
    journal = "JCAP",
    volume = "09",
    pages = "015",
    year = "2020"
}

@article{Barman:2023ymn,
    author = "Barman, Basabendu and Bernal, Nicol\'as and Xu, Yong and Zapata, {\'O}scar",
    title = "{Gravitational wave from graviton Bremsstrahlung during reheating}",
    eprint = "2301.11345",
    archivePrefix = "arXiv",
    primaryClass = "hep-ph",
    doi = "10.1088/1475-7516/2023/05/019",
    journal = "JCAP",
    volume = "05",
    pages = "019",
    year = "2023"
}

@article{Barman:2023rpg,
    author = "Barman, Basabendu and Bernal, Nicol\'as and Xu, Yong and Zapata, {\'O}scar",
    title = "{Bremsstrahlung-induced gravitational waves in monomial potentials during reheating}",
    eprint = "2305.16388",
    archivePrefix = "arXiv",
    primaryClass = "hep-ph",
    doi = "10.1103/PhysRevD.108.083524",
    journal = "Phys. Rev. D",
    volume = "108",
    number = "8",
    pages = "083524",
    year = "2023"
}

@article{Kanemura:2023pnv,
    author = "Kanemura, Shinya and Kaneta, Kunio",
    title = "{Gravitational waves from particle decays during reheating}",
    eprint = "2310.12023",
    archivePrefix = "arXiv",
    primaryClass = "hep-ph",
    reportNumber = "OU--HET--1206, OU-HET-1206",
    doi = "10.1016/j.physletb.2024.138807",
    journal = "Phys. Lett. B",
    volume = "855",
    pages = "138807",
    year = "2024"
}

@article{Bernal:2023wus,
    author = "Bernal, Nicol\'as and Cl\'ery, Simon and Mambrini, Yann and Xu, Yong",
    title = "{Probing reheating with graviton bremsstrahlung}",
    eprint = "2311.12694",
    archivePrefix = "arXiv",
    primaryClass = "hep-ph",
    reportNumber = "MITP-23-065",
    doi = "10.1088/1475-7516/2024/01/065",
    journal = "JCAP",
    volume = "01",
    pages = "065",
    year = "2024"
}

@article{Tokareva:2023mrt,
    author = "Tokareva, Anna",
    title = "{Gravitational waves from inflaton decay and bremsstrahlung}",
    eprint = "2312.16691",
    archivePrefix = "arXiv",
    primaryClass = "hep-ph",
    doi = "10.1016/j.physletb.2024.138695",
    journal = "Phys. Lett. B",
    volume = "853",
    pages = "138695",
    year = "2024"
}

@article{Choi:2024ilx,
    author = "Choi, Gongjun and Ke, Wenqi and Olive, Keith A.",
    title = "{Minimal production of prompt gravitational waves during reheating}",
    eprint = "2402.04310",
    archivePrefix = "arXiv",
    primaryClass = "hep-ph",
    reportNumber = "UMN--TH--4311/24, FTPI--MINN--24/03",
    doi = "10.1103/PhysRevD.109.083516",
    journal = "Phys. Rev. D",
    volume = "109",
    number = "8",
    pages = "083516",
    year = "2024"
}

@article{Hu:2024awd,
    author = "Hu, Weiyu and Nakayama, Kazunori and Takhistov, Volodymyr and Tang, Yong",
    title = "{Gravitational wave probe of Planck-scale physics after inflation}",
    eprint = "2403.13882",
    archivePrefix = "arXiv",
    primaryClass = "hep-ph",
    reportNumber = "KEK-QUP-2024-0006, TU-1226, KEK-TH-2607, KEK-Cosmo-0341, IPMU24-0007",
    doi = "10.1016/j.physletb.2024.138958",
    journal = "Phys. Lett. B",
    volume = "856",
    pages = "138958",
    year = "2024"
}

@article{Choi:2024acs,
    author = "Choi, Ki-Young and Lkhagvadorj, Erdenebulgan and Mahapatra, Satyabrata",
    title = "{Gravitational wave sourced by decay of massive particle from primordial black hole evaporation}",
    eprint = "2403.15269",
    archivePrefix = "arXiv",
    primaryClass = "hep-ph",
    doi = "10.1088/1475-7516/2024/07/064",
    journal = "JCAP",
    volume = "07",
    pages = "064",
    year = "2024"
}

@inproceedings{Barman:2024htg,
    author = "Barman, Basabendu and Bernal, Nicol\'as and Cl\'ery, Simon and Mambrini, Yann and Xu, Yong and Zapata, {\'O}scar",
    title = "{Probing Reheating with Gravitational Waves from Graviton Bremsstrahlung}",
    booktitle = "{58$^{th}$ Rencontres de Moriond on Electroweak Interactions and Unified Theories}",
    eprint = "2405.09620",
    archivePrefix = "arXiv",
    primaryClass = "astro-ph.CO",
    month = "5",
    year = "2024"
}

@article{Xu:2024fjl,
    author = "Xu, Yong",
    title = "{Ultra-high frequency gravitational waves from scattering, Bremsstrahlung and decay during reheating}",
    eprint = "2407.03256",
    archivePrefix = "arXiv",
    primaryClass = "hep-ph",
    reportNumber = "MITP-24-058",
    doi = "10.1007/JHEP10(2024)174",
    journal = "JHEP",
    volume = "10",
    pages = "174",
    year = "2024"
}

@article{Inui:2024wgj,
    author = "Inui, Ryoto and Mikura, Yusuke and Yokoyama, Shuichiro",
    title = "{Gravitational waves from graviton bremsstrahlung with kination phase}",
    eprint = "2408.10786",
    archivePrefix = "arXiv",
    primaryClass = "astro-ph.CO",
    doi = "10.1103/PhysRevD.111.043511",
    journal = "Phys. Rev. D",
    volume = "111",
    number = "4",
    pages = "043511",
    year = "2025"
}

@article{Jiang:2024akb,
    author = "Jiang, Yiheng and Suyama, Teruaki",
    title = "{Spectrum of high-frequency gravitational waves from graviton bremsstrahlung by the decay of inflaton: case with polynomial potential}",
    eprint = "2410.11175",
    archivePrefix = "arXiv",
    primaryClass = "astro-ph.CO",
    doi = "10.1088/1475-7516/2025/02/041",
    journal = "JCAP",
    volume = "02",
    pages = "041",
    year = "2025"
}

@article{Bernal:2024jim,
    author = "Bernal, Nicol\'as and Xu, Yong",
    title = "{Thermal gravitational waves during reheating}",
    eprint = "2410.21385",
    archivePrefix = "arXiv",
    primaryClass = "hep-ph",
    reportNumber = "MITP-24-076",
    doi = "10.1007/JHEP01(2025)137",
    journal = "JHEP",
    volume = "01",
    pages = "137",
    year = "2025"
}

@article{Bernal:2025lxp,
    author = "Bernal, Nicol{\'a}s and Wu, Quan-feng and Xu, Xun-Jie and Xu, Yong",
    title = "{Pre-thermalized gravitational waves}",
    eprint = "2503.10756",
    archivePrefix = "arXiv",
    primaryClass = "hep-ph",
    reportNumber = "MITP-25-022",
    doi = "10.1007/JHEP08(2025)125",
    journal = "JHEP",
    volume = "08",
    pages = "125",
    year = "2025"
}

@article{Caprini:2018mtu,
    author = "Caprini, Chiara and Figueroa, Daniel G.",
    title = "{Cosmological Backgrounds of Gravitational Waves}",
    eprint = "1801.04268",
    archivePrefix = "arXiv",
    primaryClass = "astro-ph.CO",
    doi = "10.1088/1361-6382/aac608",
    journal = "Class. Quant. Grav.",
    volume = "35",
    number = "16",
    pages = "163001",
    year = "2018"
}

@article{Assadullahi:2009nf,
    author = "Assadullahi, Hooshyar and Wands, David",
    title = "{Gravitational waves from an early matter era}",
    eprint = "0901.0989",
    archivePrefix = "arXiv",
    primaryClass = "astro-ph.CO",
    doi = "10.1103/PhysRevD.79.083511",
    journal = "Phys. Rev. D",
    volume = "79",
    pages = "083511",
    year = "2009"
}

@article{Durrer:2011bi,
    author = "Durrer, Ruth and Hasenkamp, Jasper",
    title = "{Testing Superstring Theories with Gravitational Waves}",
    eprint = "1105.5283",
    archivePrefix = "arXiv",
    primaryClass = "gr-qc",
    doi = "10.1103/PhysRevD.84.064027",
    journal = "Phys. Rev. D",
    volume = "84",
    pages = "064027",
    year = "2011"
}

@article{Alabidi:2013lya,
    author = "Alabidi, Laila and Kohri, Kazunori and Sasaki, Misao and Sendouda, Yuuiti",
    title = "{Observable induced gravitational waves from an early matter phase}",
    eprint = "1303.4519",
    archivePrefix = "arXiv",
    primaryClass = "astro-ph.CO",
    doi = "10.1088/1475-7516/2013/05/033",
    journal = "JCAP",
    volume = "05",
    pages = "033",
    year = "2013"
}

@article{DEramo:2019tit,
    author = "D'Eramo, Francesco and Schmitz, Kai",
    title = "{Imprint of a scalar era on the primordial spectrum of gravitational waves}",
    eprint = "1904.07870",
    archivePrefix = "arXiv",
    primaryClass = "hep-ph",
    doi = "10.1103/PhysRevResearch.1.013010",
    journal = "Phys. Rev. Research.",
    volume = "1",
    pages = "013010",
    year = "2019"
}

@article{Bernal:2019lpc,
    author = "Bernal, Nicol\'as and Hajkarim, Fazlollah",
    title = "{Primordial Gravitational Waves in Nonstandard Cosmologies}",
    eprint = "1905.10410",
    archivePrefix = "arXiv",
    primaryClass = "astro-ph.CO",
    doi = "10.1103/PhysRevD.100.063502",
    journal = "Phys. Rev. D",
    volume = "100",
    number = "6",
    pages = "063502",
    year = "2019"
}

@article{Figueroa:2019paj,
    author = "Figueroa, Daniel G. and Tanin, Erwin H.",
    title = "{Ability of LIGO and LISA to probe the equation of state of the early Universe}",
    eprint = "1905.11960",
    archivePrefix = "arXiv",
    primaryClass = "astro-ph.CO",
    doi = "10.1088/1475-7516/2019/08/011",
    journal = "JCAP",
    volume = "08",
    pages = "011",
    year = "2019"
}

@article{Bernal:2020ywq,
    author = "Bernal, Nicol\'as and Ghoshal, Anish and Hajkarim, Fazlollah and Lambiase, Gaetano",
    title = "{Primordial Gravitational Wave Signals in Modified Cosmologies}",
    eprint = "2008.04959",
    archivePrefix = "arXiv",
    primaryClass = "gr-qc",
    doi = "10.1088/1475-7516/2020/11/051",
    journal = "JCAP",
    volume = "11",
    pages = "051",
    year = "2020"
}

@article{Frey:2024jqy,
    author = "Frey, Andrew R. and Mahanta, Ratul and Maharana, Anshuman and Quevedo, Fernando and Villa, Gonzalo",
    title = "{Gravitational waves from high temperature strings}",
    eprint = "2408.13803",
    archivePrefix = "arXiv",
    primaryClass = "hep-th",
    reportNumber = "CERN-TH-2024-108",
    doi = "10.1007/JHEP12(2024)174",
    journal = "JHEP",
    volume = "12",
    pages = "174",
    year = "2024"
}

@article{Cui:2018rwi,
    author = "Cui, Yanou and Lewicki, Marek and Morrissey, David E. and Wells, James D.",
    title = "{Probing the pre-BBN universe with gravitational waves from cosmic strings}",
    eprint = "1808.08968",
    archivePrefix = "arXiv",
    primaryClass = "hep-ph",
    reportNumber = "KCL-PH-TH/2018-47",
    doi = "10.1007/JHEP01(2019)081",
    journal = "JHEP",
    volume = "01",
    pages = "081",
    year = "2019"
}

@inproceedings{Villa:2024jbf,
    author = "Villa, Gonzalo",
    title = "{Gravitational Waves from the Hagedorn Phase}",
    booktitle = "{29$^{th}$ International Symposium on Particles, String and Cosmology}",
    eprint = "2410.07350",
    archivePrefix = "arXiv",
    primaryClass = "hep-th",
    month = "10",
    year = "2024"
}

@article{Caprini:2015zlo,
    author = "Caprini, Chiara and others",
    title = "{Science with the space-based interferometer eLISA. II: Gravitational waves from cosmological phase transitions}",
    eprint = "1512.06239",
    archivePrefix = "arXiv",
    primaryClass = "astro-ph.CO",
    reportNumber = "DESY-15-246",
    doi = "10.1088/1475-7516/2016/04/001",
    journal = "JCAP",
    volume = "04",
    pages = "001",
    year = "2016"
}

@article{Caprini:2019egz,
    author = "Caprini, Chiara and others",
    title = "{Detecting gravitational waves from cosmological phase transitions with LISA: an update}",
    eprint = "1910.13125",
    archivePrefix = "arXiv",
    primaryClass = "astro-ph.CO",
    reportNumber = "DESY-19-159, IPPP/19/27, HIP-2019-14/TH, MITP/19-066, IFT-UAM/CSIC-19-139",
    doi = "10.1088/1475-7516/2020/03/024",
    journal = "JCAP",
    volume = "03",
    pages = "024",
    year = "2020"
}

@article{Athron:2023xlk,
    author = "Athron, Peter and Bal\'azs, Csaba and Fowlie, Andrew and Morris, Lachlan and Wu, Lei",
    title = "{Cosmological phase transitions: From perturbative particle physics to gravitational waves}",
    eprint = "2305.02357",
    archivePrefix = "arXiv",
    primaryClass = "hep-ph",
    doi = "10.1016/j.ppnp.2023.104094",
    journal = "Prog. Part. Nucl. Phys.",
    volume = "135",
    pages = "104094",
    year = "2024"
}

@article{Croon:2024mde,
    author = "Croon, Djuna and Weir, David J.",
    title = "{Gravitational Waves from Phase Transitions}",
    eprint = "2410.21509",
    archivePrefix = "arXiv",
    primaryClass = "hep-ph",
    reportNumber = "HIP-2024-23/TH",
    doi = "10.1080/00107514.2024.2423496",
    journal = "Contemp. Phys.",
    volume = "65",
    pages = "75",
    year = "2024"
}

@article{Kamionkowski:1993fg,
    author = "Kamionkowski, Marc and Kosowsky, Arthur and Turner, Michael S.",
    title = "{Gravitational radiation from first order phase transitions}",
    eprint = "astro-ph/9310044",
    archivePrefix = "arXiv",
    reportNumber = "IASSNS-HEP-93-44, FERMILAB-PUB-93-235-A",
    doi = "10.1103/PhysRevD.49.2837",
    journal = "Phys. Rev. D",
    volume = "49",
    pages = "2837--2851",
    year = "1994"
}

@article{Apreda:2001us,
    author = "Apreda, Riccardo and Maggiore, Michele and Nicolis, Alberto and Riotto, Antonio",
    title = "{Gravitational waves from electroweak phase transitions}",
    eprint = "gr-qc/0107033",
    archivePrefix = "arXiv",
    reportNumber = "UGVA-DPT-07-1096",
    doi = "10.1016/S0550-3213(02)00264-X",
    journal = "Nucl. Phys. B",
    volume = "631",
    pages = "342--368",
    year = "2002"
}

@article{Grojean:2006bp,
    author = "Grojean, Christophe and Servant, Geraldine",
    title = "{Gravitational Waves from Phase Transitions at the Electroweak Scale and Beyond}",
    eprint = "hep-ph/0607107",
    archivePrefix = "arXiv",
    reportNumber = "CERN-PH-TH-2006-125",
    doi = "10.1103/PhysRevD.75.043507",
    journal = "Phys. Rev. D",
    volume = "75",
    pages = "043507",
    year = "2007"
}

@article{Ashoorioon:2009nf,
    author = "Ashoorioon, A. and Konstandin, T.",
    title = "{Strong electroweak phase transitions without collider traces}",
    eprint = "0904.0353",
    archivePrefix = "arXiv",
    primaryClass = "hep-ph",
    reportNumber = "UAB-FT-666, MCTP-09-10",
    doi = "10.1088/1126-6708/2009/07/086",
    journal = "JHEP",
    volume = "07",
    pages = "086",
    year = "2009"
}

@article{Kakizaki:2015wua,
    author = "Kakizaki, Mitsuru and Kanemura, Shinya and Matsui, Toshinori",
    title = "{Gravitational waves as a probe of extended scalar sectors with the first order electroweak phase transition}",
    eprint = "1509.08394",
    archivePrefix = "arXiv",
    primaryClass = "hep-ph",
    reportNumber = "UT-HET-106",
    doi = "10.1103/PhysRevD.92.115007",
    journal = "Phys. Rev. D",
    volume = "92",
    number = "11",
    pages = "115007",
    year = "2015"
}

@article{Vaskonen:2016yiu,
    author = "Vaskonen, Ville",
    title = "{Electroweak baryogenesis and gravitational waves from a real scalar singlet}",
    eprint = "1611.02073",
    archivePrefix = "arXiv",
    primaryClass = "hep-ph",
    doi = "10.1103/PhysRevD.95.123515",
    journal = "Phys. Rev. D",
    volume = "95",
    number = "12",
    pages = "123515",
    year = "2017"
}

@article{Dorsch:2016nrg,
    author = "Dorsch, G. C. and Huber, S. J. and Konstandin, T. and No, J. M.",
    title = "{A Second Higgs Doublet in the Early Universe: Baryogenesis and Gravitational Waves}",
    eprint = "1611.05874",
    archivePrefix = "arXiv",
    primaryClass = "hep-ph",
    reportNumber = "DESY-16-213",
    doi = "10.1088/1475-7516/2017/05/052",
    journal = "JCAP",
    volume = "05",
    pages = "052",
    year = "2017"
}

@article{Beniwal:2017eik,
    author = "Beniwal, Ankit and Lewicki, Marek and Wells, James D. and White, Martin and Williams, Anthony G.",
    title = "{Gravitational wave, collider and dark matter signals from a scalar singlet electroweak baryogenesis}",
    eprint = "1702.06124",
    archivePrefix = "arXiv",
    primaryClass = "hep-ph",
    reportNumber = "ADP-17-08-T1014, ADP--17--08-T1014",
    doi = "10.1007/JHEP08(2017)108",
    journal = "JHEP",
    volume = "08",
    pages = "108",
    year = "2017"
}

@article{Ellis:2018mja,
    author = "Ellis, John and Lewicki, Marek and No, Jos\'e Miguel",
    title = "{On the Maximal Strength of a First-Order Electroweak Phase Transition and its Gravitational Wave Signal}",
    eprint = "1809.08242",
    archivePrefix = "arXiv",
    primaryClass = "hep-ph",
    reportNumber = "KCL-PH-TH/2018-46, CERN-TH/2018-197, IFT-UAM/CSIC-18-94, CERN-TH-2018-197",
    doi = "10.1088/1475-7516/2019/04/003",
    journal = "JCAP",
    volume = "04",
    pages = "003",
    year = "2019"
}

@article{Chatterjee:2022pxf,
    author = "Chatterjee, Arindam and Datta, AseshKrishna and Roy, Subhojit",
    title = "{Electroweak phase transition in the $\mathbb{Z}_{3}$-invariant NMSSM: Implications of LHC and Dark matter searches and prospects of detecting the gravitational waves}",
    eprint = "2202.12476",
    archivePrefix = "arXiv",
    primaryClass = "hep-ph",
    reportNumber = "HRI-RECAPP-2022-001",
    doi = "10.1007/JHEP06(2022)108",
    journal = "JHEP",
    volume = "06",
    pages = "108",
    year = "2022"
}

@article{Schwaller:2015tja,
    author = "Schwaller, Pedro",
    title = "{Gravitational Waves from a Dark Phase Transition}",
    eprint = "1504.07263",
    archivePrefix = "arXiv",
    primaryClass = "hep-ph",
    reportNumber = "CERN-PH-TH-2015-093",
    doi = "10.1103/PhysRevLett.115.181101",
    journal = "Phys. Rev. Lett.",
    volume = "115",
    number = "18",
    pages = "181101",
    year = "2015"
}

@article{Jaeckel:2016jlh,
    author = "Jaeckel, Joerg and Khoze, Valentin V. and Spannowsky, Michael",
    title = "{Hearing the signal of dark sectors with gravitational wave detectors}",
    eprint = "1602.03901",
    archivePrefix = "arXiv",
    primaryClass = "hep-ph",
    reportNumber = "IPPP-16-12, DCPT-16-24",
    doi = "10.1103/PhysRevD.94.103519",
    journal = "Phys. Rev. D",
    volume = "94",
    number = "10",
    pages = "103519",
    year = "2016"
}

@article{Breitbach:2018ddu,
    author = "Breitbach, Moritz and Kopp, Joachim and Madge, Eric and Opferkuch, Toby and Schwaller, Pedro",
    title = "{Dark, Cold, and Noisy: Constraining Secluded Hidden Sectors with Gravitational Waves}",
    eprint = "1811.11175",
    archivePrefix = "arXiv",
    primaryClass = "hep-ph",
    reportNumber = "CERN-TH-2018-255, MITP/18-115",
    doi = "10.1088/1475-7516/2019/07/007",
    journal = "JCAP",
    volume = "07",
    pages = "007",
    year = "2019"
}

@article{Dev:2019njv,
    author = "Dev, P. S. Bhupal and Ferrer, Francesc and Zhang, Yiyang and Zhang, Yongchao",
    title = "{Gravitational Waves from First-Order Phase Transition in a Simple Axion-Like Particle Model}",
    eprint = "1905.00891",
    archivePrefix = "arXiv",
    primaryClass = "hep-ph",
    doi = "10.1088/1475-7516/2019/11/006",
    journal = "JCAP",
    volume = "11",
    pages = "006",
    year = "2019"
}

@article{Dent:2022bcd,
    author = "Dent, James B. and Dutta, Bhaskar and Ghosh, Sumit and Kumar, Jason and Runburg, Jack",
    title = "{Sensitivity to dark sector scales from gravitational wave signatures}",
    eprint = "2203.11736",
    archivePrefix = "arXiv",
    primaryClass = "hep-ph",
    reportNumber = "MI-HET-774, KIAS-P22016",
    doi = "10.1007/JHEP08(2022)300",
    journal = "JHEP",
    volume = "08",
    pages = "300",
    year = "2022"
}

@article{Morgante:2022zvc,
    author = "Morgante, Enrico and Ramberg, Nicklas and Schwaller, Pedro",
    title = "{Gravitational waves from dark SU(3) Yang-Mills theory}",
    eprint = "2210.11821",
    archivePrefix = "arXiv",
    primaryClass = "hep-ph",
    doi = "10.1103/PhysRevD.107.036010",
    journal = "Phys. Rev. D",
    volume = "107",
    number = "3",
    pages = "036010",
    year = "2023"
}

@article{Pasechnik:2023hwv,
    author = "Pasechnik, Roman and Reichert, Manuel and Sannino, Francesco and Wang, Zhi-Wei",
    title = "{Gravitational waves from composite dark sectors}",
    eprint = "2309.16755",
    archivePrefix = "arXiv",
    primaryClass = "hep-ph",
    doi = "10.1007/JHEP02(2024)159",
    journal = "JHEP",
    volume = "02",
    pages = "159",
    year = "2024"
}

@article{Koutroulis:2023wit,
    author = "Koutroulis, Fotis and McCullough, Matthew and Merchand, Marco and Pokorski, Stefan and Sakurai, Kazuki",
    title = "{Phases of Pseudo-Nambu-Goldstone bosons}",
    eprint = "2309.15749",
    archivePrefix = "arXiv",
    primaryClass = "hep-ph",
    reportNumber = "CERN-TH-2023-172",
    doi = "10.1007/JHEP05(2024)095",
    journal = "JHEP",
    volume = "05",
    pages = "095",
    year = "2024"
}

@article{DiBari:2023upq,
    author = "Di Bari, Pasquale and Rahat, Moinul Hossain",
    title = "{Split Majoron model confronts the NANOGrav signal and cosmological tensions}",
    eprint = "2307.03184",
    archivePrefix = "arXiv",
    primaryClass = "hep-ph",
    doi = "10.1103/PhysRevD.110.055019",
    journal = "Phys. Rev. D",
    volume = "110",
    number = "5",
    pages = "055019",
    year = "2024"
}

@article{Feng:2024pab,
    author = "Feng, Wan-Zhe and Li, Jinzheng and Nath, Pran",
    title = "{Cosmologically consistent analysis of gravitational waves from hidden sectors}",
    eprint = "2403.09558",
    archivePrefix = "arXiv",
    primaryClass = "hep-ph",
    doi = "10.1103/PhysRevD.110.015020",
    journal = "Phys. Rev. D",
    volume = "110",
    number = "1",
    pages = "015020",
    year = "2024"
}

@article{Banik:2024zwj,
    author = "Banik, Amitayus and Cui, Yanou and Tsai, Yu-Dai and Tsai, Yuhsin",
    title = "{The Sound of Dark Sectors in Pulsar Timing Arrays}",
    eprint = "2412.16282",
    archivePrefix = "arXiv",
    primaryClass = "hep-ph",
    reportNumber = "LA-UR-24-33252",
    month = "12",
    year = "2024"
}

@article{Balan:2025uke,
    author = "Balan, Sowmiya and Bringmann, Torsten and Kahlhoefer, Felix and Matuszak, Jonas and Tasillo, Carlo",
    title = "{Sub-GeV dark matter and nano-Hertz gravitational waves from a classically conformal dark sector}",
    eprint = "2502.19478",
    archivePrefix = "arXiv",
    primaryClass = "hep-ph",
    doi = "10.1088/1475-7516/2025/08/062",
    journal = "JCAP",
    volume = "08",
    pages = "062",
    year = "2025"
}

@article{Barenboim:2016mjm,
    author = "Barenboim, Gabriela and Park, Wan-Il",
    title = "{Gravitational waves from first order phase transitions as a probe of an early matter domination era and its inverse problem}",
    eprint = "1605.03781",
    archivePrefix = "arXiv",
    primaryClass = "astro-ph.CO",
    reportNumber = "FTUV-16-05-07, IFIC-16-25",
    doi = "10.1016/j.physletb.2016.06.009",
    journal = "Phys. Lett. B",
    volume = "759",
    pages = "430--438",
    year = "2016"
}

@article{Guo:2020grp,
    author = "Guo, Huai-Ke and Sinha, Kuver and Vagie, Daniel and White, Graham",
    title = "{Phase Transitions in an Expanding Universe: Stochastic Gravitational Waves in Standard and Non-Standard Histories}",
    eprint = "2007.08537",
    archivePrefix = "arXiv",
    primaryClass = "hep-ph",
    doi = "10.1088/1475-7516/2021/01/001",
    journal = "JCAP",
    volume = "01",
    pages = "001",
    year = "2021"
}

@article{Buen-Abad:2023hex,
    author = "Buen-Abad, Manuel A. and Chang, Jae Hyeok and Hook, Anson",
    title = "{Gravitational wave signatures from reheating}",
    eprint = "2305.09712",
    archivePrefix = "arXiv",
    primaryClass = "hep-ph",
    doi = "10.1103/PhysRevD.108.036006",
    journal = "Phys. Rev. D",
    volume = "108",
    number = "3",
    pages = "036006",
    year = "2023"
}

@article{Dent:2024bhi,
    author = "Dent, James B. and Dutta, Bhaskar and Rai, Mudit",
    title = "{Imprints of early universe cosmology on gravitational waves}",
    eprint = "2411.09757",
    archivePrefix = "arXiv",
    primaryClass = "hep-ph",
    doi = "10.1007/JHEP03(2025)098",
    journal = "JHEP",
    volume = "03",
    pages = "098",
    year = "2025"
}

@article{Xiao:2024rsj,
    author = "Xiao, Yang and Guo, Huai-Ke and Hu, Jia-Hang and Yang, Jin Min and Zhang, Yang",
    title = "{Growth of the gravitational wave spectrum from sound waves in a universe with a generic expansion rate}",
    eprint = "2410.23666",
    archivePrefix = "arXiv",
    primaryClass = "gr-qc",
    doi = "10.1103/xyvn-hsqz",
    journal = "Phys. Rev. D",
    volume = "112",
    number = "8",
    pages = "L081302",
    year = "2025"
}

@article{Allahverdi:2020bys,
    author = "Allahverdi, Rouzbeh and others",
    title = "{The First Three Seconds: a Review of Possible Expansion Histories of the Early Universe}",
    eprint = "2006.16182",
    archivePrefix = "arXiv",
    primaryClass = "astro-ph.CO",
    reportNumber = "FERMILAB-PUB-20-242-A, KCL-PH-TH/2020-33, KEK-Cosmo-257,
  KEK-TH-2231, IPMU20-0070, PI/UAN-2020-674FT, RUP-20-22",
    doi = "10.21105/astro.2006.16182",
    journal = "Open J.Astrophys.",
    volume = "4",
    month = "6",
    year = "2021"
}

@article{Batell:2024dsi,
    author = "Batell, Brian and others",
    title = "{Conversations and deliberations: Non-standard cosmological epochs and expansion histories}",
    eprint = "2411.04780",
    archivePrefix = "arXiv",
    primaryClass = "astro-ph.CO",
    doi = "10.1142/S0217751X25300042",
    journal = "Int. J. Mod. Phys. A",
    volume = "40",
    number = "17",
    pages = "2530004",
    year = "2025"
}

@article{Kofman:1994rk,
    author = "Kofman, Lev and Linde, Andrei D. and Starobinsky, Alexei A.",
    title = "{Reheating after inflation}",
    eprint = "hep-th/9405187",
    archivePrefix = "arXiv",
    reportNumber = "UH-IFA-94-35, SU-ITP-94-13, YITP-U-94-15",
    doi = "10.1103/PhysRevLett.73.3195",
    journal = "Phys. Rev. Lett.",
    volume = "73",
    pages = "3195--3198",
    year = "1994"
}

@article{Kofman:1997yn,
    author = "Kofman, Lev and Linde, Andrei D. and Starobinsky, Alexei A.",
    title = "{Towards the theory of reheating after inflation}",
    eprint = "hep-ph/9704452",
    archivePrefix = "arXiv",
    reportNumber = "IFA-97-28, SU-ITP-97-18",
    doi = "10.1103/PhysRevD.56.3258",
    journal = "Phys. Rev. D",
    volume = "56",
    pages = "3258--3295",
    year = "1997"
}

@article{Kawasaki:2000en,
    author = "Kawasaki, M. and Kohri, Kazunori and Sugiyama, Naoshi",
    title = "{MeV scale reheating temperature and thermalization of neutrino background}",
    eprint = "astro-ph/0002127",
    archivePrefix = "arXiv",
    doi = "10.1103/PhysRevD.62.023506",
    journal = "Phys. Rev. D",
    volume = "62",
    pages = "023506",
    year = "2000"
}

@article{Hannestad:2004px,
    author = "Hannestad, Steen",
    title = "{What is the lowest possible reheating temperature?}",
    eprint = "astro-ph/0403291",
    archivePrefix = "arXiv",
    doi = "10.1103/PhysRevD.70.043506",
    journal = "Phys. Rev. D",
    volume = "70",
    pages = "043506",
    year = "2004"
}

@article{Cyburt:2015mya,
    author = "Cyburt, Richard H. and Fields, Brian D. and Olive, Keith A. and Yeh, Tsung-Han",
    title = "{Big Bang Nucleosynthesis: 2015}",
    eprint = "1505.01076",
    archivePrefix = "arXiv",
    primaryClass = "astro-ph.CO",
    reportNumber = "UMN-TH-3432-15, FTPI-MINN-15-19",
    doi = "10.1103/RevModPhys.88.015004",
    journal = "Rev. Mod. Phys.",
    volume = "88",
    pages = "015004",
    year = "2016"
}

@article{deSalas:2015glj,
    author = "de Salas, P. F. and Lattanzi, M. and Mangano, G. and Miele, G. and Pastor, S. and Pisanti, O.",
    title = "{Bounds on very low reheating scenarios after Planck}",
    eprint = "1511.00672",
    archivePrefix = "arXiv",
    primaryClass = "astro-ph.CO",
    reportNumber = "IFIC-15-70",
    doi = "10.1103/PhysRevD.92.123534",
    journal = "Phys. Rev. D",
    volume = "92",
    number = "12",
    pages = "123534",
    year = "2015"
}

@article{Coleman:1977py,
    author = "Coleman, Sidney R.",
    title = "{The Fate of the False Vacuum. 1. Semiclassical Theory}",
    reportNumber = "HUTP-77-A004",
    doi = "10.1103/PhysRevD.16.1248",
    journal = "Phys. Rev. D",
    volume = "15",
    pages = "2929--2936",
    year = "1977",
    note = "[Erratum: Phys.Rev.D 16, 1248 (1977)]"
}

@article{Callan:1977pt,
    author = "Callan, Jr., Curtis G. and Coleman, Sidney R.",
    title = "{The Fate of the False Vacuum. 2. First Quantum Corrections}",
    reportNumber = "HUTP-77-A032",
    doi = "10.1103/PhysRevD.16.1762",
    journal = "Phys. Rev. D",
    volume = "16",
    pages = "1762--1768",
    year = "1977"
}

@article{Linde:1980tt,
    author = "Linde, Andrei D.",
    title = "{Fate of the False Vacuum at Finite Temperature: Theory and Applications}",
    reportNumber = "LEBEDEV-80-92",
    doi = "10.1016/0370-2693(81)90281-1",
    journal = "Phys. Lett. B",
    volume = "100",
    pages = "37--40",
    year = "1981"
}

@article{Linde:1981zj,
    author = "Linde, Andrei D.",
    title = "{Decay of the False Vacuum at Finite Temperature}",
    reportNumber = "LEBEDEV-81-265",
    doi = "10.1016/0550-3213(83)90072-X",
    journal = "Nucl. Phys. B",
    volume = "216",
    pages = "421",
    year = "1983",
    note = "[Erratum: Nucl.Phys.B 223, 544 (1983)]"
}

@article{Bassett:2005xm,
    author = "Bassett, Bruce A. and Tsujikawa, Shinji and Wands, David",
    title = "{Inflation dynamics and reheating}",
    eprint = "astro-ph/0507632",
    archivePrefix = "arXiv",
    doi = "10.1103/RevModPhys.78.537",
    journal = "Rev. Mod. Phys.",
    volume = "78",
    pages = "537--589",
    year = "2006"
}

@article{Allahverdi:2010xz,
    author = "Allahverdi, Rouzbeh and Brandenberger, Robert and Cyr-Racine, Francis-Yan and Mazumdar, Anupam",
    title = "{Reheating in Inflationary Cosmology: Theory and Applications}",
    eprint = "1001.2600",
    archivePrefix = "arXiv",
    primaryClass = "hep-th",
    doi = "10.1146/annurev.nucl.012809.104511",
    journal = "Ann. Rev. Nucl. Part. Sci.",
    volume = "60",
    pages = "27--51",
    year = "2010"
}

@article{Amin:2014eta,
    author = "Amin, Mustafa A. and Hertzberg, Mark P. and Kaiser, David I. and Karouby, Johanna",
    title = "{Nonperturbative Dynamics Of Reheating After Inflation: A Review}",
    eprint = "1410.3808",
    archivePrefix = "arXiv",
    primaryClass = "hep-ph",
    doi = "10.1142/S0218271815300037",
    journal = "Int. J. Mod. Phys. D",
    volume = "24",
    pages = "1530003",
    year = "2014"
}

@article{Lozanov:2019jxc,
    author = "Lozanov, Kaloian D.",
    title = "{Lectures on Reheating after Inflation}",
    eprint = "1907.04402",
    archivePrefix = "arXiv",
    primaryClass = "astro-ph.CO",
    month = "7",
    year = "2019"
}

@article{Barman:2025lvk,
    author = "Barman, Basabendu and Bernal, Nicol{\'a}s and Rubio, Javier",
    title = "{Two or three things particle physicists (mis)understand about (pre)heating}",
    eprint = "2503.19980",
    archivePrefix = "arXiv",
    primaryClass = "hep-ph",
    reportNumber = "IPARCOS-UCM-25-019",
    doi = "10.1016/j.nuclphysb.2025.116996",
    journal = "Nucl. Phys. B",
    volume = "1018",
    pages = "116996",
    year = "2025"
}

@article{Drees:2015exa,
    author = "Drees, Manuel and Hajkarim, Fazlollah and Schmitz, Ernany Rossi",
    title = "{The Effects of QCD Equation of State on the Relic Density of WIMP Dark Matter}",
    eprint = "1503.03513",
    archivePrefix = "arXiv",
    primaryClass = "hep-ph",
    doi = "10.1088/1475-7516/2015/06/025",
    journal = "JCAP",
    volume = "06",
    pages = "025",
    year = "2015"
}

@article{Bernal:2024yhu,
    author = "Bernal, Nicol\'as and Deka, Kuldeep and Losada, Marta",
    title = "{Thermal dark matter with low-temperature reheating}",
    eprint = "2406.17039",
    archivePrefix = "arXiv",
    primaryClass = "hep-ph",
    doi = "10.1088/1475-7516/2024/09/024",
    journal = "JCAP",
    volume = "09",
    pages = "024",
    year = "2024"
}

@article{Bernal:2024ndy,
    author = "Bernal, Nicol\'as and Fong, Chee Sheng and Zapata, {\'O}scar",
    title = "{Probing low-reheating scenarios with minimal freeze-in dark matter}",
    eprint = "2412.04550",
    archivePrefix = "arXiv",
    primaryClass = "hep-ph",
    doi = "10.1007/JHEP02(2025)161",
    journal = "JHEP",
    volume = "02",
    pages = "161",
    year = "2025"
}

@article{Bernal:2025fdr,
    author = "Bernal, Nicol\'as and Deka, Kuldeep and Losada, Marta",
    title = "{Dark matter ultraviolet freeze-in in general reheating scenarios}",
    eprint = "2501.04774",
    archivePrefix = "arXiv",
    primaryClass = "hep-ph",
    doi = "10.1103/PhysRevD.111.055034",
    journal = "Phys. Rev. D",
    volume = "111",
    number = "5",
    pages = "055034",
    year = "2025"
}

@article{Co:2020xaf,
    author = "Co, Raymond T. and Gonz\'alez, Eric and Harigaya, Keisuke",
    title = "{Increasing Temperature toward the Completion of Reheating}",
    eprint = "2007.04328",
    archivePrefix = "arXiv",
    primaryClass = "astro-ph.CO",
    reportNumber = "LCTP-20-15",
    doi = "10.1088/1475-7516/2020/11/038",
    journal = "JCAP",
    volume = "11",
    pages = "038",
    year = "2020"
}

@article{Giudice:2000ex,
    author = "Giudice, Gian Francesco and Kolb, Edward W. and Riotto, Antonio",
    title = "{Largest temperature of the radiation era and its cosmological implications}",
    eprint = "hep-ph/0005123",
    archivePrefix = "arXiv",
    reportNumber = "SNS-PH-00-05, FERMILAB-PUB-00-075-A, CERN-TH-2000-107",
    doi = "10.1103/PhysRevD.64.023508",
    journal = "Phys. Rev. D",
    volume = "64",
    pages = "023508",
    year = "2001"
}

@article{Starobinsky:1980te,
    author = "Starobinsky, Alexei A.",
    editor = "Khalatnikov, I. M. and Mineev, V. P.",
    title = "{A New Type of Isotropic Cosmological Models Without Singularity}",
    doi = "10.1016/0370-2693(80)90670-X",
    journal = "Phys. Lett. B",
    volume = "91",
    pages = "99--102",
    year = "1980"
}

@article{Drees:2021wgd,
    author = "Drees, Manuel and Xu, Yong",
    title = "{Small field polynomial inflation: reheating, radiative stability and lower bound}",
    eprint = "2104.03977",
    archivePrefix = "arXiv",
    primaryClass = "hep-ph",
    doi = "10.1088/1475-7516/2021/09/012",
    journal = "JCAP",
    volume = "09",
    pages = "012",
    year = "2021"
}

@article{Bernal:2021qrl,
    author = "Bernal, Nicol\'as and Xu, Yong",
    title = "{Polynomial inflation and dark matter}",
    eprint = "2106.03950",
    archivePrefix = "arXiv",
    primaryClass = "hep-ph",
    reportNumber = "PI/UAN-2021-690FT",
    doi = "10.1140/epjc/s10052-021-09694-5",
    journal = "Eur. Phys. J. C",
    volume = "81",
    number = "10",
    pages = "877",
    year = "2021"
}

@article{Drees:2022aea,
    author = "Drees, Manuel and Xu, Yong",
    title = "{Large field polynomial inflation: parameter space, predictions and (double) eternal nature}",
    eprint = "2209.07545",
    archivePrefix = "arXiv",
    primaryClass = "astro-ph.CO",
    doi = "10.1088/1475-7516/2022/12/005",
    journal = "JCAP",
    volume = "12",
    pages = "005",
    year = "2022"
}

@article{Bernal:2024ykj,
    author = "Bernal, Nicol\'as and Harz, Julia and Mojahed, Martin A. and Xu, Yong",
    title = "{Graviton- and inflaton-mediated dark matter production after large field polynomial inflation}",
    eprint = "2406.19447",
    archivePrefix = "arXiv",
    primaryClass = "hep-ph",
    reportNumber = "MITP-24-056",
    doi = "10.1103/PhysRevD.111.043517",
    journal = "Phys. Rev. D",
    volume = "111",
    number = "4",
    pages = "043517",
    year = "2025"
}

@article{Kallosh:2013hoa,
    author = "Kallosh, Renata and Linde, Andrei",
    title = "{Universality Class in Conformal Inflation}",
    eprint = "1306.5220",
    archivePrefix = "arXiv",
    primaryClass = "hep-th",
    doi = "10.1088/1475-7516/2013/07/002",
    journal = "JCAP",
    volume = "07",
    pages = "002",
    year = "2013"
}

@article{Kallosh:2013maa,
    author = "Kallosh, Renata and Linde, Andrei",
    title = "{Non-minimal Inflationary Attractors}",
    eprint = "1307.7938",
    archivePrefix = "arXiv",
    primaryClass = "hep-th",
    doi = "10.1088/1475-7516/2013/10/033",
    journal = "JCAP",
    volume = "10",
    pages = "033",
    year = "2013"
}

@article{Turner:1983he,
    author = "Turner, Michael S.",
    title = "{Coherent Scalar Field Oscillations in an Expanding Universe}",
    reportNumber = "EFI-83-29-CHICAGO",
    doi = "10.1103/PhysRevD.28.1243",
    journal = "Phys. Rev. D",
    volume = "28",
    pages = "1243",
    year = "1983"
}

@article{Garcia:2020wiy,
    author = "Garc\'ia, Marcos A. G. and Kaneta, Kunio and Mambrini, Yann and Olive, Keith A.",
    title = "{Inflaton Oscillations and Post-Inflationary Reheating}",
    eprint = "2012.10756",
    archivePrefix = "arXiv",
    primaryClass = "hep-ph",
    reportNumber = "UMN-TH-4006/20, FTPI-MINN-20/37, IFT-UAM/CSIC-20-185, KIAS-P20071",
    doi = "10.1088/1475-7516/2021/04/012",
    journal = "JCAP",
    volume = "04",
    pages = "012",
    year = "2021"
}

@article{Xu:2023lxw,
    author = "Xu, Yong",
    title = "{Constraining axion and ALP dark matter from misalignment during reheating}",
    eprint = "2308.15322",
    archivePrefix = "arXiv",
    primaryClass = "hep-ph",
    reportNumber = "MITP-23-047",
    doi = "10.1103/PhysRevD.108.083536",
    journal = "Phys. Rev. D",
    volume = "108",
    number = "8",
    pages = "083536",
    year = "2023"
}

@article{Barman:2024mqo,
    author = "Barman, Basabendu and Bernal, Nicol\'as and Xu, Yong",
    title = "{Resonant reheating}",
    eprint = "2404.16090",
    archivePrefix = "arXiv",
    primaryClass = "hep-ph",
    reportNumber = "MITP-24-045",
    doi = "10.1088/1475-7516/2024/08/014",
    journal = "JCAP",
    volume = "08",
    pages = "014",
    year = "2024"
}

@article{Shtanov:1994ce,
    author = "Shtanov, Y. and Traschen, Jennie H. and Brandenberger, Robert H.",
    title = "{Universe reheating after inflation}",
    eprint = "hep-ph/9407247",
    archivePrefix = "arXiv",
    reportNumber = "BROWN-HET-957",
    doi = "10.1103/PhysRevD.51.5438",
    journal = "Phys. Rev. D",
    volume = "51",
    pages = "5438--5455",
    year = "1995"
}

@article{Ichikawa:2008ne,
    author = "Ichikawa, Kazuhide and Suyama, Teruaki and Takahashi, Tomo and Yamaguchi, Masahide",
    title = "{Primordial Curvature Fluctuation and Its Non-Gaussianity in Models with Modulated Reheating}",
    eprint = "0807.3988",
    archivePrefix = "arXiv",
    primaryClass = "astro-ph",
    doi = "10.1103/PhysRevD.78.063545",
    journal = "Phys. Rev. D",
    volume = "78",
    pages = "063545",
    year = "2008"
}

@article{Bernal:2022wck,
    author = "Bernal, Nicol\'as and Xu, Yong",
    title = "{WIMPs during reheating}",
    eprint = "2209.07546",
    archivePrefix = "arXiv",
    primaryClass = "hep-ph",
    reportNumber = "PI/UAN-2022-722FT",
    doi = "10.1088/1475-7516/2022/12/017",
    journal = "JCAP",
    volume = "12",
    pages = "017",
    year = "2022"
}

@article{Spokoiny:1993kt,
    author = "Spokoiny, Boris",
    title = "{Deflationary universe scenario}",
    eprint = "gr-qc/9306008",
    archivePrefix = "arXiv",
    reportNumber = "KUNS-1201",
    doi = "10.1016/0370-2693(93)90155-B",
    journal = "Phys. Lett. B",
    volume = "315",
    pages = "40--45",
    year = "1993"
}

@article{Ferreira:1997hj,
    author = "Ferreira, Pedro G. and Joyce, Michael",
    title = "{Cosmology with a primordial scaling field}",
    eprint = "astro-ph/9711102",
    archivePrefix = "arXiv",
    reportNumber = "CFPA-97-TH-20",
    doi = "10.1103/PhysRevD.58.023503",
    journal = "Phys. Rev. D",
    volume = "58",
    pages = "023503",
    year = "1998"
}

@article{Barman:2022tzk,
    author = "Barman, Basabendu and Bernal, Nicol\'as and Xu, Yong and Zapata, {\'O}scar",
    title = "{Ultraviolet freeze-in with a time-dependent inflaton decay}",
    eprint = "2202.12906",
    archivePrefix = "arXiv",
    primaryClass = "hep-ph",
    reportNumber = "PI/UAN-2022-710FT",
    doi = "10.1088/1475-7516/2022/07/019",
    journal = "JCAP",
    volume = "07",
    number = "07",
    pages = "019",
    year = "2022"
}

@article{Chowdhury:2023jft,
    author = "Chowdhury, Debtosh and Hait, Arpan",
    title = "{Thermalization in the presence of a time-dependent dissipation and its impact on dark matter production}",
    eprint = "2302.06654",
    archivePrefix = "arXiv",
    primaryClass = "hep-ph",
    doi = "10.1007/JHEP09(2023)085",
    journal = "JHEP",
    volume = "09",
    pages = "085",
    year = "2023"
}

@article{Cosme:2024ndc,
    author = "Cosme, Catarina and Costa, Francesco and Lebedev, Oleg",
    title = "{Temperature evolution in the Early Universe and freeze-in at stronger coupling}",
    eprint = "2402.04743",
    archivePrefix = "arXiv",
    primaryClass = "hep-ph",
    doi = "10.1088/1475-7516/2024/06/031",
    journal = "JCAP",
    volume = "06",
    pages = "031",
    year = "2024"
}

@article{Guada:2020xnz,
    author = "Guada, Victor and Nemev\v{s}ek, Miha and Pintar, Matev\v{z}",
    title = "{FindBounce: Package for multi-field bounce actions}",
    eprint = "2002.00881",
    archivePrefix = "arXiv",
    primaryClass = "hep-ph",
    doi = "10.1016/j.cpc.2020.107480",
    journal = "Comput. Phys. Commun.",
    volume = "256",
    pages = "107480",
    year = "2020"
}

@article{Ellis:2020nnr,
    author = "Ellis, John and Lewicki, Marek and Vaskonen, Ville",
    title = "{Updated predictions for gravitational waves produced in a strongly supercooled phase transition}",
    eprint = "2007.15586",
    archivePrefix = "arXiv",
    primaryClass = "astro-ph.CO",
    reportNumber = "KCL-PH-TH/2020-40, CERN-TH-2020-129",
    doi = "10.1088/1475-7516/2020/11/020",
    journal = "JCAP",
    volume = "11",
    pages = "020",
    year = "2020"
}

@article{Freese:2022qrl,
    author = "Freese, Katherine and Winkler, Martin Wolfgang",
    title = "{Have pulsar timing arrays detected the hot big bang: Gravitational waves from strong first order phase transitions in the early Universe}",
    eprint = "2208.03330",
    archivePrefix = "arXiv",
    primaryClass = "astro-ph.CO",
    reportNumber = "UTTG 11-2022, NORDITA 2022-056",
    doi = "10.1103/PhysRevD.106.103523",
    journal = "Phys. Rev. D",
    volume = "106",
    number = "10",
    pages = "103523",
    year = "2022"
}

@article{Winkler:2024olr,
    author = "Winkler, Martin Wolfgang and Freese, Katherine",
    title = "{Origin of the stochastic gravitational wave background: First-order phase transition versus black hole mergers}",
    eprint = "2401.13729",
    archivePrefix = "arXiv",
    primaryClass = "astro-ph.CO",
    reportNumber = "NORDITA-2024-002, UT-WI-02-2024",
    doi = "10.1103/PhysRevD.111.083509",
    journal = "Phys. Rev. D",
    volume = "111",
    number = "8",
    pages = "083509",
    year = "2025"
}

@article{Guth:1982pn,
    author = "Guth, Alan H. and Weinberg, Erick J.",
    title = "{Could the Universe Have Recovered from a Slow First Order Phase Transition?}",
    reportNumber = "MIT-CTP-950",
    doi = "10.1016/0550-3213(83)90307-3",
    journal = "Nucl. Phys. B",
    volume = "212",
    pages = "321--364",
    year = "1983"
}

@article{Kierkla:2022odc,
    author = "Kierkla, Maciej and Karam, Alexandros and Swiezewska, Bogumila",
    title = "{Conformal model for gravitational waves and dark matter: a status update}",
    eprint = "2210.07075",
    archivePrefix = "arXiv",
    primaryClass = "astro-ph.CO",
    doi = "10.1007/JHEP03(2023)007",
    journal = "JHEP",
    volume = "03",
    pages = "007",
    year = "2023"
}

@article{Kosowsky:1992rz,
    author = "Kosowsky, Arthur and Turner, Michael S. and Watkins, Richard",
    title = "{Gravitational waves from first order cosmological phase transitions}",
    reportNumber = "FERMILAB-PUB-91-333-A-REV, FERMILAB-PUB-91-333-A",
    doi = "10.1103/PhysRevLett.69.2026",
    journal = "Phys. Rev. Lett.",
    volume = "69",
    pages = "2026--2029",
    year = "1992"
}

@article{Kosowsky:1992vn,
    author = "Kosowsky, Arthur and Turner, Michael S.",
    title = "{Gravitational radiation from colliding vacuum bubbles: envelope approximation to many bubble collisions}",
    eprint = "astro-ph/9211004",
    archivePrefix = "arXiv",
    reportNumber = "FERMILAB-PUB-92-295-A",
    doi = "10.1103/PhysRevD.47.4372",
    journal = "Phys. Rev. D",
    volume = "47",
    pages = "4372--4391",
    year = "1993"
}

@article{Caprini:2007xq,
    author = "Caprini, Chiara and Durrer, Ruth and Servant, Geraldine",
    title = "{Gravitational wave generation from bubble collisions in first-order phase transitions: An analytic approach}",
    eprint = "0711.2593",
    archivePrefix = "arXiv",
    primaryClass = "astro-ph",
    reportNumber = "CERN-PH-TH-2007-206, SACLAY-T07-142",
    doi = "10.1103/PhysRevD.77.124015",
    journal = "Phys. Rev. D",
    volume = "77",
    pages = "124015",
    year = "2008"
}

@article{Huber:2008hg,
    author = "Huber, Stephan J. and Konstandin, Thomas",
    title = "{Gravitational Wave Production by Collisions: More Bubbles}",
    eprint = "0806.1828",
    archivePrefix = "arXiv",
    primaryClass = "hep-ph",
    doi = "10.1088/1475-7516/2008/09/022",
    journal = "JCAP",
    volume = "09",
    pages = "022",
    year = "2008"
}

@article{Hindmarsh:2013xza,
    author = "Hindmarsh, Mark and Huber, Stephan J. and Rummukainen, Kari and Weir, David J.",
    title = "{Gravitational waves from the sound of a first order phase transition}",
    eprint = "1304.2433",
    archivePrefix = "arXiv",
    primaryClass = "hep-ph",
    reportNumber = "HIP-2013-07-TH",
    doi = "10.1103/PhysRevLett.112.041301",
    journal = "Phys. Rev. Lett.",
    volume = "112",
    pages = "041301",
    year = "2014"
}

@article{Weir:2016tov,
    author = "Weir, David J.",
    title = "{Revisiting the envelope approximation: gravitational waves from bubble collisions}",
    eprint = "1604.08429",
    archivePrefix = "arXiv",
    primaryClass = "astro-ph.CO",
    doi = "10.1103/PhysRevD.93.124037",
    journal = "Phys. Rev. D",
    volume = "93",
    number = "12",
    pages = "124037",
    year = "2016"
}

@article{Giblin:2014qia,
    author = "Giblin, John T. and Mertens, James B.",
    title = "{Gravitional radiation from first-order phase transitions in the presence of a fluid}",
    eprint = "1405.4005",
    archivePrefix = "arXiv",
    primaryClass = "astro-ph.CO",
    doi = "10.1103/PhysRevD.90.023532",
    journal = "Phys. Rev. D",
    volume = "90",
    number = "2",
    pages = "023532",
    year = "2014"
}

@article{Hindmarsh:2015qta,
    author = "Hindmarsh, Mark and Huber, Stephan J. and Rummukainen, Kari and Weir, David J.",
    title = "{Numerical simulations of acoustically generated gravitational waves at a first order phase transition}",
    eprint = "1504.03291",
    archivePrefix = "arXiv",
    primaryClass = "astro-ph.CO",
    reportNumber = "HIP-2015-13-TH",
    doi = "10.1103/PhysRevD.92.123009",
    journal = "Phys. Rev. D",
    volume = "92",
    number = "12",
    pages = "123009",
    year = "2015"
}

@article{Hindmarsh:2016lnk,
    author = "Hindmarsh, Mark",
    title = "{Sound shell model for acoustic gravitational wave production at a first-order phase transition in the early Universe}",
    eprint = "1608.04735",
    archivePrefix = "arXiv",
    primaryClass = "astro-ph.CO",
    doi = "10.1103/PhysRevLett.120.071301",
    journal = "Phys. Rev. Lett.",
    volume = "120",
    number = "7",
    pages = "071301",
    year = "2018"
}

@article{Hindmarsh:2017gnf,
    author = "Hindmarsh, Mark and Huber, Stephan J. and Rummukainen, Kari and Weir, David J.",
    title = "{Shape of the acoustic gravitational wave power spectrum from a first order phase transition}",
    eprint = "1704.05871",
    archivePrefix = "arXiv",
    primaryClass = "astro-ph.CO",
    reportNumber = "HIP-2017-02-TH, HIP-2017-02/TH",
    doi = "10.1103/PhysRevD.96.103520",
    journal = "Phys. Rev. D",
    volume = "96",
    number = "10",
    pages = "103520",
    year = "2017",
    note = "[Erratum: Phys.Rev.D 101, 089902 (2020)]"
}

@article{Hindmarsh:2019phv,
    author = "Hindmarsh, Mark and Hijazi, Mulham",
    title = "{Gravitational waves from first order cosmological phase transitions in the Sound Shell Model}",
    eprint = "1909.10040",
    archivePrefix = "arXiv",
    primaryClass = "astro-ph.CO",
    reportNumber = "NORDITA-2019-083, HIP-2019-29/TH",
    doi = "10.1088/1475-7516/2019/12/062",
    journal = "JCAP",
    volume = "12",
    pages = "062",
    year = "2019"
}

@article{Kosowsky:2001xp,
    author = "Kosowsky, Arthur and Mack, Andrew and Kahniashvili, Tinatin",
    title = "{Gravitational radiation from cosmological turbulence}",
    eprint = "astro-ph/0111483",
    archivePrefix = "arXiv",
    reportNumber = "RAP-334",
    doi = "10.1103/PhysRevD.66.024030",
    journal = "Phys. Rev. D",
    volume = "66",
    pages = "024030",
    year = "2002"
}

@article{Dolgov:2002ra,
    author = "Dolgov, Alexander D. and Grasso, Dario and Nicolis, Alberto",
    title = "{Relic backgrounds of gravitational waves from cosmic turbulence}",
    eprint = "astro-ph/0206461",
    archivePrefix = "arXiv",
    doi = "10.1103/PhysRevD.66.103505",
    journal = "Phys. Rev. D",
    volume = "66",
    pages = "103505",
    year = "2002"
}

@article{Gogoberidze:2007an,
    author = "Gogoberidze, Grigol and Kahniashvili, Tina and Kosowsky, Arthur",
    title = "{The Spectrum of Gravitational Radiation from Primordial Turbulence}",
    eprint = "0705.1733",
    archivePrefix = "arXiv",
    primaryClass = "astro-ph",
    doi = "10.1103/PhysRevD.76.083002",
    journal = "Phys. Rev. D",
    volume = "76",
    pages = "083002",
    year = "2007"
}

@article{Caprini:2009yp,
    author = "Caprini, Chiara and Durrer, Ruth and Servant, Geraldine",
    title = "{The stochastic gravitational wave background from turbulence and magnetic fields generated by a first-order phase transition}",
    eprint = "0909.0622",
    archivePrefix = "arXiv",
    primaryClass = "astro-ph.CO",
    doi = "10.1088/1475-7516/2009/12/024",
    journal = "JCAP",
    volume = "12",
    pages = "024",
    year = "2009"
}

@article{Niksa:2018ofa,
    author = {Niksa, Peter and Schlederer, Martin and Sigl, G\"unter},
    title = "{Gravitational Waves produced by Compressible MHD Turbulence from Cosmological Phase Transitions}",
    eprint = "1803.02271",
    archivePrefix = "arXiv",
    primaryClass = "astro-ph.CO",
    doi = "10.1088/1361-6382/aac89c",
    journal = "Class. Quant. Grav.",
    volume = "35",
    number = "14",
    pages = "144001",
    year = "2018"
}

@article{Espinosa:2010hh,
    author = "Espinosa, Jose R. and Konstandin, Thomas and No, Jose M. and Servant, Geraldine",
    title = "{Energy Budget of Cosmological First-order Phase Transitions}",
    eprint = "1004.4187",
    archivePrefix = "arXiv",
    primaryClass = "hep-ph",
    reportNumber = "CERN-PH-TH-2010-027",
    doi = "10.1088/1475-7516/2010/06/028",
    journal = "JCAP",
    volume = "06",
    pages = "028",
    year = "2010"
}

@article{Planck:2018vyg,
    author = "Aghanim, N. and others",
    collaboration = "Planck",
    title = "{Planck 2018 results. VI. Cosmological parameters}",
    eprint = "1807.06209",
    archivePrefix = "arXiv",
    primaryClass = "astro-ph.CO",
    doi = "10.1051/0004-6361/201833910",
    journal = "Astron. Astrophys.",
    volume = "641",
    pages = "A6",
    year = "2020",
    note = "[Erratum: Astron.Astrophys. 652, C4 (2021)]"
}

@article{Weir:2017wfa,
    author = "Weir, David J.",
    title = "{Gravitational waves from a first order electroweak phase transition: a brief review}",
    eprint = "1705.01783",
    archivePrefix = "arXiv",
    primaryClass = "hep-ph",
    reportNumber = "HIP-2017-06-TH, HIP-2017-06/TH",
    doi = "10.1098/rsta.2017.0126",
    journal = "Phil. Trans. Roy. Soc. Lond. A",
    volume = "376",
    number = "2114",
    pages = "20170126",
    year = "2018",
    note = "[Erratum: Phil.Trans.Roy.Soc.Lond.A 381, 20230212 (2023)]"
}

@article{Steinhardt:1981ct,
    author = "Steinhardt, Paul Joseph",
    title = "{Relativistic Detonation Waves and Bubble Growth in False Vacuum Decay}",
    reportNumber = "UPR-0181T",
    doi = "10.1103/PhysRevD.25.2074",
    journal = "Phys. Rev. D",
    volume = "25",
    pages = "2074",
    year = "1982"
}

@article{Ellis:2020awk,
    author = "Ellis, John and Lewicki, Marek and No, Jos\'e Miguel",
    title = "{Gravitational waves from first-order cosmological phase transitions: lifetime of the sound wave source}",
    eprint = "2003.07360",
    archivePrefix = "arXiv",
    primaryClass = "hep-ph",
    reportNumber = "KCL-PH-TH/2020-04, CERN-TH-2020-016, IFT-UAM/CSIC-20-35",
    doi = "10.1088/1475-7516/2020/07/050",
    journal = "JCAP",
    volume = "07",
    pages = "050",
    year = "2020"
}

@article{Gowling:2021gcy,
    author = "Gowling, Chloe and Hindmarsh, Mark",
    title = "{Observational prospects for phase transitions at LISA: Fisher matrix analysis}",
    eprint = "2106.05984",
    archivePrefix = "arXiv",
    primaryClass = "astro-ph.CO",
    doi = "10.1088/1475-7516/2021/10/039",
    journal = "JCAP",
    volume = "10",
    pages = "039",
    year = "2021"
}

@article{Guo:2024gmu,
    author = "Guo, Huai-ke and Hajkarim, Fazlollah and Sinha, Kuver and White, Graham and Xiao, Yang",
    title = "{A precise fitting formula for gravitational wave spectra from the sound shell model}",
    eprint = "2407.02580",
    archivePrefix = "arXiv",
    primaryClass = "hep-ph",
    doi = "10.1088/1475-7516/2025/02/056",
    journal = "JCAP",
    volume = "02",
    pages = "056",
    year = "2025"
}

@article{Ben-Dayan:2019gll,
    author = "Ben-Dayan, Ido and Keating, Brian and Leon, David and Wolfson, Ira",
    title = "{Constraints on scalar and tensor spectra from $N_{eff}$}",
    eprint = "1903.11843",
    archivePrefix = "arXiv",
    primaryClass = "astro-ph.CO",
    doi = "10.1088/1475-7516/2019/06/007",
    journal = "JCAP",
    volume = "06",
    pages = "007",
    year = "2019"
}

@article{Anderson:1991zb,
    author = "Anderson, Greg W. and Hall, Lawrence J.",
    title = "{The Electroweak phase transition and baryogenesis}",
    reportNumber = "LBL-31169, UCB-PTH-91-41",
    doi = "10.1103/PhysRevD.45.2685",
    journal = "Phys. Rev. D",
    volume = "45",
    pages = "2685--2698",
    year = "1992"
}

@article{Espinosa:1992kf,
    author = "Espinosa, J. R. and Quirós, M. and Zwirner, F.",
    title = "{On the nature of the electroweak phase transition}",
    eprint = "hep-ph/9212248",
    archivePrefix = "arXiv",
    reportNumber = "CERN-TH-6577-92, IEM-FT-58-92",
    doi = "10.1016/0370-2693(93)90450-V",
    journal = "Phys. Lett. B",
    volume = "314",
    pages = "206--216",
    year = "1993"
}

@article{Dolan:1973qd,
    author = "Dolan, L. and Jackiw, R.",
    title = "{Symmetry Behavior at Finite Temperature}",
    reportNumber = "MIT-CTP-406",
    doi = "10.1103/PhysRevD.9.3320",
    journal = "Phys. Rev. D",
    volume = "9",
    pages = "3320--3341",
    year = "1974"
}

@inproceedings{Quiros:1999jp,
    author = "Quirós, Mariano",
    title = "{Finite temperature field theory and phase transitions}",
    booktitle = "{ICTP Summer School in High-Energy Physics and Cosmology}",
    eprint = "hep-ph/9901312",
    archivePrefix = "arXiv",
    reportNumber = "IEM-FT-187-99",
    pages = "187--259",
    month = "1",
    year = "1999"
}

@book{Laine:2016hma,
    author = "Laine, Mikko and Vuorinen, Aleksi",
    title = "{Basics of Thermal Field Theory}",
    eprint = "1701.01554",
    archivePrefix = "arXiv",
    primaryClass = "hep-ph",
    doi = "10.1007/978-3-319-31933-9",
    publisher = "Springer",
    volume = "925",
    year = "2016"
}

@article{Carrington:1991hz,
    author = "Carrington, M. E.",
    title = "{The Effective potential at finite temperature in the Standard Model}",
    reportNumber = "TPI-MINN-91-48-T-REV, TPI-MINN-91-48-T",
    doi = "10.1103/PhysRevD.45.2933",
    journal = "Phys. Rev. D",
    volume = "45",
    pages = "2933--2944",
    year = "1992"
}

@article{Arnold:1992rz,
    author = "Arnold, Peter Brockway and Espinosa, Olivier",
    title = "{The Effective potential and first order phase transitions: Beyond leading-order}",
    eprint = "hep-ph/9212235",
    archivePrefix = "arXiv",
    reportNumber = "UW-PT-92-18, USM-TH-60",
    doi = "10.1103/PhysRevD.47.3546",
    journal = "Phys. Rev. D",
    volume = "47",
    pages = "3546",
    year = "1993",
    note = "[Erratum: Phys.Rev.D 50, 6662 (1994)]"
}

@article{Parwani:1991gq,
    author = "Parwani, Rajesh R.",
    title = "{Resummation in a hot scalar field theory}",
    eprint = "hep-ph/9204216",
    archivePrefix = "arXiv",
    reportNumber = "ITP-SB-91-64",
    doi = "10.1103/PhysRevD.45.4695",
    journal = "Phys. Rev. D",
    volume = "45",
    pages = "4695",
    year = "1992",
    note = "[Erratum: Phys.Rev.D 48, 5965 (1993)]"
}

@article{Cline:2011mm,
    author = "Cline, James M. and Kainulainen, Kimmo and Trott, Michael",
    title = "{Electroweak Baryogenesis in Two Higgs Doublet Models and B meson anomalies}",
    eprint = "1107.3559",
    archivePrefix = "arXiv",
    primaryClass = "hep-ph",
    doi = "10.1007/JHEP11(2011)089",
    journal = "JHEP",
    volume = "11",
    pages = "089",
    year = "2011"
}

@article{Laine:2017hdk,
    author = "Laine, M. and Meyer, M. and Nardini, G.",
    title = "{Thermal phase transition with full 2-loop effective potential}",
    eprint = "1702.07479",
    archivePrefix = "arXiv",
    primaryClass = "hep-ph",
    doi = "10.1016/j.nuclphysb.2017.04.023",
    journal = "Nucl. Phys. B",
    volume = "920",
    pages = "565--600",
    year = "2017"
}

@article{Kainulainen:2019kyp,
    author = "Kainulainen, Kimmo and Keus, Venus and Niemi, Lauri and Rummukainen, Kari and Tenkanen, Tuomas V. I. and Vaskonen, Ville",
    title = "{On the validity of perturbative studies of the electroweak phase transition in the Two Higgs Doublet model}",
    eprint = "1904.01329",
    archivePrefix = "arXiv",
    primaryClass = "hep-ph",
    doi = "10.1007/JHEP06(2019)075",
    journal = "JHEP",
    volume = "06",
    pages = "075",
    year = "2019"
}

@article{Kainulainen:2021eki,
    author = "Kainulainen, Kimmo and Koskivaara, Olli",
    title = "{Non-equilibrium dynamics of a scalar field with quantum backreaction}",
    eprint = "2105.09598",
    archivePrefix = "arXiv",
    primaryClass = "hep-ph",
    doi = "10.1007/JHEP12(2021)190",
    journal = "JHEP",
    volume = "12",
    pages = "190",
    year = "2021"
}

@article{Delaunay:2007wb,
    author = "Delaunay, Cedric and Grojean, Christophe and Wells, James D.",
    title = "{Dynamics of Non-renormalizable Electroweak Symmetry Breaking}",
    eprint = "0711.2511",
    archivePrefix = "arXiv",
    primaryClass = "hep-ph",
    reportNumber = "CERN-PH-TH-2007-219, MCTP-07-31, SACLAY-T07-141",
    doi = "10.1088/1126-6708/2008/04/029",
    journal = "JHEP",
    volume = "04",
    pages = "029",
    year = "2008"
}

@article{Barni:2024lkj,
    author = "Barni, Giulio and Blasi, Simone and Vanvlasselaer, Miguel",
    title = "{The hydrodynamics of inverse phase transitions}",
    eprint = "2406.01596",
    archivePrefix = "arXiv",
    primaryClass = "hep-ph",
    doi = "10.1088/1475-7516/2024/10/042",
    journal = "JCAP",
    volume = "10",
    pages = "042",
    year = "2024"
}

@article{Kolesova:2023mno,
    author = "Kolesova, H. and Laine, M.",
    title = "{Update on gravitational wave signals from post-inflationary phase transitions}",
    eprint = "2311.03718",
    archivePrefix = "arXiv",
    primaryClass = "gr-qc",
    doi = "10.1016/j.physletb.2024.138553",
    journal = "Phys. Lett. B",
    volume = "851",
    pages = "138553",
    year = "2024"
}

@article{Traschen:1990sw,
    author = "Traschen, Jennie H. and Brandenberger, Robert H.",
    title = "{Particle Production During Out-of-equilibrium Phase Transitions}",
    reportNumber = "BROWN-HET-731",
    doi = "10.1103/PhysRevD.42.2491",
    journal = "Phys. Rev. D",
    volume = "42",
    pages = "2491--2504",
    year = "1990"
}

@article{Dolgov:1989us,
    author = "Dolgov, A. D. and Kirilova, D. P.",
    title = "{On Particle Creation by a Time Dependent Scalar Field}",
    reportNumber = "JINR-E2-89-321",
    journal = "Sov. J. Nucl. Phys.",
    volume = "51",
    pages = "172--177",
    year = "1990"
}

@article{Sharma:2023mao,
    author = "Sharma, Ramkishor and Dahl, Jani and Brandenburg, Axel and Hindmarsh, Mark",
    title = "{Shallow relic gravitational wave spectrum with acoustic peak}",
    eprint = "2308.12916",
    archivePrefix = "arXiv",
    primaryClass = "gr-qc",
    reportNumber = "NORDITA-2023-051, HIP-2023-13/TH",
    doi = "10.1088/1475-7516/2023/12/042",
    journal = "JCAP",
    volume = "12",
    pages = "042",
    year = "2023"
}

@article{RoperPol:2023dzg,
    author = "Roper Pol, Alberto and Procacci, Simona and Caprini, Chiara",
    title = "{Characterization of the gravitational wave spectrum from sound waves within the sound shell model}",
    eprint = "2308.12943",
    archivePrefix = "arXiv",
    primaryClass = "gr-qc",
    doi = "10.1103/PhysRevD.109.063531",
    journal = "Phys. Rev. D",
    volume = "109",
    number = "6",
    pages = "063531",
    year = "2024"
}

@article{Caprini:2024gyk,
    author = "Caprini, Chiara and Jinno, Ryusuke and Konstandin, Thomas and Roper Pol, Alberto and Rubira, Henrique and Stomberg, Isak",
    title = "{Gravitational waves from first-order phase transitions: from weak to strong}",
    eprint = "2409.03651",
    archivePrefix = "arXiv",
    primaryClass = "gr-qc",
    doi = "10.1007/JHEP07(2025)217",
    journal = "JHEP",
    volume = "07",
    pages = "217",
    year = "2025"
}

@article{Correia:2025qif,
    author = "Correia, Jos{\'e} and Hindmarsh, Mark and Rummukainen, Kari and Weir, David J.",
    title = "{Gravitational waves from strong first-order phase transitions}",
    eprint = "2505.17824",
    archivePrefix = "arXiv",
    primaryClass = "astro-ph.CO",
    doi = "10.1103/8wmq-f635",
    journal = "Phys. Rev. D",
    volume = "112",
    number = "12",
    pages = "123546",
    year = "2025"
}

@article{Lewicki:2019gmv,
    author = "Lewicki, Marek and Vaskonen, Ville",
    title = "{On bubble collisions in strongly supercooled phase transitions}",
    eprint = "1912.00997",
    archivePrefix = "arXiv",
    primaryClass = "astro-ph.CO",
    reportNumber = "KCL-PH-TH/2019-88",
    doi = "10.1016/j.dark.2020.100672",
    journal = "Phys. Dark Univ.",
    volume = "30",
    pages = "100672",
    year = "2020"
}

@article{Lewicki:2020jiv,
    author = "Lewicki, Marek and Vaskonen, Ville",
    title = "{Gravitational wave spectra from strongly supercooled phase transitions}",
    eprint = "2007.04967",
    archivePrefix = "arXiv",
    primaryClass = "astro-ph.CO",
    reportNumber = "KCL-PH-TH/2020-36",
    doi = "10.1140/epjc/s10052-020-08589-1",
    journal = "Eur. Phys. J. C",
    volume = "80",
    number = "11",
    pages = "1003",
    year = "2020"
}

@article{Lewicki:2022pdb,
    author = "Lewicki, Marek and Vaskonen, Ville",
    title = "{Gravitational waves from bubble collisions and fluid motion in strongly supercooled phase transitions}",
    eprint = "2208.11697",
    archivePrefix = "arXiv",
    primaryClass = "astro-ph.CO",
    doi = "10.1140/epjc/s10052-023-11241-3",
    journal = "Eur. Phys. J. C",
    volume = "83",
    number = "2",
    pages = "109",
    year = "2023"
}

@article{Lewicki:2020azd,
    author = "Lewicki, Marek and Vaskonen, Ville",
    title = "{Gravitational waves from colliding vacuum bubbles in gauge theories}",
    eprint = "2012.07826",
    archivePrefix = "arXiv",
    primaryClass = "astro-ph.CO",
    doi = "10.1140/epjc/s10052-021-09232-3",
    journal = "Eur. Phys. J. C",
    volume = "81",
    number = "5",
    pages = "437",
    year = "2021",
    note = "[Erratum: Eur.Phys.J.C 81, 1077 (2021)]"
}

@article{Ellis:2019oqb,
    author = "Ellis, John and Lewicki, Marek and No, Jos{\'e} Miguel and Vaskonen, Ville",
    title = "{Gravitational wave energy budget in strongly supercooled phase transitions}",
    eprint = "1903.09642",
    archivePrefix = "arXiv",
    primaryClass = "hep-ph",
    reportNumber = "KCL-PH-TH/2019-32, CERN-TH-2019-032, IFT-UAM/CSIC-19-32",
    doi = "10.1088/1475-7516/2019/06/024",
    journal = "JCAP",
    volume = "06",
    pages = "024",
    year = "2019"
}

@article{RoperPol:2019wvy,
    author = "Roper Pol, Alberto and Mandal, Sayan and Brandenburg, Axel and Kahniashvili, Tina and Kosowsky, Arthur",
    title = "{Numerical simulations of gravitational waves from early-universe turbulence}",
    eprint = "1903.08585",
    archivePrefix = "arXiv",
    primaryClass = "astro-ph.CO",
    reportNumber = "NORDITA-2019-024",
    doi = "10.1103/PhysRevD.102.083512",
    journal = "Phys. Rev. D",
    volume = "102",
    number = "8",
    pages = "083512",
    year = "2020"
}

@article{Kahniashvili:2020jgm,
    author = "Kahniashvili, Tina and Brandenburg, Axel and Gogoberidze, Grigol and Mandal, Sayan and Roper Pol, Alberto",
    title = "{Circular polarization of gravitational waves from early-Universe helical turbulence}",
    eprint = "2011.05556",
    archivePrefix = "arXiv",
    primaryClass = "astro-ph.CO",
    reportNumber = "NORDITA-2020-102",
    doi = "10.1103/PhysRevResearch.3.013193",
    journal = "Phys. Rev. Res.",
    volume = "3",
    number = "1",
    pages = "013193",
    year = "2021"
}

@article{RoperPol:2021xnd,
    author = "Roper Pol, Alberto and Mandal, Sayan and Brandenburg, Axel and Kahniashvili, Tina",
    title = "{Polarization of gravitational waves from helical MHD turbulent sources}",
    eprint = "2107.05356",
    archivePrefix = "arXiv",
    primaryClass = "gr-qc",
    reportNumber = "NORDITA-2021-062",
    doi = "10.1088/1475-7516/2022/04/019",
    journal = "JCAP",
    volume = "04",
    number = "04",
    pages = "019",
    year = "2022"
}

@article{Auclair:2022jod,
    author = "Auclair, Pierre and Caprini, Chiara and Cutting, Daniel and Hindmarsh, Mark and Rummukainen, Kari and Steer, Dani{\`e}le A. and Weir, David J.",
    title = "{Generation of gravitational waves from freely decaying turbulence}",
    eprint = "2205.02588",
    archivePrefix = "arXiv",
    primaryClass = "astro-ph.CO",
    reportNumber = "HIP-2021-35/TH",
    doi = "10.1088/1475-7516/2022/09/029",
    journal = "JCAP",
    volume = "09",
    pages = "029",
    year = "2022"
}

@article{Gouttenoire:2021jhk,
    author = "Gouttenoire, Yann and Servant, Geraldine and Simakachorn, Peera",
    title = "{Kination cosmology from scalar fields and gravitational-wave signatures}",
    eprint = "2111.01150",
    archivePrefix = "arXiv",
    primaryClass = "hep-ph",
    reportNumber = "DESY 21-134",
    month = "11",
    year = "2021"
}

@article{LISACosmologyWorkingGroup:2022jok,
    author = "Auclair, Pierre and others",
    collaboration = "LISA Cosmology Working Group",
    title = "{Cosmology with the Laser Interferometer Space Antenna}",
    eprint = "2204.05434",
    archivePrefix = "arXiv",
    primaryClass = "astro-ph.CO",
    reportNumber = "LISA CosWG-22-03, FERMILAB-PUB-22-349-SCD",
    doi = "10.1007/s41114-023-00045-2",
    journal = "Living Rev. Rel.",
    volume = "26",
    number = "1",
    pages = "5",
    year = "2023"
}

@article{Sesana:2019vho,
    author = "Sesana, Alberto and others",
    title = "{Unveiling the gravitational universe at $\mu$-Hz frequencies}",
    eprint = "1908.11391",
    archivePrefix = "arXiv",
    primaryClass = "astro-ph.IM",
    doi = "10.1007/s10686-021-09709-9",
    journal = "Exper. Astron.",
    volume = "51",
    number = "3",
    pages = "1333--1383",
    year = "2021"
}

@article{Thrane:2013oya,
    author = "Thrane, Eric and Romano, Joseph D.",
    title = "{Sensitivity curves for searches for gravitational-wave backgrounds}",
    eprint = "1310.5300",
    archivePrefix = "arXiv",
    primaryClass = "astro-ph.IM",
    doi = "10.1103/PhysRevD.88.124032",
    journal = "Phys. Rev. D",
    volume = "88",
    number = "12",
    pages = "124032",
    year = "2013"
}

@article{Caprini:2009fx,
    author = "Caprini, Chiara and Durrer, Ruth and Konstandin, Thomas and Servant, Geraldine",
    title = "{General Properties of the Gravitational Wave Spectrum from Phase Transitions}",
    eprint = "0901.1661",
    archivePrefix = "arXiv",
    primaryClass = "astro-ph.CO",
    doi = "10.1103/PhysRevD.79.083519",
    journal = "Phys. Rev. D",
    volume = "79",
    pages = "083519",
    year = "2009"
}

@article{Cai:2019cdl,
    author = "Cai, Rong-Gen and Pi, Shi and Sasaki, Misao",
    title = "{Universal infrared scaling of gravitational wave background spectra}",
    eprint = "1909.13728",
    archivePrefix = "arXiv",
    primaryClass = "astro-ph.CO",
    reportNumber = "IPMU19-0135, YITP-19-88",
    doi = "10.1103/PhysRevD.102.083528",
    journal = "Phys. Rev. D",
    volume = "102",
    number = "8",
    pages = "083528",
    year = "2020"
}

@article{Hook:2020phx,
    author = "Hook, Anson and Marques-Tavares, Gustavo and Racco, Davide",
    title = "{Causal gravitational waves as a probe of free streaming particles and the expansion of the Universe}",
    eprint = "2010.03568",
    archivePrefix = "arXiv",
    primaryClass = "hep-ph",
    doi = "10.1007/JHEP02(2021)117",
    journal = "JHEP",
    volume = "02",
    pages = "117",
    year = "2021"
}

@article{Domenech:2020kqm,
    author = "Dom{\`e}nech, Guillem and Pi, Shi and Sasaki, Misao",
    title = "{Induced gravitational waves as a probe of thermal history of the universe}",
    eprint = "2005.12314",
    archivePrefix = "arXiv",
    primaryClass = "gr-qc",
    reportNumber = "YITP-20-70, IPMU20-0053",
    doi = "10.1088/1475-7516/2020/08/017",
    journal = "JCAP",
    volume = "08",
    pages = "017",
    year = "2020"
}

@article{Baldes:2018emh,
    author = "Baldes, Iason and Garc\'ia-Cely, Camilo",
    title = "{Strong gravitational radiation from a simple dark matter model}",
    eprint = "1809.01198",
    archivePrefix = "arXiv",
    primaryClass = "hep-ph",
    reportNumber = "DESY 18-155, DESY-18-155",
    doi = "10.1007/JHEP05(2019)190",
    journal = "JHEP",
    volume = "05",
    pages = "190",
    year = "2019"
}

@article{Madge:2018gfl,
    author = "Madge, Eric and Schwaller, Pedro",
    title = "{Leptophilic dark matter from gauged lepton number: Phenomenology and gravitational wave signatures}",
    eprint = "1809.09110",
    archivePrefix = "arXiv",
    primaryClass = "hep-ph",
    reportNumber = "MITP/18-088",
    doi = "10.1007/JHEP02(2019)048",
    journal = "JHEP",
    volume = "02",
    pages = "048",
    year = "2019"
}

@article{Fabbrichesi:2020wbt,
    author = "Fabbrichesi, Marco and Gabrielli, Emidio and Lanfranchi, Gaia",
    title = "{The Dark Photon}",
    eprint = "2005.01515",
    archivePrefix = "arXiv",
    primaryClass = "hep-ph",
    doi = "10.1007/978-3-030-62519-1",
    month = "5",
    year = "2020"
}
\end{document}